\documentclass[aps,prb,superscriptaddress,letterpaper,amsmath,amssymb,twocolumn,footinbib,floatfix]{revtex4}

\usepackage{graphicx}
\usepackage[ansinew]{inputenc}
\usepackage{array}
\usepackage{color}
\usepackage{amsmath}
\usepackage{amsxtra}
\usepackage{amstext}
\usepackage{amssymb}
\usepackage{latexsym}
\usepackage{soul}

\usepackage{IEEEtrantools}
\usepackage{bm}
\usepackage[breaklinks=true]{hyperref}

\makeatletter
\newcommand\ztag[1]{%
\def\@currentlabel{#1}%
\gdef\tmp{%
\addtocounter{equation}{-1}%
\def\theequation{#1}}%
\aftergroup\aftergroup\aftergroup\aftergroup\aftergroup\aftergroup
\aftergroup\aftergroup\aftergroup\aftergroup\aftergroup\aftergroup
\aftergroup\aftergroup\aftergroup\aftergroup\aftergroup\aftergroup
\aftergroup\aftergroup\aftergroup\aftergroup\aftergroup\aftergroup
\aftergroup\aftergroup\aftergroup\aftergroup\aftergroup\aftergroup
\aftergroup
\tmp}
\makeatother

\newcommand\vb[1]{\mathbf{#1}}

\makeatletter
\newcommand\RedeclareMathOperator{%
  \@ifstar{\def\rmo@s{m}\rmo@redeclare}{\def\rmo@s{o}\rmo@redeclare}%
}
\newcommand\rmo@redeclare[2]{%
  \begingroup \escapechar\m@ne\xdef\@gtempa{{\string#1}}\endgroup
  \expandafter\@ifundefined\@gtempa
     {\@latex@error{\noexpand#1undefined}\@ehc}%
     \relax
  \expandafter\rmo@declmathop\rmo@s{#1}{#2}}
\newcommand\rmo@declmathop[3]{%
  \DeclareRobustCommand{#2}{\qopname\newmcodes@#1{#3}}%
}
\@onlypreamble\RedeclareMathOperator
\makeatother

\RedeclareMathOperator{\Re}{Re}
\RedeclareMathOperator{\Im}{Im}
\begin{document}

%===========================
%   Title, Author, Affiliation
%===========================

\title{Terahertz Spectroscopy of Semiconductor Microcavity Lasers I: Photon Lasers}

\author{M.~Em.~\surname{Spotnitz}}
\affiliation{Department of Physics, The University of Arizona, Tucson, AZ 85721}
\affiliation{Wyant College of Optical Sciences, The University of Arizona, Tucson, AZ 85721}

\author{N.H.~\surname{Kwong}}
\affiliation{Wyant College of Optical Sciences, The University of Arizona, Tucson, AZ 85721}

\author{R.~\surname{Binder}}
\affiliation{Wyant College of Optical Sciences, The University of Arizona, Tucson, AZ 85721}
\affiliation{Department of Physics, The University of Arizona, Tucson, AZ 85721}

% Activate to display a given date or no date (if empty), otherwise the current date is printed
\date{\today{}}

%======================================================
%   Abstract
%======================================================
\begin{abstract}
Semiconductor microcavities  can exhibit various macroscopic quantum phenomena, including Bose-Einstein condensation of polaritons,
 Bardeen-Cooper-Schrieffer (BCS) states of polaritons, and photon lasing (lasing with negligible Coulombic exciton effects). An important aspect of
 possible experimental identification of these states is a gap in the excitation spectrum (the BCS gap in the case of a polaritonic BCS state).
 Similar to the polaritonic BCS gap, a light-induced gap can  exist in photon lasers. Although polaritonic BCS states have been observed on the basis of
 spectroscopy in the vicinity of the laser frequency, the direct observation of polaritonic BCS gaps using light spectrally centered at or around the emission frequency has not been achieved.
 It has been conjectured that low-frequency (terahertz) spectroscopy should be able to identify such gaps. In this first of two studies, a theory aimed at identifying features of light-induced gaps in the
 linear
 terahertz spectroscopy of photon lasers is developed and numerically evaluated. It is shown that spectral features in the intraband conductivity, and therefore in the system's
 transmissivity and absorptivity can be related to the light-induced gap. For sufficiently small Drude damping this includes spectral regions of THz gain.
  A future study will generalize the present formalism to include Coulomb effects.
\end{abstract}

\maketitle

\section{Introduction}

Semiconductor microcavity lasers are examples of open-dissipative-pumped systems that undergo phase transitions to states with spontaneously broken symmetry. In the simple case of a so-called photon laser, i.e.\ a microcavity laser in which Coulombic exciton effects between the charge carriers (electrons and holes) are negligible, the interband polarization and light field in the cavity can be viewed as generalized order parameters of the symmetry-broken state. In addition to photon lasers, semiconductor microcavities  have been intensively studied because of a variety of intriguing polariton effects
(see, for example, Refs.\
\onlinecite{%
fan-etal.97pra,%
cao-etal.97,%
kuwata-gonokami-etal.97,%
kira-etal.99b,%
moskalenko-snoke.00,%
ciuti-etal.00,%
savvidis-etal.00,%
kwong-etal.01prl,%
baumberg-lagoudakis.05,%
balili-etal.06,%
balili-etal.07,%
keeling_collective_2007,%
schumacher-etal.07prb,%
bajoni-etal.08,%
berman-etal.08,%
berney-etal.08,%
amo-etal.09,%
timofeev-sanvitto.12,%
semkat-etal.09,%
kamide-ogawa.10,%
deng-etal.10,%
snoke-littlewood.10,%
liu-etal.15,%
schulze-etal.14,%
menard-etal.14,%
kamandardezfouli-etal.14,%
schmutzler-etal.15,%
leeuw-etal.16,%
hayenga-khajavikhan.17,%
kavokin-etal.17,%
bao-etal.19,%
carcamo-etal.20}).
In the low density regime, polaritons, quasi-particles comprised of excitons (bound electron-hole pairs) and photons (the cavity light field), have been found to undergo Bose-Einstein condensation
(for reviews, see, e.g., Refs.\ \onlinecite{deng-etal.10,moskalenko-snoke.00})
which can be viewed as a limiting case of lasing processes in a semiconductor microcavity. In the intermediate density regime, the concepts of excitonic and polaritonic Bardeen-Cooper-Schrieffer (BCS) states have long been discussed
(e.g.\
\onlinecite{%
comte-nozieres.82,%
keeling-etal.05,%
kremp-etal.08,%
semkat-etal.09,%
kamide-ogawa.10,%
byrnes-etal.10,%
combescot-shiau.15,%
hu-liu.20}).
As in the case of the photon laser, above the lasing threshold the interband polarization and the light field in the cavity can be viewed as generalized order parameters, because both are non-zero macroscopic fields that appear spontaneously (in practice as a result of an instability triggered by a fluctuation). One of the key features of a BCS state, be it in superconductivity or BCS generalizations to excitons or polaritons, is the existence of a gap in the excitation spectrum.

A recent experimental observation of a polariton laser in the BCS regime has been reported in Ref.\ \onlinecite{hu-etal.21}. An estimate of the polaritonic BCS gap given in that publication was later confirmed on the basis of a rigorous linear-response theory of the condensed many-particle state.\cite{binder-kwong.21} The value of the polariton BCS gap in the range of experimentally accessible pump powers was between approximately 1 and 10 meV. Due to the reflectivity stop band of the high-quality microcavity, whose width is on the order of 10 meV, direct interband spectroscopy (using light spectrally centered in the vicinity of the lasing frequency) is presently unable to observe polaritonic BCS gaps of less than 10 meV. Therefore, it has been conjectured in Ref.\ \onlinecite{hu-etal.21} that low-frequency terahertz (THz) spectroscopy should be more suitable for the observation of a polaritonic BCS gap, because the microcavity will not act as a resonator for light of that frequency. In other words, the THz spectroscopy can be performed on the semiconductor quantum well inside the cavity without any effect of the cavity on the THz field.

The formal and physical analogy between the theory of superconductivity and a semiconductor excited by a coherent optical field with frequency in the interband continuum (i.e.\ light frequency larger than the fundamental bandgap of the semiconductor, $E_g$), has been discussed as early as 1970; see Ref.\ \onlinecite{galitskii-etal.70}. In that analysis, the coherent field is a strong external field, strong enough to create gaps in the single-particle spectrum of the valence and conduction band.
While at a formal level the light-induced gaps in the bandstructure of a semiconductor with non-interacting electrons and holes are gaps in the single-particle spectrum, whereas in the BCS superconductor the gaps are in the excitation spectrum of the (correlated) many-particle state, it is helpful to view the light-induced gaps in analogy to the BCS gaps, since without interaction the excitation spectrum of the many-particle state is given by the single-particle excitation spectrum. These light-induced energy gaps in semiconductors excited by strong optical fields may be called Galitskii-Elesin gaps.
\cite{%
galitskii-etal.70,%
nishimura-nishimura.73,%
keldysh.95}
In semiconductor lasers, this effect is closely related to spectral hole burning.
In this paper, instead of an external coherent field, the gaps are created by the laser light field, which in turn is brought about by spontaneous symmetry breaking (the U(1) symmetry related to the phase of the coherent field).
Experimental signatures of spectral hole burning have been reported as early as 1979 in Ref.\ \onlinecite{patel-etal.79} and investigated theoretically, e.g.\ in Refs.\
\onlinecite{%
schmitt-rink-etal.88,%
paul-etal.92,%
henneberger-etal.92,%
meissner-etal.93}.

More recently, the opening of gaps in the  energy bands
has been investigated in the case of exciton polaritons in GaAs semiconductor microcavities \cite{yamaguchi-etal.15} and in bulk GaAs\cite{murotani-etal.19}.
The use of dressed electron-hole-photon states reveals the connection among these closely related effects, as well as other connected phenomena such as Rabi splitting or the Mollow triplet in resonance fluorescence.\cite{quochi-etal.98,berney-etal.08,byrnes-etal.10,horikiri-etal.16,yamaguchi-etal.15,binder-kwong.21}

\begin{figure}
	\centering
	\includegraphics[width=3in]{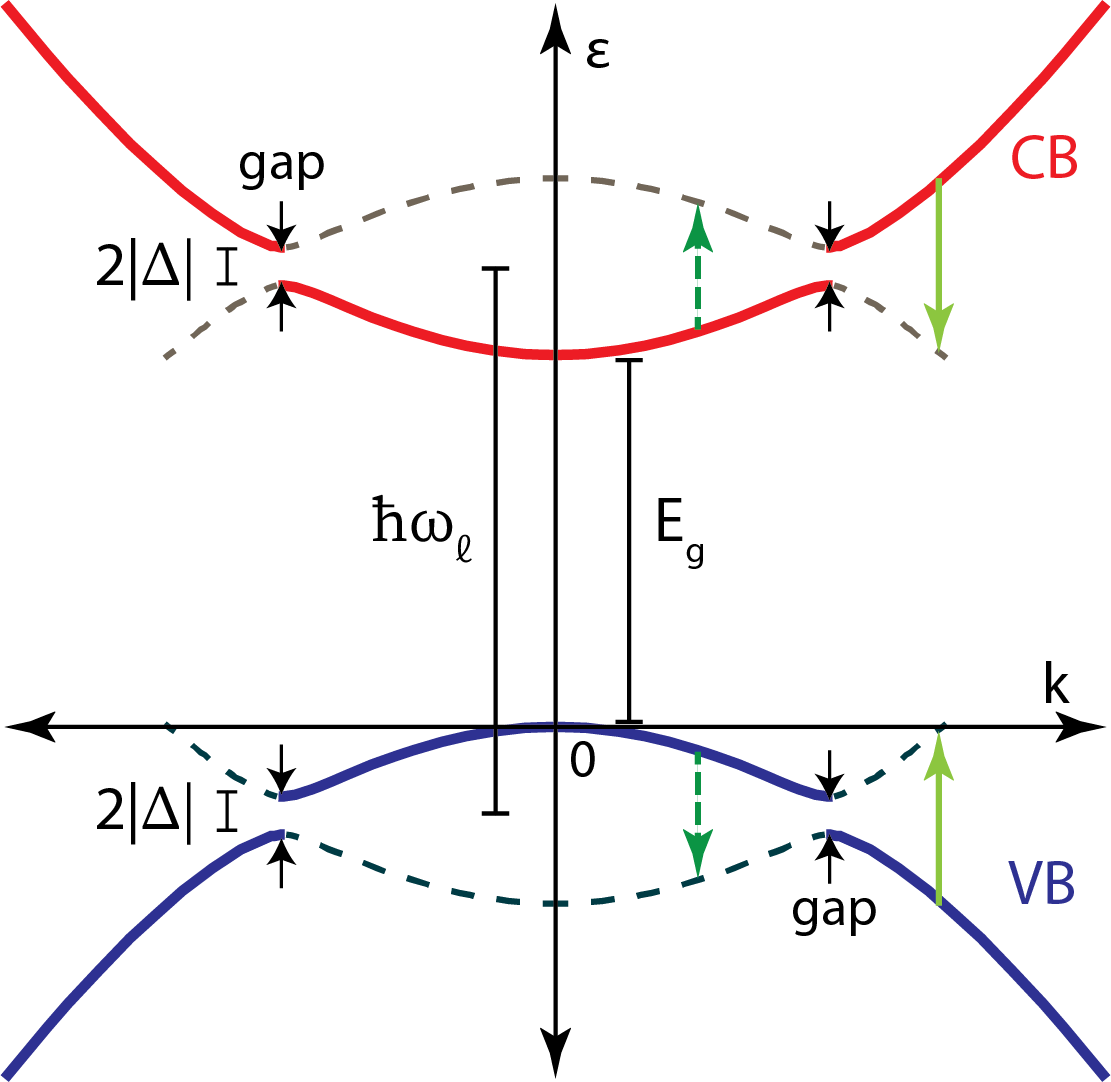}
	\caption{
		(Color online.)
		Schematic of the parabolic two-band band structure
		with the conduction band (CB) in red and the valence band (VB) in blue, renormalized by the light field (dressed bands). In this paper, we call the branches shown as solid lines original branches, and those shown as dashed lines light-induced branches. In the limit of vanishing light field, the original branches become the undressed bands and the light-induced branches vanish (cf.\ the spectral function in Eq.\ \eqref{eq:A_SS}).
		The light-induced gap of size $2 |\Delta_{\ell}|$  and the laser transition frequency  $\omega_{\ell}$ are indicated. Also indicated by green vertical arrows are two examples of vertical (in k-space), coherent transitions between the original and light-induced branches. In the case discussed in the main part of this paper (cf.\ Sec.\ \ref{sec:results}) THz absorption (dashed green upward arrow) can be overcompensated by THz gain (solid green downward arrow). But the resulting effective gain may be less than the THz absorption due to the Drude term (not indicated in this figure).}
	\label{fig:bandstruct}
\end{figure}

\begin{figure}
	\centering
	\includegraphics{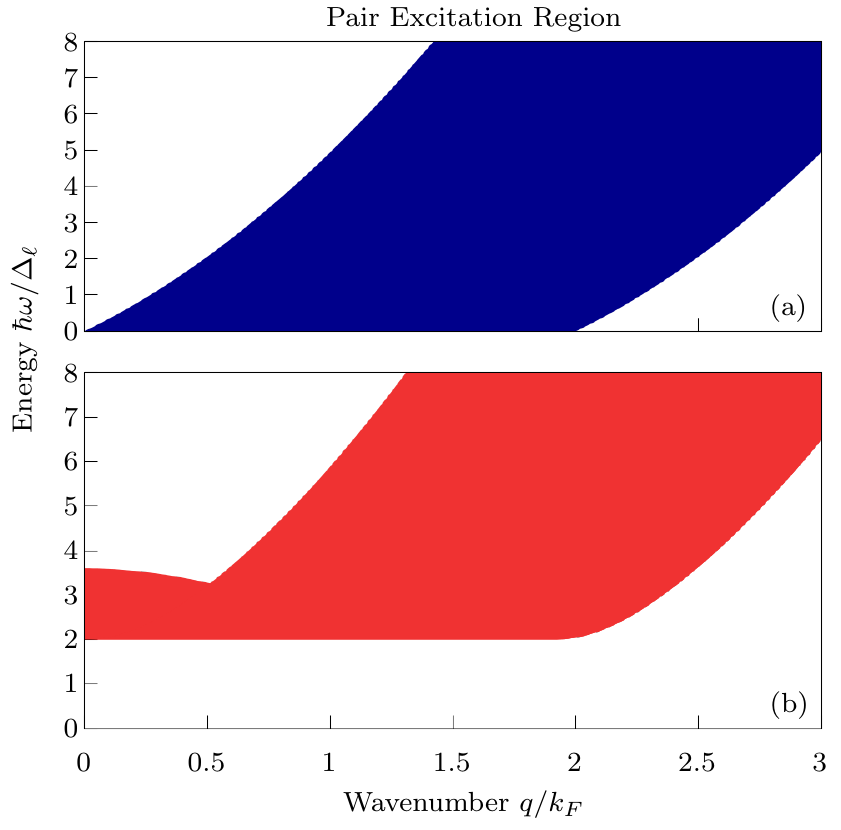}
	\caption{
		(Color online.)
		Schematic diagrams of the intraband pair-excitation region (PER). (a) The well-known PER for an electron gas (i.e. a single parabolic band without light-induced gaps) in the $T\to0$ limit in which the states up to the Fermi wave vector $k_F$ are occupied \cite{mahan.00}. (b)
		Schematic of the PER corresponding to Fig.\ \protect\ref{fig:bandstruct}, under the simplifying assumption
		that the electrons (and similarly the holes) occupy only the original branch below the light-induced gap
up to the wave vector at which the light-induced gap occurs, and that only transitions from the lower to the upper branches occur.
	}
	\label{fig:pair-excitation}
\end{figure}

In the following we will refer to light-induced gaps, schematically shown in Fig.\ \ref{fig:bandstruct},
as Galitskii-Elesin gaps if brought about by an external coherent light field, and as BCS-like gaps if brought about by the internal laser light field within the semiconductor microcavity, i.e.\ after spontaneous symmetry breaking above the laser threshold. In the idealized case of carriers occupying only states below the chemical potential up to the wave vector at which the light-induced gap occurs, and that only transitions from the lower to the upper branches occur,
the intraband pair excitation region assumes the form shown in Fig.\ \ref{fig:pair-excitation}b.
This is similar to  the well-known pair excitation region for a zero-temperature plasma
shown in Fig.\ \ref{fig:pair-excitation}a
(see for example Fig.\ 5.12 of Ref.\ \onlinecite{mahan.00}), but with a vertical shift due to the light-induced gap
(see Eqs.\ \eqref{eq:kfklper} and \eqref{eq:om1dckf} in Appx.\ \ref{sec:app:per}).
The presence of the the light induced branches (the dashed lines in Fig.\ \ref{fig:bandstruct}) enables  optical transitions with small and even zero wave vector $q$,
which, as we show below, can be probed in THz spectroscopy.

Terahertz and far-infrared radiation have been used in a wide range of spectroscopic measurements \cite{jepsen-etal.11}, including
 the observation of intra-excitonic and intraband transitions in semiconductors
\cite{gonokami-etal.04,gonokami.05,kira-etal.01,danielson-etal.07,kaindl-etal.09,kira-koch.11,teich-etal.14},
analysis of polariton condensates
\cite{menard-etal.14},
charge carrier dynamics
\cite{ulbricht-etal.11,%
kampfrath-etal.13},
nonlinear terahertz spectroscopy
\cite{maag-etal.16},
detection of Berry curvatures
\cite{virk-sipe.11},
and the spectroscopy of graphene
\cite{dawlaty-etal.08,rao-sipe.14}
and non-crystalline materials
\cite{wietzke-etal.09}.
Furthermore, terahertz gain and stimulated emission as a result of intra-excitonic transitions ($3p$-$2s$, $2p$-$1s$) have been shown in Refs.\ \onlinecite{kira-koch.04,huber-etal.06}, and THz excitation and possible THz lasing via stimulated emission from the $2p$ exciton to the $1s$ exciton-polariton
or, alternatively, between the the upper and lower $1s$-exciton polariton branches in semiconductor microcavities
have been studied in Refs.\
\onlinecite{%
	kavokin-etal.10,%
	valle-kavokin.11,%
	savenko-etal.11,%
	kavokin-etal.12,%
	tomaino-etal.12,%
	deliberato-etal.13,%
	schmutzler-etal.14,%
	huppert-etal.14,%
	lemenager-etal.14,%
	barachati-etal.15%
}.
	THz emission due to the Rabi splitting of a two-level system (a simpler version of the effect seen in the present paper) and optical triplet harmonics produced by optical pumping were
 proposed in Ref.\ \onlinecite{kibis-etal.09}.
Similar studies of THz emission for the cases of asymmetric quantum wells and dots were performed in Refs.\ \onlinecite{%
		shammah-etal.14,%
		chestnov-etal.17,%
		deliberato.18,%
		mandal-etal.19%
	}.
In the present paper, the intraband matrix element which permits dipole-allowed THz transitions derives not from an engineered asymmetrical microstructure nor from different excitonic states, but from the inherently asymmetric wavefunctions of those electron states with nonzero quasimomentum.

The objective of this first study, and a planned subsequent study, is to lay out the theory of intraband (as opposed to intra-exciton) THz spectroscopy for microcavity lasers. In the first study, we focus on the photon laser.
The photon laser is not only a mathematically convenient model; the limiting case of a semiconductor microcavity without Coulomb interaction has been related to a photonic BEC in Ref.\ \onlinecite{kamide-ogawa.10}.
Even though Coulombic exciton effects are neglected (only Coulombic scattering and relaxation effects are accounted for at a phenomenological level), the theory to be laid out below is relatively complex, owing to the interplay of the laser (condensed phase) with the THz linear response theory. However, because of the absence of Coulomb effects the results are still relatively intuitive and mathematically transparent. In particular, we show that spectral features in the laser's low-frequency conductivity (as well as transmissivity, reflectivity and absorptivity/gain) can be uniquely related to the BCS-like gap that the laser induces upon itself. Since we are using a single-time equation of motion technique, rather than two-time Green's functions that readily yield spectral functions, we clarify the connection between our results and the single-particle spectral function through a second formulation of our theory where we use an analog to the Nambu-space theory. We identify features in our theory with the polariton-BCS single-particle energy structure and spectral function \cite{yamaguchi-etal.15}, evaluated for the case of vanishing Coulomb potential. In a subsequent study we plan to extend the present formalism to include Coulomb and hence polariton effects.

This paper is organized as follows. In Sec.\ \ref{sec:theoretical-basis}, we formulate the theoretical basis for the dynamical behavior of a two-band semiconductor quantum well inside a single-mode microcavity, including both the laser field (at interband transition frequencies) and a weak THz probe field.
In Sec.\ \ref{sec:linear-terahertz-response}, we provide the general linear response theory for the THz field, formulated in terms of the populations and optical coherences in the undressed band picture. Further analysis is provided
in Sec.\ \ref{sec:nambu-basis}, where we present the linear THz response in a basis similar to the Nambu space in superconductivity, which can also be called  the dressed-band picture. This is helpful for interpreting the results from the general theory given in Sec.\ \ref{sec:linear-terahertz-response}.
In Sec.\ \ref{sec:results}, we present numerical results, analysis, and interpretation for the case of the photon laser, the case of an external coherent field, and the case of a microcavity laser in quasi-thermal equilibrium. In Sec.\ \ref{sec:conclusion}, we summarize our findings.

%===========================
%   THEORETICAL BASIS
%===========================
\section{Theoretical Basis}

\label{sec:theoretical-basis}

%===========================
%  Model
%===========================
Our model system is a semiconductor quantum well in a microcavity with a direct band gap and parabolic bands around $\vb{k}=0$. The parameters of the lowest conduction band and the highest heavy-hole valence band of GaAs, each having double spin degeneracy, are used here. The dynamical degrees of freedom are the conduction-band electrons, the valence-band holes, and the laser field in the cavity. The linear response of this system to an applied THz probe is considered. Only the coupling between the charged particles and the laser and THz fields are included in the interaction Hamiltonian. The two photon fields are treated as distinct, since they are spectrally well-separated. The laser field is assumed to be classical, neglecting the effects of its quantum fluctuations. The Coulomb interactions between the charged particles are also neglected. In the derived dynamical equations, the effects of dephasing, pumping, and dissipation are included via phenomenological gain/loss terms.

%===========================
% The Hamiltonian
%===========================
Our model Hamiltonian for the electrons, holes, and laser field photons is
\begin{align}
\hat{H} &= \sum_{\alpha, \vb{k}} \varepsilon_{\alpha \vb{k}} a_{\alpha \vb{k}}^{\dagger} a_{\alpha \vb{k}}
+ \sum_{\lambda \vb{q}} \hbar \omega_{\lambda\vb{q}} c_{\lambda \vb{q}}^{\dagger} c_{\lambda \vb{q}} \label{modelH}\\
&\quad + \sum_{\lambda e h \vb{q}, \vb{k}}
\left[ \Gamma_{eh}^{\lambda} (\vb{k} ,\vb{q}) c_{\lambda \vb{q}} a_{e, \mathbf{k}}^{\dagger} a_{h, \mathbf{q}-\mathbf{k}}^{\dagger} + h.c. \right] \nonumber \\
&\quad + \sum_{\nu \alpha \vb{q}, \vb{k}} g_{\alpha}^{\nu} ( \vb{k}, \vb{q} ) A_{T\nu} ( \vb{q} , t ) a_{\alpha , \mathbf{q}+\mathbf{k}}^{\dagger} a_{\alpha , \mathbf{k}}   \nonumber
\end{align}
where $a_{e \vb{k}}$, $a_{h \vb{k}}$, and $c_{\lambda \vb{q}}$ are the annihilation operators for a conduction band electron, a valence band hole, and a laser field photon respectively. $\vb{k}$ and $\vb{q}$ are 2D wavevectors in the quantum well's plane (all wavevectors in this paper are in-plane unless specified otherwise). $\lambda$ labels the optical photon spin. We consider interband transitions only between the highest heavy-hole valence band and the lowest conduction band. So the band subscripts label the degenerate spin orbitals: $e = \pm 1/2, h = \pm 3/2$. The subscript $\alpha$ runs through both electron and hole bands. Parabolic bands are used for the charges: $\varepsilon_{e \vb{k}} = \frac {\hbar^2 k^2} {2 m_e} + E_g$ and $\varepsilon_{h \vb{k}} = \frac {\hbar^2 k^2} {2 m_h}$, where $m_\alpha$ is the effective mass in band $\alpha$
(both $m_e$ and $m_h$ are positive on our case),
and $E_g$ is the band gap. $\omega_{\lambda \vb{q}}$ is the cavity resonance frequency.

We invoke the rotating wave approximation in setting the interband eh-laser interaction term.
While treated as an input parameter in the numerical calculations,
the interband coupling strength $\Gamma_{eh}^{\lambda} ( \vb{k} , \vb{q} )$
can be
 given by
$\Gamma_{eh}^{\lambda} ( \vb{k} , \vb{q} )
= -\left\vert \vb{d}_{\mathrm{c}\mathrm{v}} \left(\vb{k} , \vb{q} \right) \cdot \bm{\epsilon}_{\ell \lambda} \right\vert \Psi_{\mathrm{cav}} \left(z_{\mathrm{QW}} \right) \sqrt{2 \pi \hbar \omega_{\lambda \vb{q}}/\epsilon_{b}}$.
 The interband dipole moment is
$\vb{d}_{\mathrm{c}\mathrm{v}} \left(\vb{k} , \vb{q} \right) = i q_{e} \left\langle\mathrm{c}, \vb{k} + \vb{q} \right\vert \hat{\vb{p}}\left\vert \mathrm{v}, \vb{k} \right\rangle / ( m_0 \omega_{\lambda \vb{q}} )$, where $m_0$ is the free-space electron mass, $q_e$ is the magnitude  of the electron's charge ($q_e > 0$), and the states $\left\vert \mathrm{c}, \vb{k} \right\rangle$ ($\left\vert \mathrm{v}, \vb{k} \right\rangle$) in the electron momentum matrix element are conduction (valence) band Bloch wave functions. $\Psi_{\mathrm{cav}} \left(z_{\mathrm{QW}} \right)$ is the cavity photon mode wave function along the $z$ direction evaluated at the position of the quantum well $z_{\mathrm{QW}}$, $\epsilon_{b}$ is the background dielectric function inside the cavity, and $\bm{\epsilon}_{\ell \lambda}$ is the polarization unit vector of the optical field.
(Some nuances of the relation between the interband dipole and momentum matrix elements are discussed in Ref.\ \onlinecite{gu-etal.13}; see also Ref.\ \onlinecite{mahon-etal.19}.)
 The specific form of the interband matrix element governing the $e = \pm 1/2$ to $h = \pm 3/2$ transitions yields (see, e.g., Ref.\ \onlinecite{hu-etal.21})
$\vb{d}_{\mathrm{c}\mathrm{v}} \left(\vb{k} , \vb{q} \right) \cdot \bm{\epsilon}_{\ell \lambda} =
d_{\mathrm{c}\mathrm{v}} \left(\vb{k} , \vb{q} \right)
\delta_{|e+h|,1}
 \delta_{e+h,\lambda}$
where the angular momentum labels in the conduction-valence band picture are related to those in the electron-hole picture via
 $\mathrm{c}=e$ and $\mathrm{v}=-h$. In other words, for a given conduction band, the corresponding valence band and laser-photon polarization are fixed. We may call this form of selection rules ``circular selection rules.''
In our numerical evaluations, we approximate $\Gamma_{eh}^{\lambda} ( \vb{k} , \vb{q} )$ by its value at $\vb{k} = \vb{q} = \vb{0}$.

The THz probe is treated in the Hamiltonian Eq.\ \eqref{modelH} as a classical applied vector potential $\vb{A}_{T\nu} (\vb{q} , t )$.
$\nu$ denotes the THz field polarization. We use a gauge in which the scalar potential is zero \cite{mahon-etal.19} so that $\vb{E}_{T\nu} ( \vb{x} , t ) = - \tfrac{1}{c} \partial \vb{A}_{T\nu} ( \vb{x} , t ) / \partial t$, with $\vb{x}$ being the 3D spatial coordinates. The probe induces intraband transitions with the coupling $g_{\alpha}^{\nu}(\vb{k},\vb{q})$. With the approximation of isotropy and $\vb{q}\ll\vb{k}$, the coupling strength is evaluated as
\begin{equation}
g_{\alpha}^{\nu}(\vb{k},\vb{q}) \simeq
g_{\alpha}^{\nu} \left(\vb{k} + \tfrac{1}{2}\vb{q} \right) = - \frac{ s_\alpha q_{e} }{ m_{\alpha} c} \hbar \left(\vb{k}+  \tfrac{1}{2}\vb{q} \right) \cdot \bm{\epsilon}_{\nu},
\label{eq:gnuform}
\end{equation}
 where $\bm{\epsilon}_{\nu}$ is the polarization unit vector for the THz field and $s_\alpha$ is the sign of the particle's charge: $s_e = -1 , s_h = 1$.
Our study is restricted to so-called $s$-polarization, where the THz field is polarized normal to the plane of incidence, and thus in the plane of the quantum well and perpendicular to the 2D vector $\vb{q}$.

%===========================
%   Equations of Motion
%===========================

Using the Hamiltonian in Eq.\ \eqref{modelH} and the Heisenberg picture, we derive the single-time
 equations of motion of the interband polarization
$ p_{eh} (\vb{k}_1, \vb{k}_2 , t ) \equiv \left\langle a_{h, \vb{k}_2} (t) a_{e, \vb{k}_1} (t) \right\rangle$, the intraband density matrices $f_{\alpha} (\vb{k}_1, \vb{k}_2, t ) \equiv \left\langle a_{\alpha, \vb{k}_2}^{\dagger} (t) a_{\alpha, \vb{k}_1}  (t) \right\rangle$, and the laser field amplitude (the squared magnitude of which is the 2D photon density in the designated mode) $E_{\ell \lambda} ( \vb{q}, t ) \equiv ( 1 /
\sqrt{\cal{A}}
) \left\langle c_{\lambda \vb{q}}  (t) \right\rangle$,
where $\cal{A}$ is the system's cross-sectional area.
Our classical approximation of the photon field leads to the factorization of expectation values of products of photon and fermion operators, which closes the set of equations. The Hamiltonian Eq.\ \eqref{modelH} does not account for pumping, cavity loss via emission of the laser field,
 and for carrier scattering, relaxation and dephasing (via electron-electron and electron-phonon interaction). We include these effects phenomenologically
 in the same way as was done in Ref.\ \onlinecite{hu-etal.21} by extending the equations of motion for the interband polarization, carrier distributions and cavity laser field by appropriate incoherent terms that contain incoherent rates $\gamma$. We then obtain the equation of motion for the interband polarization,
\begin{IEEEeqnarray*}{l}
\left( i \hbar  \frac{\partial}{\partial t} - (\varepsilon_{e \vb{k}_1} + \varepsilon_{h \vb{k}_2}) \right) p_{eh} (\vb{k}_1, \vb{k}_2, t) \yesnumber \label{p_eh-1.equ}\\
\phantom{(i \hbar} = \sum_{\vb{k}^{\prime} \lambda} \left[ \Gamma_{eh}^{\lambda} (\vb{k}_1, \vb{k}_1 + \vb{k}^{\prime})  E_{\ell \lambda} (\vb{k}_1 + \vb{k}^{\prime}, t)   \right. \nonumber \\
\phantom{(i \hbar) = \sum_{\vb{k}^{\prime}\lambda} \left[ \Gamma_{eh}^{\lambda} \right. } \times
\left( \delta_{\vb{k}_2, \vb{k}^{\prime}} - f_{h} (\vb{k}_2, \vb{k}^{\prime}, t) \right)  \nonumber \\
\phantom{(i \hbar) = \sum_{\vb{k}^{\prime}}} \left. - \Gamma_{eh}^{\lambda} (\vb{k}^{\prime}, \vb{k}_2 + \vb{k}^{\prime}) f_{e} (\vb{k}_1, \vb{k}^{\prime}, t) E_{\ell \lambda} (\vb{k}_2 + \vb{k}^{\prime}, t) \right] \nonumber \\
\phantom{(i \hbar =} + \sum_{\vb{k}^{\prime}\nu} \left[ g_{h}^{\nu}\left( \vb{k}^{\prime},\vb{k}_{2}-\vb{k}^{\prime} \right) A_{T\nu} (\vb{k}_2 - \vb{k}^{\prime}, t) p_{eh} (\vb{k}_1, \vb{k}^{\prime}, t) \right. \nonumber\\
\phantom{(i \hbar) = \sum_{\vb{k}^{\prime}}} \left. + g_{e}^{\nu}\left(\vb{k}^{\prime},\vb{k}_{1}-\vb{k}^{\prime}\right)  A_{T\nu} (\vb{k}_1 - \vb{k}^{\prime}, t)   p_{eh} (\vb{k}^{\prime}, \vb{k}_2, t)  \right] \nonumber\\
\phantom{(i \hbar) =} + i \hbar \left. \frac {\partial p_{eh} (\vb{k}_1, \vb{k}_2, t)    } {\partial t} \right|_{\rm incoh},   \nonumber
\end{IEEEeqnarray*}
the equation for the electron distribution function,
\begin{IEEEeqnarray*}{l}
\left( i \hbar  \frac{\partial}{\partial t} - (\varepsilon_{e \vb{k}_1} - \varepsilon_{e \vb{k}_2}) \right) f_{e} (\vb{k}_1, \vb{k}_2, t) \yesnumber \label{f_e-1.equ}\\
\phantom{(i \hbar)} = \sum_{\vb{k}^{\prime} \lambda} \left[ \Gamma_{eh}^{\lambda} (\vb{k}_1, \vb{k}_1 + \vb{k}^{\prime}) p_{eh}^{\ast} (\vb{k}_2, \vb{k}^{\prime}, t) E_{\ell \lambda} (\vb{k}_1 + \vb{k}^{\prime}, t) \right. \nonumber \\
\phantom{(i \hbar) = \sum_{\vb{k}^{\prime}}} \left. - \Gamma_{eh}^{\lambda \ast} (\vb{k}_2, \vb{k}_2 + \vb{k}^{\prime}) p_{eh} (\vb{k}_1, \vb{k}^{\prime}, t) E_{\ell \lambda}^{\ast} (\vb{k}_2 + \vb{k}^{\prime}, t) \right] \nonumber \\
\phantom{(i \hbar) = } + \sum_{\vb{k}^{\prime} \nu} \left[
 g_{e}^{\nu}\left(\vb{k}^{\prime},\vb{k}_{1}-\vb{k}^{\prime}\right)  A_{T\nu} (\vb{k}_1 - \vb{k}^{\prime}, t)   f_{e} (\vb{k}^{\prime}, \vb{k}_2, t)
 \right.  \nonumber\\
\phantom{(i \hbar) = \sum_{\vb{k}^{\prime} \lambda} \left[\right.} \left.
- g_{e}^{\nu}\left(\vb{k}_{2},\vb{k}^{\prime}-\vb{k}_{2}\right)   A_{T\nu} (\vb{k}^{\prime} - \vb{k}_2, t) f_{e} (\vb{k}_1, \vb{k}^{\prime}, t)
\right] \nonumber\\
\phantom{(i \hbar) = }  + i \hbar \left. \frac {\partial f_{e} (\vb{k}_1, \vb{k}_2, t)    } {\partial t} \right|_{\rm incoh} ,  \nonumber
\end{IEEEeqnarray*}
the equation for the hole distribution function,
\begin{IEEEeqnarray*}{l}
\left( i \hbar  \frac{\partial}{\partial t} - (\varepsilon_{h \vb{k}_1} - \varepsilon_{h \vb{k}_2}) \right) f_{h} (\vb{k}_1, \vb{k}_2, t) \yesnumber \label{f_h-1.equ}\\
\phantom{(i \hbar)} = \sum_{\vb{k}^{\prime} \lambda} \left[
\Gamma_{eh}^{\lambda} (\vb{k}^{\prime}, \vb{k}_1 + \vb{k}^{\prime}) p_{eh}^{\ast} (\vb{k}^{\prime}, \vb{k}_2, t) E_{\ell \lambda} (\vb{k}_1 + \vb{k}^{\prime}, t) \right. \nonumber \\
\phantom{(i \hbar) = \sum_{\vb{k}^{\prime}}}
 - \Gamma_{eh}^{\lambda \ast} (\vb{k}^{\prime}, \vb{k}_2 + \vb{k}^{\prime}) p_{eh} (\vb{k}^{\prime}, \vb{k}_1, t) E_{\ell \lambda}^{\ast} (\vb{k}_2 + \vb{k}^{\prime}, t) \nonumber \\
\phantom{(i \hbar) =} + \sum_{\vb{k}^{\prime}\nu} \left[
 g_{h}^{\nu}\left(\vb{k}^{\prime},\vb{k}_{1}-\vb{k}^{\prime}\right) A_{T\nu} (\vb{k}_1 - \vb{k}^{\prime}, t) f_{h} (\vb{k}^{\prime}, \vb{k}_2, t)
 \right. \nonumber\\
\phantom{(i \hbar) = \sum_{\vb{k}^{\prime} \lambda}} \left.
 - g_{h}^{\nu}\left(\vb{k}_{2},\vb{k}^{\prime}-\vb{k}_{2}\right) A_{T\nu} (\vb{k}^{\prime} - \vb{k}_2, t) f_{h} (\vb{k}_1, \vb{k}^{\prime}, t)
 \right] \nonumber\\
\phantom{(i \hbar) =}
 + i \hbar \left. \frac {\partial f_{h} (\vb{k}_1, \vb{k}_2, t)    } {\partial t} \right|_{\rm incoh} ,  \nonumber
\end{IEEEeqnarray*}
and the single-mode equation for the cavity field,
\footnote{An alternative to the single-mode equation for the cavity field based on the
	propagation of the light through the entire microcavity structure has been
	given in Ref.\ \protect\onlinecite{carcamo-etal.20}.}
\begin{IEEEeqnarray}{rCl}
\left[ i \hbar \frac{\partial}{\partial t} - \hbar \omega_{\lambda\vb{q}} \right]  E_{\ell \lambda} (\vb{q}, t)
&=& \sum_{eh \vb{k}} \Gamma_{eh}^{\lambda \ast} (\vb{k}, \vb{q}) p_{eh} (\vb{k}, \vb{q}-\vb{k}, t) \nonumber\\
&& + i \hbar \left. \frac {\partial   E_{\ell \lambda} (\vb{q}, t)     } {\partial t} \right|_{\rm incoh} .
\yesnumber \label{E_ell-1.equ}
\end{IEEEeqnarray}
The specific form of the incoherent (dephasing, relaxation, and pump) contributions will be chosen appropriately for the laser and discussed in Appx.\ \ref{appx:laserdist}.

In Eqs.\ \eqref{modelH} and \eqref{p_eh-1.equ}--\eqref{E_ell-1.equ}, the general form of the intraband matrix element, $g_{\alpha}^{\nu}(\vb{k},\vb{q})$, is used. Hence, these equations are valid for general systems that may be anisotropic. As mentioned above, the specific form of the intraband matrix element using effective masses and isotropic parabolic bands, $g_{\alpha}^{\nu}(\vb{k}+\tfrac{1}{2}\vb{q})$ in Eq.\ \eqref{eq:gnuform}, is used below and in the numerical evaluation.

\begin{figure}
\centering
\includegraphics{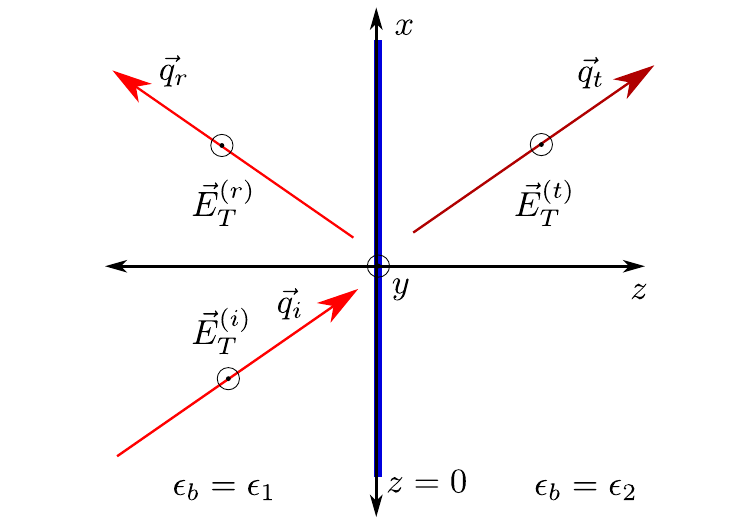}
\caption{Geometry of the THz probe wave, used in the derivation of Eqs.\ \eqref{eq:Tqxomega} and \eqref{eq:beta_def}, and in Appx.\ \ref{appx:THz_transmission}. The quantum well is taken to be at $z=0$. The cavity material in the vicinity of the QW is taken to be dielectric with relative permittivity $\epsilon_{b}$.}
\label{fig:wavevecs}
\end{figure}

Eqs.\ \eqref{p_eh-1.equ}--\eqref{E_ell-1.equ} describe the (two-dimensional) dynamics inside the quantum well. We choose a simple setting, shown in Fig.\ \ref{fig:wavevecs}, for the propagation of the external THz probe. The zero-width quantum well lies between two dielectric media. The medium on the incident (transmitted) side has a dielectric constant of $\epsilon_1$ ($\epsilon_2$). A coordinate system is set up in which the $z$ axis is normal to the quantum well's plane, the plane of incidence is the $x$-$z$ plane, and the quantum well is placed at $z=0$.
The THz probe field propagates according to Maxwell's equations with the intraband current in the quantum well as a source field:
\begin{multline}
 \left[\frac{n^{2}(z)}{c^{2}} \frac{\partial^{2}}{\partial t^{2}}+\nabla \times \nabla \times\right] \vb{E}_{T} \left(\mathbf{r} , t \right) \label{ETHzwaveeq} \\
 =  \frac{4 \pi}{c^2} \delta (z) \sum_{\alpha, \vb{q}}  \frac{\partial}{\partial t} \vb{J}_{\alpha} \left(\vb{q} , t \right) e^{i \vb{q} \cdot \vb{r}_{\|}}
\end{multline}
where $\vb{r} = (x, y, z)$ and $\vb{r}_{\|} = (x, y)$. In our gauge, $E_{T\nu} \left(\mathbf{r} , t \right) = - \tfrac{1}{c} \partial A_{T\nu} \left(\mathbf{r} , t \right) / \partial t $. $n(z)$ is the background refractive index of the cavity material, $n(z) = \sqrt{\epsilon_1}$ for $z < 0$, and $= \sqrt{\epsilon_2}$ for $z > 0$.
The vector components of the two-dimensional current $J_{\alpha\nu} = \vb{J}_{\alpha} \cdot \bm{\epsilon}_{\nu}$ are limited
to the plane of the quantum well (note the delta function in $z$) and consist of a paramagnetic part $J_\alpha^{p \nu}$ and a diamagnetic part $J_\alpha^{d \nu}$,
\begin{align}
J_{\alpha\nu} \left(\vb{q} , t \right) &= J_{\alpha\nu}^{p} \left(\vb{q} , t \right) + J_{\alpha\nu}^{d} \left(\vb{q} , t \right) \label{current-def.equ} \\
J_{\alpha\nu}^{p} \left(\vb{q} , t \right) &= \frac{1}{\cal{A}} \sum_{\vb{k}} \frac {s_\alpha \hbar q_e} {m_\alpha} \left( \vb{k} + \tfrac{1}{2} \vb{q} \right) \cdot \bm{\epsilon}_{\nu} f_{\alpha} \left( \vb{k} + \vb{q} , \vb{k} , t \right) \nonumber\\
J_{\alpha\nu}^{d} \left(\vb{q} , t \right) &=
- \frac{1}{\cal{A}} \sum_{\vb{k} \vb{q}^{\prime}} \frac{q_e^2}{m_{\alpha} c} A_{T\nu} \left(\vb{q}^{\prime} \right) f_\alpha \left( \vb{k} +\vb{q} , \vb{k} +\vb{q}^{\prime} , t \right) \nonumber
\end{align}
The use of the effective mass $m_{\alpha}$ in the diamagnetic current has been shown in Ref.\ \onlinecite{steiner.08} to be a consequence of including the THz field in the Hamiltonian interband transition term in the more fundamental Bloch wave function model.

%%% Perturbation expansion
Eqs.\ \eqref{p_eh-1.equ}--\eqref{current-def.equ} form a closed set of equations for the combined system of the laser and the THz probe. Of interest is the laser's linear response to the THz field. To obtain this,
Eqs.\ \eqref{p_eh-1.equ}--\eqref{E_ell-1.equ} are expanded up to first order in the THz field. The zeroth order steady state fields, which describe the unperturbed steady-state laser, are discussed in Appx.\ \ref{appx:laserdist}. The properties of the first order equations and solutions are further developed in the next several sections. Included here are some remarks on the steady-state laser solution and the relation of the first-order current to the THz reflectivity, transmissivity, and absorptivity.

We denote the perturbation order of the solution by a superscript.
The laser emission being primarily normal to the QW, the zeroth-order optical field is set to have zero in-plane momentum, $E_{\ell \lambda}^{(0)} (\vb{q} , t) = \delta_{\vb{q} \vb{0}} E_{\ell \lambda}^{(0)} ( t )$  This implies a corresponding momentum restriction on the zeroth order density matrix and interband polarization, $f_{\alpha}^{(0)} (\vb{k}_1, \vb{k}_2, t) = \delta_{\vb{k}_2 , \vb{k}_1} f_{\alpha}^{(0)} (\vb{k}_1, t) , \alpha = e , h$, and $p_{eh}^{(0)} (\vb{k}_1, \vb{k}_2, t) = \delta_{\vb{k}_2 , - \vb{k}_1} p_{eh}^{(0)} (\vb{k}_1, t)$. In the  expressions for the zeroth order fields, we remove the momentum labels made redundant by the delta functions. When the laser reaches a monochromatic steady state, the occupation $f_{\alpha}^{(0)} (\vb{k})$ is time-independent while the interband polarization $p_{eh}^{(0)} (\vb{k}, t)$ and the light field $E_{\ell \lambda}^{(0)} ( t )$ oscillate at a lasing frequency $\omega_{\ell}$.

The first-order current $J_{\alpha}^{\nu(1)} \left(\vb{q} , t \right)$ induced by the THz field is given by Eq.\ \eqref{current-def.equ} with the density matrix $f_{\alpha}$ set equal to its first-order component $f^{(1)}_{\alpha}$.
In the frequency domain, the THz field and the total current induced by it are related by the conductivity matrix
\begin{multline}
	\sum_{e} J_{e\nu}^{(1)}(\vb{q},\omega)  + \sum_{h} J_{h\nu}^{(1)}(\vb{q},\omega) \\
	=\sum_{\nu'} \sigma_{T\nu}^{\nu^{\prime}}(\vb{q},\omega) \tilde{E}_{T\nu^{\prime}}(\vb{q},\omega)
	\label{eq:thz-cond-def}
\end{multline}
$\tilde{E}_{T\nu} (\vb{q}, \omega )$ is the time-frequency Fourier transform of $E_{T\nu} (\vb{q} , t)$, which is the THz electric field in the quantum well's plane,
\begin{equation}
	E_{T\nu} (\vb{q} , t) = \int d \vb{r}_{\|} e^{-i \vb{q} \cdot \vb{r}_{\|}} E_{T\nu} (\vb{r}_{\|} , z = 0 , t). \nonumber
\end{equation}
Using spatial symmetry arguments
	and those expressions given in Sec.\ \ref{sec:linear-terahertz-response} which determine the conductivity matrix,
one can show that the off-diagonal elements of $\sigma_{T\nu}^{\nu^{\prime}}$ vanish.
Therefore, in the remainder of the paper, only the diagonal terms are considered. For brevity they are denoted by $\sigma_{T\nu}^{\nu} \equiv \sigma_{T\nu}$.
Like the current density, the conductivity can be written as a sum of a paramagnetic term $\sigma_{T\nu}^{p} \left(\vb{q}, \omega \right) = [ \sum_{e} J_{e\nu}^{p (1)} \left(\vb{q} , \omega \right) + \sum_{h} J_{h\nu}^{p (1)} \left(\vb{q} , \omega \right) ] / \tilde{E}_{T\nu} \left(\vb{q} , \omega \right)$ and a diamagnetic term $\sigma_{T\nu}^{d} \left(\vb{q}, \omega \right) = [ \sum_{e} J_{e\nu}^{d (1)} \left(\vb{q} , \omega \right) + \sum_{h} J_{h\nu}^{d (1)} \left(\vb{q} , \omega \right) ]/ \tilde{E}_{T\nu} \left(\vb{q} , \omega \right)$.

The outgoing (reflected and transmitted) THz waves are given in terms of the conductivity. We quote the result here, the derivation being given in Appx.\ \ref{appx:THz_transmission}. For an $s$-polarized probe, the electric field points along the $y$ axis in our coordinate system ($\bm{\epsilon}_{\nu} = \hat{y}$). The intensities (magnitudes of Poynting vectors) of the incident, reflected, and transmitted THz beams are denoted by $I^{(i)}_y$, $I^{(r)}_y$, and $I^{(t)}_y$, respectively. The transmissivity $| T |^2$, reflectivity $| R |^2$, and absorptivity $A$ are given by
\begin{IEEEeqnarray}{rCl}
| T \left( q_x, \omega \right) |^2 &\equiv& \frac{I^{(t)}_{y}}{I^{(i)}_{y}} = \sqrt{\frac {\epsilon_2} {\epsilon_1}} \left| \frac{2}{1 + \beta \left( q_x, \omega \right)} \right|^2 \label{eq:Tqxomega} \\
| R \left( q_x, \omega \right) |^2 &\equiv& \frac{I^{(r)}_{y}}{I^{(i)}_{y}} = \left| \frac{1 - \beta \left( q_x, \omega \right)}{1 + \beta \left( q_x, \omega \right)} \right|^2 \yesnumber \label{eq:Rqxomega} \\
A \left( q_x, \omega \right) &=& 1 - |T \left( q_x, \omega \right)|^2 - |R \left( q_x, \omega \right)|^2  \yesnumber \label{eq:Aqxomega} \\
&=& \frac{4 \left( {\rm Re} [\beta \left( q_x, \omega \right)] - \sqrt{\epsilon_2 / \epsilon_1} \right)}{\left\vert 1 + \beta \left( q_x, \omega \right) \right\vert^2} \nonumber
\end{IEEEeqnarray}
where
\begin{IEEEeqnarray}{rCl}
\beta \left( q_x, \omega \right) &=& \frac{1}{q_{ i z}} \left( q_{t z} + \frac{4 \pi \omega}{c^2} \sigma_{Ty} \left( q_x, \omega \right) \right) .
\label{eq:beta_def} \yesnumber
\end{IEEEeqnarray}
$q_x$ is the $x$ component of the wave vector common to the incident and the two outgoing waves, and $q_{iz}$ and $q_{tz}$ are the $z$ components of the wave vectors of the incident wave and transmitted wave respectively.

%===========================
% THz response $\vb{q}, \omega$ symmetries
%===========================
Because the model system is symmetric under coordinate inversions in the plane of the quantum well, i.e.\ symmetric under the transformation $(x,y)\to(-x,-y)$, the conductivity has the symmetry $\sigma_{T\nu}(\vb{q},\omega)=\sigma_{T\nu}(-\vb{q},\omega)$.
The THz conductivity $\sigma_{T\nu}(\vb{q},\omega)$ is the Fourier transform of the conductivity in real space, $\sigma_{T\nu}(\vb{r},t)$. The real-space conductivity is real, as it is the ratio of two real quantities, $J_{\alpha\nu}(\vb{r},t)$ and $E_{T\nu}(\vb{r},t)$. Therefore, the Fourier transform of the conductivity obeys the relation $\sigma_{T\nu}(\vb{q},\omega)=\sigma_{T\nu}^{\ast}(-\vb{q},-\omega)$. Combining these two symmetries gives a symmetry in $\omega$ space, $\sigma_{T\nu}(\vb{q},\omega)=\sigma_{T\nu}^{\ast}(\vb{q},-\omega)$. In polar coordinates in the complex plane, this is $|\sigma_{T\nu}|(\vb{q},\omega)=|\sigma_{T\nu}|(\vb{q},-\omega)$ and $\arg\sigma_{T\nu}(\vb{q},\omega)=-\arg\sigma_{T\nu}(\vb{q},-\omega)$.
The quantities $R$ and $T$ are functionals of $\sigma_{T\nu}$, and so obey the same symmetries.
Since $A$, $|R|^2$, and $|T|^2$ are real quantities, they all have the symmetry of the form $A(\vb{q},\omega)=A (\vb{q},-\omega)$.

%===========================
%   First order equations
%===========================
\section{Linear Terahertz Response}
\label{sec:linear-terahertz-response}

The first order (in the THz field) terms in the expansion of Eqs.\ \eqref{p_eh-1.equ}--\eqref{E_ell-1.equ} dictate the time evolution of the material's linear response to the THz probe.
Being interested in the spectral decomposition of the response, we lay out below the first order equations in frequency space.
Since the zeroth order quantities are assumed to be monochromatic (see Eq.\ \eqref{eq:laserosc}), with lasing frequency $\omega_{\ell}$, it is convenient to adopt the following convention in the time Fourier transform of the variables,
\begin{IEEEeqnarray*}{rCl}
p^{(1)}_{eh} \left(\vb{k}_1, \vb{k}_2, t \right) &=& \int \frac{d \omega}{2\pi} e^{-i \left(\omega + \omega_{\ell} \right) t} \tilde{p}^{(1)}_{eh}  (\vb{k}_1, \vb{k}_2, \omega) \\
E^{(1)}_{\ell \lambda}  \left(\vb{q}, t\right) &=& \int \frac{\mathrm{d} \omega}{2\pi} e^{-i \left(\omega + \omega_{\ell} \right) t} \tilde{E}_{\ell \lambda}^{(1)}  (\vb{q}, \omega) \\
f_{\alpha}^{(1)} \left(\vb{k}_1, \vb{k}_2, t\right) &=& \int \frac{d \omega}{2\pi} e^{-i \omega t} \tilde{f}_{\alpha}^{(1)}   (\vb{k}_1, \vb{k}_2, \omega), \ \  \alpha = e, h \\
E_{T\nu} \left(\vb{q}, t\right) &=& \int \frac{d \omega}{2\pi} e^{-i \omega t} \tilde{E}_{T\nu} (\vb{q}, \omega) .
\end{IEEEeqnarray*}
The frequency-space THz electric field is related to the vector potential by $\tilde{A}_{T\nu} (\vb{q} , \omega) = \frac{c}{i \omega} \tilde{E}_{T\nu} (\vb{q} , \omega)$.
The first-order equations of the frequency-space variables are the
the equation for the first-order interband polarization,
\begin{IEEEeqnarray*}{l}
\left[\hbar \omega - \left( \Delta \varepsilon (\vb{k}, \vb{q} - \vb{k} ) - \hbar \omega_{\ell} \right) + i \gamma_{p} \right]
       \tilde{p}_{e h}^{(1)} (\vb{k}, \vb{q} - \vb{k}, \omega ) \label{eq:pehFT}\\
    \phantom{[\hbar \omega} =
    \sum_{\lambda}
 \left\{
 - \left[ \tilde{f}_e^{(1)} (\vb{k}, \vb{k} - \vb{q}, \omega ) \Gamma_{eh}^{\lambda} (\vb{k} - \vb{q}, \vb{0}) \right.
 \right.
  \yesnumber \label{eq:P1ehfinc}
   \\
\phantom{[\hbar \omega = \sum_{\lambda}\{-[}
  \left. + \tilde{f}_h^{(1)} (\vb{q} - \vb{k}, -\vb{k}, \omega )
 \Gamma_{eh}^{\lambda} (\vb{k}, \vb{0}) \right] \tilde{E}_{\ell\lambda}^{(0)}
     \\
\phantom{[\hbar \omega =}  \left.+ \left[ 1 - f_h^{(0)} (\vb{q} - \vb{k} ) - f_e^{(0)} (\vb{k}) \right]
 \Gamma_{eh}^{\lambda} (\vb{k}, \vb{q} ) \tilde{E}_{\ell \lambda}^{(1)} (\vb{q}, \omega) \right\}
       \\
\phantom{[\hbar \omega =} +  \sum_{\nu} \left\{ \left[ \tilde{p}^{(0)}_{eh} \left(\vb{k} - \vb{q}\right)
 g_{e}^{\nu}\left(\vb{k} - \tfrac{1}{2} \vb{q}\right) \right. \right.
       \\
\phantom{[\hbar \omega = +\sum_{\nu}[} \left.\left. + \tilde{p}^{(0)}_{eh} \left(\vb{k}\right)
 g_{h}^{\nu}\left(\tfrac{1}{2}\vb{q}-\vb{k}\right) \right] \frac{c}{i \omega}
 \tilde{E}_{T\nu} \left(\vb{q}, \omega\right)
\right\},
\end{IEEEeqnarray*}
the equation for the first-order electron distribution function,
\begin{IEEEeqnarray*}{l}
\left[\hbar \omega -  \Delta \varepsilon_{e} (\vb{k}, \vb{k} - \vb{q} ) + i\gamma_f \right] \tilde{f}_{e}^{(1)} (\vb{k}, \vb{k} - \vb{q}, \omega ) \yesnumber \label{eq:f1eFT} \\
\phantom{[\hbar \omega}
 =  \sum_{\lambda} \left\{ \tilde{p}^{(0) \ast}_{eh} \left( \vb{k} - \vb{q} \right) \Gamma_{eh}^{\lambda} \left( \vb{k}, \vb{q} \right) \tilde{E}_{\ell \lambda}^{(1)} \left (\vb{q}, \omega \right)   \right.\\
\phantom{[\hbar \omega = \sum_{\lambda} \left\{\right.}
 - \tilde{p}^{(0)}_{eh} \left (\vb{k} \right) \Gamma_{eh}^{\lambda \ast} \left( \vb{k} - \vb{q}, - \vb{q} \right) \tilde{E}^{(1) \ast}_{\ell \lambda} \left( - \vb{q}, -\omega \right)
 \\
\phantom{[\hbar \omega = \sum_{\lambda} \left\{\right.}
 + \tilde{p}_{e h}^{(1) \ast} \left( \vb{k} - \vb{q}, - \vb{k}, - \omega \right)  \Gamma_{eh}^{\lambda} \left( \vb{k}, \vb{0} \right) \tilde{E}_{\ell\lambda}^{(0)}  \\
\phantom{[\hbar \omega = \sum_{\lambda} \left\{\right.}   \left.
 - \tilde{p}_{e h}^{(1)} \left( \vb{k}, \vb{q} - \vb{k}, \omega \right) \Gamma_{eh}^{\lambda \ast} \left( \vb{k} - \vb{q}, \vb{0} \right) \tilde{E}_{\ell\lambda}^{(0)\ast}       \right\} \\
\phantom{[\hbar \omega =}
 + \sum_{\nu} \left\{  \left[ f_e^{(0)} \left( \vb{k} - \vb{q} \right) - f_{e}^{(0)} \left( \vb{k} \right) \right]  \right. \nonumber \\
\phantom{[\hbar \omega = +\sum_{\nu} \left\{\right.}
\left.
\times g_{e}^{\nu}\left(\vb{k} - \tfrac{1}{2}\vb{q} \right) \frac{c}{i \omega} \tilde{E}_{T\nu} \left( \vb{q}, \omega \right)
\right\},
\end{IEEEeqnarray*}
the equation for the first-order hole distribution function,
\begin{IEEEeqnarray*}{l}
\left[\hbar \omega -  \Delta \varepsilon_{h} \left(  \vb{q} - \vb{k}, -\vb{k} \right) + i \gamma_f \right] \tilde{f}_{h}^{(1)} \left(  \vb{q} - \vb{k}, -\vb{k}, \omega \right) \yesnumber \label{eq:f1hFT} \\
\phantom{[\hbar \omega} =
\sum_{\lambda} \left\{
\tilde{p}^{(0) \ast}_{eh} \left(\vb{k} \right)
\Gamma_{eh}^{\lambda} \left( \vb{k}, \vb{q} \right) \tilde{E}_{\ell \lambda}^{(1)} \left (\vb{q}, \omega \right)
\right.
\\
\phantom{[\hbar \omega =\sum_{\lambda}} - \tilde{p}^{(0)}_{eh} \left (\vb{k} - \vb{q} \right) \Gamma_{eh}^{\lambda \ast} \left( \vb{k} - \vb{q}, - \vb{q} \right) \tilde{E}^{(1) \ast}_{\ell \lambda} \left( - \vb{q}, -\omega \right) \\
\phantom{[\hbar \omega =\sum_{\lambda}}+ \tilde{p}_{e h}^{(1) \ast} \left( \vb{k} - \vb{q}, - \vb{k},  - \omega \right) \Gamma_{eh}^{\lambda} \left( \vb{k} - \vb{q}, \vb{0} \right) \tilde{E}_{\ell\lambda}^{(0)}  \\
\phantom{[\hbar \omega =\sum_{\lambda}\left\{\right.}\left. - \tilde{p}_{e h}^{(1)} \left( \vb{k},  \vb{q} - \vb{k}, \omega \right) \Gamma_{eh}^{\lambda \ast} \left( \vb{k}, \vb{0} \right)  \tilde{E}_{\ell\lambda}^{(0)\ast} \right\} \\
\phantom{[\hbar \omega =} + \sum_{\nu} \left\{ \left[ f_{h}^{(0)} \left( -\vb{k} \right) - f_{h}^{(0)} \left(\vb{q} - \vb{k} \right) \right] \right. \nonumber \\
\phantom{[\hbar \omega = + \sum_{\nu} \left\{ \right.}
\left.
\times g_{h}^{\nu}\left(\tfrac{1}{2}\vb{q}-\vb{k}\right) \frac{c}{i \omega} \tilde{E}_{T\nu} \left( \vb{q}, \omega \right)
\right\}
\end{IEEEeqnarray*}
and the equation for the first-order cavity field,
\begin{multline}
\left[ \hbar \omega - \hbar \omega_{\lambda\vb{0}} + \hbar \omega_{\ell} + i \gamma_{E} \right] \tilde{E}_{\ell \lambda}^{(1)} \left( \vb{q}, \omega \right)  \label{eq:E1P1kqmk} \\
= \sum_{\vb{k} e h} \Gamma_{eh}^{\lambda \ast} (\vb{k}, \vb{q} ) \tilde{p}_{e h}^{(1)} \left( \vb{k}, \vb{q} - \vb{k}, \omega \right) .
\end{multline}
Here, $\Delta \varepsilon \left( \vb{k}, \vb{q} - \vb{k} \right) = \varepsilon_{e \vb{k}} + \varepsilon_{h \left( \vb{q} - \vb{k} \right) }$ and $\Delta \varepsilon_{\alpha} \left( \vb{k}_1, \vb{k}_2 \right) = \varepsilon_{\alpha \vb{k}_1} - \varepsilon_{\alpha \vb{k}_2 }$, with $\alpha = e, h$.

Eqs.\ \eqref{eq:pehFT}--\eqref{eq:f1hFT} can be formally solved as a system of matrix equations.
To simplify the algebra, the angular momenta $e$, $h$, and $\lambda$ are chosen to be one of the two sets of values that satisfy the circular selection rules.
Then, by eliminating the $\tilde{f}_{\alpha}^{(1)}$ from Eq.\ \eqref{eq:P1ehfinc} using Eqs.\ \eqref{eq:f1eFT}--\eqref{eq:f1hFT}, $\tilde{p}_{eh}^{(1)}$, $\tilde{E}_{\ell\lambda}^{(1)}$, and $ \tilde{E}_{T\nu}$ can be related by the $2\times2$ matrix equation:
\begin{IEEEeqnarray}{rCl}
\hat{p}_{eh}^{(1)} &=& M^{-1} N \hat{E}_{\ell\lambda}^{(1)} + M^{-1} \sum_{\nu}  Q^{\nu} \hat{E}_{T\nu}, \label{eq:PEtmat}
\end{IEEEeqnarray}
where
\begin{IEEEeqnarray}{rCl}
\hat{p}_{eh}^{(1)} &=& \begin{pmatrix} \tilde{p}_{e h}^{(1)} (\vb{k}, \vb{q} - \vb{k}, \omega) \\
           \tilde{p}_{e h}^{(1) \ast} (\vb{k}-\vb{q}, -\vb{k}, -\omega) \end{pmatrix}, \nonumber\\[4pt]
\hat{E}_{\ell\lambda}^{(1)} &=& \begin{pmatrix} \tilde{E}_{\ell \lambda}^{(1)}  (\vb{q}, \omega) \\
\tilde{E}^{(1) \ast}_{\ell \lambda} (-\vb{q}, -\omega) \end{pmatrix}, \nonumber\\[4pt]
\text{and} \quad
\hat{E}_{T\nu} &=& \begin{pmatrix} \tilde{E}_{T\nu} \left(\vb{q}, \omega\right) \\
  \tilde{E}_{T\nu}^{\ast} \left(-\vb{q}, -\omega\right)  \end{pmatrix} . \yesnumber\label{eq:ETmatdef}
\end{IEEEeqnarray}
The $2\times2$ matrices $M$, $N$, and $Q^{\nu}$ are functions only of the zeroth order quantities, and their formulae are given in Eqs.\ \eqref{eq:Mmat}--\eqref{eq:Qmat}, respectively, in Appx.\ \ref{appx:ehbasismatrices}.
Since the THz field $E_{T\nu}(\vb{r},t)$ is real, expressions with the Fourier transformed THz field can be written using the symmetry $\tilde{E}_{T\nu}(\vb{q},\omega)=\tilde{E}_{T\nu}^{\ast}(-\vb{q},-\omega)$.

Eq.\ \eqref{eq:E1P1kqmk} can also be written as a $2\times2$ matrix equation that relates $\hat{E}_{\ell\lambda}^{(1)}$ and $\hat{p}_{eh}^{(1)}$:
\begin{equation}
J \hat{E}_{\ell\lambda}^{(1)} - \sum_{\vb{k}} C \hat{p}_{eh}^{(1)} = 0
\label{eq:jecp}
\end{equation}
where $J$ and $C$ are defined in Eqs.\ \eqref{eq:Jdef} and \eqref{eq:Cdef}, respectively.
Applying $\sum_{\vb{k}} C$ to Eq.~\eqref{eq:PEtmat}
gives
\begin{equation*}
\left( J - \sum_{\vb{k}} C M^{-1} N \right) \hat{E}_{\ell\lambda}^{(1)} = \sum_{\nu} \left(\sum_{\vb{k}} C M^{-1} Q^{\nu}\right) \hat{E}_{T\nu},
\end{equation*}
so
\begin{multline}
\hat{E}_{\ell\lambda}^{(1)} = \sum_{\nu^{\prime}} \left( J - \sum_{\vb{k}} C M^{-1} N \right)^{-1} \\
\times  \left( \sum_{\vb{k}} C M^{-1} Q^{\nu^{\prime}} \right) \hat{E}_{T\nu^{\prime}}
\equiv \sum_{\nu^{\prime}} D^{\nu^{\prime}} \left(\vb{q}, \omega\right) \hat{E}_{T\nu^{\prime}}. \label{eq:Dmatdef}
\end{multline}
Substituting Eq.\ \eqref{eq:Dmatdef} into Eq.\ \eqref{eq:PEtmat} yields
\begin{multline}
\hat{p}_{eh}^{(1)} = \sum_{\nu^{\prime}} \left( M^{-1} N D^{\nu^{\prime}} + M^{-1} Q^{\nu^{\prime}} \right)\hat{E}_{T\nu^{\prime}} \\
\equiv \sum_{\nu^{\prime}} X^{\nu^{\prime}} \left(\vb{k}, \vb{q}, \omega\right) \hat{E}_{T\nu^{\prime}} .
\end{multline}
Finally, with the definition,
\begin{equation}
\hat{f}^{(1)}
\equiv
\begin{pmatrix}
\tilde{f}^{(1)}_e (\vb{k}, \vb{k} - \vb{q}, \omega)
\\
\tilde{f}^{(1)}_h (\vb{q} - \vb{k}, - \vb{k}, \omega)
\end{pmatrix} \label{eq:f1matdef}
\end{equation}
the $\tilde{f}_{\alpha}^{(1)}$ in Eqs.\ \eqref{eq:f1eFT} and \eqref{eq:f1hFT} can be written in matrix form as
\begin{multline}
\hat{f}^{(1)} = \sum_{\nu^{\prime}} \Xi   \left[ G D^{\nu^{\prime}}  + H X^{\nu^{\prime}}   + L \right]  \hat{E}_{T\nu^{\prime}} \\
\equiv \sum_{\nu^{\prime}} V^{\nu^{\prime}}(\vb{k}, \vb{q}, \omega) \hat{E}_{T\nu^{\prime}} .
\label{eq:f1ETmatform}
\end{multline}
The formulae for the component matrices of $V$ are given in Eqs.\ \eqref{eq:Gmat}--\eqref{eq:Ximat}.

The conductivity is calculated from Eqs.\ \eqref{current-def.equ} and \eqref{eq:thz-cond-def} as
\begin{IEEEeqnarray*}{rCl}
\sigma_{T\nu}^{\mathrm{p}} \left(\vb{q},\omega\right) &=& \frac{c}{\tilde{E}_{T\nu} \left(\vb{q},\omega\right) \mathcal{A}}  \yesnumber \label{eq:condthz} \\
&\times &  \sum_{\vb{k}} \left[
\sum_{h} g_{h}^{\nu} \left(\vb{k}-\tfrac{1}{2}\vb{q}\right) \tilde{f}_{h}^{(1)} \left(\vb{q}-\vb{k},-\vb{k},\omega\right)\right. \\
&& \phantom{E_T} \left.  - \sum_{e} g_{e}^{\nu} \left(\vb{k}-\tfrac{1}{2}\vb{q}\right) \tilde{f}_{e}^{(1)} \left(\vb{k}, \vb{k} - \vb{q}, \omega\right) \right]
\end{IEEEeqnarray*}
and
\begin{multline}
\sigma_{T\nu}^{\mathrm{d}} \left(\vb{q},\omega\right) =  \frac{i q_{e}^{2}}{\omega_{\vb{q}} + i \gamma_{D}}  \\
 \times  \frac{1}{\mathcal{A}} \sum_{\vb{k}} \left[ \sum_e \frac{f_{e}^{(0)} \left(\vb{k}\right)}{m_e} + \sum_h \frac{f_{h}^{(0)} \left(-\vb{k}\right)}{m_h} \right]. \label{eq:Drudecond}
\end{multline}
The THz polarization $\nu$ enters the paramagnetic THz conductivity $\sigma_{T\nu}^{\mathrm{p}} \left(\vb{q},\omega\right)$ explicitly in Eq.\ \eqref{eq:condthz} through the factors $g_{\alpha}^{\nu}$.
Additionally, a sum over all THz polarizations occurs in Eq.\ \eqref{eq:f1ETmatform} and enters Eq.\ \eqref{eq:condthz} through the $f_{\alpha}^{(1)}$.
However, as stated below Eq.\ \eqref{eq:thz-cond-def}, the $\nu^{\prime} \neq \nu$ contributions vanish from the integral over $\vb{k}$ due to the spatial symmetries of the integrand, leaving only $\nu$-diagonal contributions to the conductivity.
Eq.\ \eqref{eq:Drudecond} is the diamagnetic or Drude conductivity,
with the Drude scattering rate $\gamma_{D}$ included phenomenologically.\cite{kaindl-etal.09,ulbricht-etal.11}

In general, Eq.\ \eqref{eq:condthz} has to be evaluated numerically, and we show results in Sec.\ \ref{sec:results}.
However, in the limiting case of
an e-h plasma at zero temperature and not coupled to a cavity, so that the laser field $\tilde{E}_{\ell\lambda}^{(0)}$ and interband polarization $\tilde{p}_{eh}^{(0)}$ are absent and only the Lindhard contributions to $\tilde{f}_{\alpha}^{(1)}$ remain (see Eq.\ \eqref{eq:lindrespmatdef}),
one can obtain an analytical result for the paramagnetic conductivity. This is useful in order to check numerical results against analytical results in at least one limiting case. The analytical result for the Lindhard term is given in Appx.\ \ref{appx:Lindhard}.

%===========================
%  Nambu Basis
%===========================

\section{THz Response in the Nambu Basis}

\label{sec:nambu-basis}

The linear response theory developed in Sec.\ \ref{sec:linear-terahertz-response} provides a general framework for the THz response, and we will show below in Sec.\ \ref{sec:results} that it indeed predicts features that can be interpreted as signatures of light-induced and BCS-like gaps schematically shown in Fig.\ \ref{fig:bandstruct}. Since, however, this theory is a single-time theory, one might ask whether the relation between the results of the single-time formalism
and Fig.\ \ref{fig:bandstruct} can be further substantiated. The usual approach to single-particle spectral functions involved two-time Green's functions, and their evaluation for the case of excitonic BCS states and light-induced gaps has been discussed, e.g.,
in Refs.\ \onlinecite{jahnke-henneberger.92,kremp-etal.08,yamaguchi-etal.15}. In those works, the relation to conventional concepts of superconductivity are evident.
The purpose of the present section is to elucidate the relationship between the single-time response theory and conventional concepts of the theory by reformulating the single-time theory. The reformulation utilizes an approach analogous to the Nambu space (or Nambu spinor representation) approach frequently used in condensed matter theories. Using this approach, we can relate the frequency and wave vector-dependent intraband response function $f_{e}^{(1)} \left(\vb{k}, \vb{k}-\vb{q}, t\right)$ to the single-particle spectral function, and hence identify features in the frequency-dependent conductivity as resulting from gaps in the spectral function.

The THz linear response equations, Eq.\ \eqref{eq:P1ehfinc} for $\tilde{p}_{eh}^{(1)}$ and Eqs.\ \eqref{eq:f1eFT}--\eqref{eq:f1hFT} for  $f_{\alpha}^{(1)}, \alpha = e , h$ in Section  \ref{sec:linear-terahertz-response},  are written in the frequency domain. For clarity, we temporarily transform them back to the time domain.
If it is approximated that $\gamma_p = \gamma_f \equiv \gamma$, then the equations can be written as a single $2\times2$ matrix equation,
\begin{IEEEeqnarray}{l}
\left[ i \hbar \frac{\partial}{\partial t}  + i \gamma \right]  \tilde{D}^{(1)} \left(\vb{k}, \vb{q}, t\right) \yesnumber \label{eq:nb2} \\
\phantom{i \hbar \frac{\partial}{\partial t}} = h \left(\vb{k}\right) \tilde{D}^{(1)} \left(\vb{k}, \vb{q}, t\right)  - \tilde{D}^{(1)} \left(\vb{k}, \vb{q}, t\right) h \left(\vb{k} - \vb{q}\right)  \nonumber  \\
\phantom{i \hbar \frac{\partial}{\partial t} = }+ M_{\ell} \left(\vb{k} , \vb{q} , t\right) + \sum_{\nu} M_{T\nu} \left(\vb{k} , \vb{q} , t\right) \nonumber
\end{IEEEeqnarray}
The $p^{(1)}_{e h}$ and $f^{(1)}_{\alpha}$ are now represented by the matrix
\begin{equation}
\tilde{D}^{(1)} \left( \vb{k}, \vb{q}, t \right) \equiv \begin{pmatrix}
f_{e}^{(1)} \left(\vb{k}, \vb{k}-\vb{q}, t\right) & \tilde{p}_{eh}^{(1)} \left( \vb{k}, \vb{q} - \vb{k}, t\right)  \\
\tilde{p}_{eh}^{(1) \ast} \left( \vb{k} - \vb{q}, - \vb{k}, t\right) & - f_{h}^{(1)} \left(  \vb{q} - \vb{k}, -\vb{k}, t \right)
\end{pmatrix} .
\label{eq:nb1}
\end{equation}
where $p_{eh}^{(1)} \left( \vb{k}_1, \vb{k}_2, t\right)  = \tilde{p}_{eh}^{(1)} \left( \vb{k}_1, \vb{k}_2, t\right)  e^{-i \omega_{\ell} t}$, $\omega_{\ell}$ being the laser frequency.
The matrix $h \left(\vb{k}\right)$ is
\begin{equation}
h \left(\vb{k}\right) = \begin{pmatrix} \xi_{e} \left(\vb{k}\right) & \Delta \left(\vb{k}\right) \\
\Delta^{\ast} \left(\vb{k}\right) & -\xi_{h} \left(-\vb{k}\right) \end{pmatrix}
\label{eq:nb3}
\end{equation}
where
\begin{IEEEeqnarray}{rCl}
\xi_{\alpha} &=& \varepsilon_{\alpha} \left(\vb{k}\right) - \frac{\hbar \omega_{\ell}}{2}, \quad \alpha = e, h \label{eq:nbenrg} \\
\Delta \left(\vb{k}\right) &=& \Gamma_{eh}^{\lambda} \left(\vb{k}, \vb{0}\right) \tilde{E}_{\ell\lambda}^{(0)} \yesnumber\label{eq:deltak}
\end{IEEEeqnarray}
and $E_{\ell \lambda}^{(0)} \left(t\right) = \tilde{E}_{\ell\lambda}^{(0)} e^{-i \omega_{\ell} t}$.
If the $\vb{k}$ dependence of the coupling strength is neglected, so that $\Gamma_{eh}^{\lambda} = \Gamma_{eh}^{\lambda} (\vb{0},\vb{0} )$, then the Rabi energy is $\Delta_{\ell} \equiv \left\vert\Gamma_{eh}^{\lambda} \tilde{E}_{\ell\lambda}^{(0)} \right\vert$.
In this section, we choose the zero-point in the electron bandstructure to be half-way between the valence and conduction band, not, as in Fig.\ \ref{fig:bandstruct}, at the top of the valence band. Hence, in this section
$\varepsilon_{e \vb{k}} = \frac {\hbar^2 k^2} {2 m_e} + E_g / 2$  and  $\varepsilon_{h \vb{k}} = \frac {\hbar^2 k^2} {2 m_h} + E_g / 2$.

The structure of Eq.\ \eqref{eq:nb2} shows that $\tilde{D}^{(1)}$ can be regarded as a density matrix and $h ( \vb{k} )$ as a Hamiltonian governing the evolution of the THz fluctuations that $\tilde{D}^{(1)}$ represents.  $\tilde{D}^{(1)}$ and $h ( \vb{k} )$ are similar to the $2\times2$ Green's function and Hamiltonian matrices, respectively, introduced in Nambu space and used, for example, in the BCS theory of superconductivity \cite{rammer.07}.
$\Delta (\vb{k})$ is the analog of the BCS pairing gap function, and the laser photon energy $\hbar \omega_{\ell}$ here takes the place of twice the electron chemical potential in BCS.

$M_{\ell}$ and $M_{T\nu}$ are matrices linear in $\tilde{E}_{\ell \lambda}^{(1)}$ and $E_{T\nu}$, respectively.
These two matrices do not contain $f_{\alpha}^{(1)}$ nor $\tilde{p}_{eh}^{(1)}$ explicitly.
The THz field source matrix is
\begin{IEEEeqnarray*}{l}
 M_{T\nu} \left(\vb{k} , \vb{q} , t\right) = \left[ \hat{p}^{(0)}_{e h} \left(\vb{k},t\right) - \hat{f}^{(0)} \left(\vb{k},t\right)\right] \hat{g}^{\nu}  \left(\vb{k},\vb{q},t\right) \label{eq:THzsource} \yesnumber \\
\phantom{ M_{T\nu} \left(E_{T\nu}\right)}+ \hat{g}^{\nu}   \left(\vb{k},\vb{q},t\right) \left[ \hat{p}^{(0)}_{e h} \left(\vb{k}-\vb{q},t\right) + \hat{f}^{(0)} \left(\vb{k}-\vb{q},t\right) \right] \nonumber
\end{IEEEeqnarray*}
\begin{IEEEeqnarray*}{LrCl}
\text{where}\quad & \hat{f}^{(0)} \left(\vb{k},t\right) &\equiv& \begin{pmatrix} f_{e}^{(0)} \left(\vb{k},t\right) & 0 \\ 0 & f_{h}^{(0)} \left(-\vb{k},t\right) \end{pmatrix} , \\
& \hat{p}^{(0)}_{e h} \left(\vb{k},t\right) &\equiv& \begin{pmatrix} 0 & \tilde{p}^{(0)}_{e h} \left(\vb{k},t\right) \\ - \tilde{p}^{(0) \ast}_{e h} \left(\vb{k},t\right) & 0 \end{pmatrix},  \\
& \hat{g}^{\nu}   \left(\vb{k},\vb{q},t\right) &\equiv& \begin{pmatrix} h_{Te}^{\nu}   \left(\vb{k},\vb{q},t\right) & 0 \\ 0& h_{Th}^{\nu} \left(\vb{q}-\vb{k},\vb{q},t\right) \end{pmatrix} , \\
\text{and}\quad
& h_{T\alpha}^{\nu} \left(\vb{k},\vb{q},t\right) &\equiv& \,
g_{\alpha}^{\nu} (\vb{k} - \tfrac{1}{2}\vb{q}) A_{T\nu} (\vb{q}, t). \nonumber
\end{IEEEeqnarray*}
$\alpha = e , h$ and $p^{(0)}_{eh} \left(\vb{k}, t\right) = \tilde{p}^{(0)}_{eh} \left(\vb{k}\right) e^{-i \omega_{\ell} t}$.

The matrix $M_{\ell} ( \vb{k} , \vb{q} , t)$ is proportional to $E^{(1)}_{\ell \lambda} ( \vb{q} , t)$. $E^{(1)}_{\ell \lambda} ( \vb{q} , t)$ is calculated through the matrix $D$,
as defined in Eq.\ \eqref{eq:Dmatdef}.
As explained in Appx.\ \ref{appx:ehbasismatrices}, $D(\vb{q},\omega)$ is relatively small for THz realistic $\vb{q}$, and vanishes entirely for $\vb{q}=0$. The same is true for $M_{\ell} (\vb{k} , \vb{q} , t)$, as it is linear in $D(\vb{q},\omega)$.
Further below, we will, for clarity, only consider the $\vb{q} = 0$ case. Since $M_{\ell} (\vb{k} , \vb{0} , t)$ does not contribute to the THz response, we omit its explicit expression.

We solve Eq.\ \eqref{eq:nb2} by using a normal mode expansion. The eigenvalues of $h \left(\vb{k}\right)$ are
\begin{IEEEeqnarray}{rCl}
\lambda_{\pm} (\vb{k}) &=& \xi_{eh}^{-} \left(\vb{k}\right) \pm E_{eh} \left(\vb{k}\right). \label{eq:nb4}
\end{IEEEeqnarray}
where
\begin{IEEEeqnarray}{rCl}
\xi_{eh}^{\pm} \left(\vb{k}\right) &=& \frac{\xi_{e} \left(\vb{k}\right) \pm \xi_{h} \left(-\vb{k}\right) }{2}  \nonumber \\
E_{eh} \left(\vb{k}\right) &=& \sqrt{\left(\xi_{eh}^{+} \left(\vb{k}\right)\right)^{2} + \left\vert\Delta \left(\vb{k}\right)\right\vert^2 } \yesnumber\label{eq:Eehk}
\end{IEEEeqnarray}
If $\varepsilon_{e} \left(\vb{k}\right) = \varepsilon_{h} \left(-\vb{k}\right)$, which is taken to be true in this paper since it is approximated that $m_e = m_h$, then $\lambda_{\pm} \left(\vb{k}\right) = \pm E_{eh} \left(\vb{k}\right)$.
The two eigenvectors of $h \left(\vb{k}\right)$ are
\begin{IEEEeqnarray*}{rCl}
\begin{pmatrix} x_{+} \left(\vb{k}\right)  \\ y_{+} \left(\vb{k}\right)  \end{pmatrix}
&=& \begin{pmatrix} u\left(\vb{k}\right) e^{i\theta\left(\vb{k}\right)/2} \\ v\left(\vb{k}\right) e^{-i \theta \left(\vb{k}\right)/2}  \end{pmatrix} \nonumber \\
\begin{pmatrix} x_{-} \left(\vb{k}\right)  \\ y_{-} \left(\vb{k}\right)  \end{pmatrix}
&=& \begin{pmatrix} v \left(\vb{k}\right) e^{i \theta \left(\vb{k}\right)/2}  \\ -u \left(\vb{k}\right) e^{-i \theta \left(\vb{k}\right)/2} \end{pmatrix}, \yesnumber \label{eq:nb6}\\
\text{where} \quad u \left(\vb{k}\right) &=& \sqrt{\frac{1}{2} \left(1 + \frac{\xi_{eh}^{+} \left(\vb{k}\right) }{E_{eh} \left(\vb{k}\right)}\right) } \\
v \left(\vb{k}\right) &=& \sqrt{\frac{1}{2} \left(1 - \frac{\xi_{eh}^{+} \left(\vb{k}\right) }{E_{eh} \left(\vb{k}\right)}\right) }  \nonumber \\
\text{and} \quad \Delta \left(\vb{k}\right) &\equiv& \left\vert\Delta\left(\vb{k}\right)\right\vert e^{i \theta \left(\vb{k}\right)}.
\end{IEEEeqnarray*}
The eigenvectors are labeled by the same index as the eigenenergies, i.e.\
$h \begin{pmatrix} x_{\pm} \\ y_{\pm} \end{pmatrix} = \lambda_{\pm} \begin{pmatrix} x_{\pm} \\ y_{\pm} \end{pmatrix}$.
The eigenvectors are normalized to unity: $\left\vert x_{\pm}\right\vert^{2} + \left\vert y_{\pm}\right\vert^2 = 1$.
We follow the common BCS notational convention in using the symbols $u ( \vb{k} )$ and $v ( \vb{k} )$ in the eigenvectors.

The eigenenergies defined by Eq.\ \eqref{eq:nb4} are identical to those obtained elsewhere in the literature from the single particle spectral function, if the Coulomb-induced renormalization is neglected.
(cf.\ Refs.\ \onlinecite{jahnke-henneberger.92,%
kremp-etal.08,%
yamaguchi-etal.15,%
murotani-etal.19}.)
These eigenenergies are shown below, in Sec.\ \ref{sec:results}. There, we will also show the corresponding spectral function,
\begin{multline}
A_{e/h}\left(\vb{k},\omega\right) = 2|u \left(\vb{k}\right)|^2 \frac{\gamma}{(\hbar \omega - \xi_{eh}^{-} \left(\vb{k}\right) \mp E_{eh} \left(\vb{k}\right))^2 + \gamma^2} \label{eq:A_SS}  \\
 + 2|v \left(\vb{k}\right)|^2 \frac{\gamma}{(\hbar \omega - \xi_{eh}^{-} \left(\vb{k}\right) \pm E_{eh} \left(\vb{k}\right))^2 + \gamma^2}
\end{multline}
 as well at the joint density of states for vertical intraband transitions (i.e.\ transitions between the two branches of the conduction band, and equivalently between the two branches of the valence band).
The spectral function corresponding to the single-particle Hamiltonian \eqref{eq:nb3} can be obtained
through standard Green's functions techniques. The diagonal elements of the resulting $2 \times 2$ matrix for the spectral function are the $\gamma\to 0$ limit of Eq.\ \eqref{eq:A_SS}. (This expression was also given in Ref.\ \onlinecite{yamaguchi-etal.15}).

The matrix $h \left(\vb{k}\right)$ can be diagonalized by the unitary transformation
\begin{IEEEeqnarray*}{rCl}
h \left(\vb{k}\right) &=& U \left(\vb{k}\right) d \left(\vb{k}\right) U^{\dagger} \left(\vb{k}\right),
\yesnumber \label{eq:nb8} \\
\text{where} \quad d \left(\vb{k}\right) &=& \begin{pmatrix} \lambda_{+} \left(\vb{k}\right)  & 0 \\ 0 & \lambda_{-} \left(\vb{k}\right) \end{pmatrix}, \\
\text{and} \quad U \left(\vb{k}\right) &=& \begin{pmatrix} x_{+} \left(\vb{k}\right) & x_{-} \left(\vb{k}\right) \\ y_{+} \left(\vb{k}\right) & y_{-} \left(\vb{k}\right) \end{pmatrix} .
\end{IEEEeqnarray*}
Under the same transformation, Eq.~\eqref{eq:nb2} becomes
\begin{IEEEeqnarray}{l}
\left[ i \hbar \frac{\partial}{\partial t} + i \gamma \right] \mathcal{D}^{(1)} \left(\vb{k}, \vb{q}, t\right) \yesnumber \label{eq:nb9} \\
\phantom{i \hbar \frac{\partial}{\partial t}} = d \left(\vb{k}\right) \mathcal{D}^{(1)} \left(\vb{k}, \vb{q}, t\right)  - \mathcal{D}^{(1)} \left(\vb{k}, \vb{q}, t\right) d \left(\vb{k} - \vb{q}\right)   \nonumber  \\
\phantom{i \hbar \frac{\partial}{\partial} } + U^{\dagger} \left(\vb{k}\right) \left[ M_{\ell} (\vb{k} , \vb{q} , t) + \sum_{\nu} M_{T\nu} (\vb{k} , \vb{q} , t) \right] U \left(\vb{k} - \vb{q}\right) \nonumber
\end{IEEEeqnarray}
where $\mathcal{D}^{(1)} \left(\vb{k}, \vb{q}, t\right) \equiv U^{\dagger} \left(\vb{k}\right) \tilde{D}^{(1)} \left(\vb{k}, \vb{q}, t\right) U \left(\vb{k} - \vb{q}\right)$.

We solve Eq.\ \eqref{eq:nb9} in frequency space. After some algebra, the solution to the Fourier transformed Eq.\ \eqref{eq:nb9} is obtained as
\begin{IEEEeqnarray}{l}
\mathcal{D}^{(1)} \left(\vb{k}, \vb{q}, \omega\right) = \label{eq:nb13} \yesnumber \\
\begin{pmatrix} \frac{t_{11}}{ \hbar \omega -  \left( \lambda_{+} \left(\vb{k}\right) - \lambda_{+} \left(\vb{k} - \vb{q}\right)\right) + i \gamma}
& \frac{t_{12}}{\hbar \omega - \left( \lambda_{+} \left(\vb{k}\right) - \lambda_{-} \left(\vb{k} - \vb{q}\right)\right) + i \gamma} \\
 \frac{t_{21}}{\hbar \omega - \left( \lambda_{-} \left(\vb{k}\right) - \lambda_{+} \left(\vb{k} - \vb{q}\right)\right) + i \gamma}
&  \frac{t_{22}}{\hbar \omega - \left( \lambda_{-} \left(\vb{k}\right) - \lambda_{-} \left(\vb{k} - \vb{q}\right)\right) + i \gamma} \end{pmatrix}
\nonumber
\end{IEEEeqnarray}
where the quantities $t_{i j} , i,j = 1,2$ are the elements of the frequency-domain source matrix:
\begin{multline}
U^{\dagger} \left(\vb{k}\right) \left[ M_{\ell} (\vb{k} , \vb{q} , \omega) + M_{T\nu} (\vb{k} , \vb{q} , \omega) \right] U \left(\vb{k} - \vb{q}\right) \\ =
\begin{pmatrix} t_{11} & t_{12} \\ t_{21} & t_{22} \end{pmatrix} . \label{eq:nb12}
\end{multline}
Eq.\ \eqref{eq:nb12}
shows that there are resonances at energy differences between the two bands $\lambda_{+} \left(\vb{k}\right) - \lambda_{-} \left(\vb{k} - \vb{q}\right)$ in the response.
The strength of the response depends on the projection functions $u \left(\vb{k}\right)$ and $v \left(\vb{k}\right)$.

To clarify the formalism, the $\vb{q} = \vb{0}$ case is treated below. We first obtain the matrix elements $t_{i j}$ of Eq.\ \eqref{eq:nb12}.
Since $M_{\ell} ( \vb{k} , \vb{q} = \vb{0} , \omega) = 0$ as explained above, the matrix in Eq.\ \eqref{eq:nb12} is $U^{\dagger} \left(\vb{k}\right) M_{T\nu} ( \vb{k} , \vb{0} , \omega ) U \left(\vb{k} \right)$. Eq.\ \eqref{eq:THzsource}, transformed to frequency space, is simplified to
\begin{IEEEeqnarray*}{l}
M_{T\nu} \left(\vb{k}, \vb{0}, \omega\right) = \begin{pmatrix} 0 & \tilde{p}^{(0)}_{e h} \left(\vb{k}\right) h_{T}^{\nu}  \left(\vb{k},\omega\right) \\
- \tilde{p}^{(0) \ast}_{e h} \left(\vb{k}\right) h_{T}^{\nu}  \left(\vb{k},\omega\right)  & 0
\end{pmatrix} \nonumber \\
\yesnumber \label{eq:nb17}
\end{IEEEeqnarray*}
where
\begin{IEEEeqnarray*}{l}
h_{T}^{\nu} \left(\vb{k},\omega\right) \equiv h_{Te}^{\nu}  \left(\vb{k}, \vb{0}, \omega\right) +  h_{Th}^{\nu}  \left(\vb{k}, \vb{0}, \omega\right) \yesnumber
\end{IEEEeqnarray*}
Using the definition of $g_{\alpha}^{\nu}$, the relation $\tilde{A}^{\ast}_{T\nu} \left( - \vb{q} , - \omega\right) = \tilde{A}_{T\nu} \left(\vb{q} , \omega\right)$, implied by the input time-domain vector potential being real, and $\tilde{A}_{T\nu} \left(\vb{q} , \omega\right) = \frac{c}{\omega i} \tilde{E}_{T\nu} \left(\vb{q} , \omega\right)$, we obtain $h_{T}^{\nu} \left(\vb{k},\omega\right)$ explicitly as
\begin{IEEEeqnarray*}{l}
h_{T}^{\nu} \left(\vb{k},\omega\right) = \frac{\hbar q_e}{i m_r \omega} \vb{k} \cdot  \bm{\epsilon}_{\nu} \tilde{E}_{T\nu} \left(\vb{0}, \omega\right),
\yesnumber
\label{eq:nb18}
\end{IEEEeqnarray*}
where $m_r$ is the electron-hole reduced mass, $\frac{1}{m_r} = \frac{1}{m_e} + \frac{1}{m_h}$.
Explicitly, in the eigenstate basis, the source matrix is
\begin{IEEEeqnarray*}{l}
U^{\dagger} \left(\vb{k}\right) M_{T\nu} (\vb{k} , \vb{0} , \omega) U \left(\vb{k}\right)
= h_{T}^{\nu}  \left(\vb{k},\omega\right) \left\vert\tilde{p}^{(0)}_{e h} \left(\vb{k}\right)\right\vert  \yesnumber \label{eq:nb19} \\
\times \begin{pmatrix} 2 i  u \left(\vb{k}\right) v\left(\vb{k}\right) \sin \theta_0 & -u^2 \left(\vb{k}\right) e^{i \theta_0} - v^2 \left(\vb{k}\right) e^{-i \theta_0} \\ v^2 \left(\vb{k}\right) e^{i \theta_0} + u^2 \left(\vb{k}\right) e^{-i \theta_0} & - 2 i  u \left(\vb{k}\right) v\left(\vb{k}\right) \sin \theta_0 \end{pmatrix},
\nonumber
\end{IEEEeqnarray*}
where $\theta_0$ is the phase difference between $\tilde{p}^{(0)}_{e h} \left(\vb{k}\right)$ and $\Delta \left(\vb{k}\right)$:
\begin{equation*}
\theta_0 \left(\vb{k}\right) = \theta_p \left(\vb{k}\right) - \theta \left(\vb{k}\right), \qquad \tilde{p}^{(0)}_{e h} \left(\vb{k}\right) = \left\vert\tilde{p}^{(0)}_{e h} \left(\vb{k}\right)\right\vert e^{i \theta_p \left(\vb{k}\right)}
\end{equation*}
We substitute the matrix elements from Eq.\ \eqref{eq:nb19} into Eq.\ \eqref{eq:nb13} to yield $\mathcal{D}^{(1)} \left( \vb{k}, \vb{0}, \omega\right)$ and then obtain $\tilde{D}^{(1)} \left( \vb{k}, \vb{0}, \omega\right)$ as
\begin{equation}
\tilde{D}^{(1)} \left( \vb{k}, \vb{0}, \omega\right) = U^{\dag} (\vb{k}) \mathcal{D}^{(1)} \left( \vb{k}, \vb{0}, \omega\right) U (\vb{k})
\end{equation}
Carrying out the algebra gives the elements of $\tilde{D}^{(1)} \left( \vb{k}, \vb{0}, \omega\right)$ as
\begin{IEEEeqnarray*}{l}
f_{e}^{(1)} \left(\vb{k}, \vb{k}, \omega\right)  = \sum_{\nu} h_{T}^{\nu}  \left(\vb{k},\omega\right) \left\vert\tilde{p}^{(0)}_{e h} \left(\vb{k}\right)\right\vert u \left(\vb{k}\right) v \left(\vb{k}\right)  \yesnumber \label{eq:nb22} \\
\phantom{f_{e}^{(1)}(\vb{k})}
 \cdot  \left[ \frac{2 i \left( u^2 \left(\vb{k}\right) - v^2 \left(\vb{k}\right) \right) \sin \theta_0 \left(\vb{k}\right) }{\hbar \omega + i \gamma} \right.  \nonumber \\
\phantom{f_{e}^{(1)}(\vb{k}) \cdot \left[\right.}
 - \frac{\cos \theta_0 \left(\vb{k}\right) + i \left( u^2 \left(\vb{k}\right) - v^2 \left(\vb{k}\right) \right)  \sin \theta_0 \left(\vb{k}\right) }{\hbar \omega - \Delta \lambda \left(\vb{k}\right) + i \gamma}  \nonumber \\
\phantom{f_{e}^{(1)}(\vb{k}) \cdot \left[\right.}
 \left. + \frac{\cos \theta_0 \left(\vb{k}\right) - i \left( u^2 \left(\vb{k}\right) - v^2 \left(\vb{k}\right) \right)  \sin \theta_0 \left(\vb{k}\right) }{\hbar \omega + \Delta \lambda \left(\vb{k}\right) + i \gamma} \right] \nonumber \\
f_{h}^{(1)} \left( \vb{k}, \vb{k}, \omega\right) = f_{e}^{(1)} \left(-\vb{k}, -\vb{k}, \omega\right) \yesnumber
\end{IEEEeqnarray*}
and
\begin{IEEEeqnarray*}{l}
\tilde{p}_{eh}^{(1)} \left(\vb{k}, -\vb{k}, \omega\right) = \sum_{\nu} h_{T}^{\nu}  \left(\vb{k},\omega\right) \left\vert\tilde{p}^{(0)}_{e h} \left(\vb{k}\right)\right\vert e^{i \theta \left(\vb{k}\right)}  \\
 \qquad \cdot \left[ \frac{4 i u^2 \left(\vb{k}\right) v^2 \left(\vb{k}\right)  \sin \theta_0 \left(\vb{k}\right) }{\hbar \omega + i \gamma} \right.  \nonumber \\
\quad + u^2 \left(\vb{k}\right) \frac{ \cos \theta_0 \left(\vb{k}\right) + i \left( u^2 \left(\vb{k}\right) - v^2 \left(\vb{k}\right) \right)  \sin \theta_0 \left(\vb{k}\right)}{\hbar \omega - \Delta \lambda \left(\vb{k}\right) + i \gamma}  \nonumber \\
\quad \left. + v^2 \left(\vb{k}\right) \frac{\cos \theta_0 \left(\vb{k}\right) - i \left( u^2 \left(\vb{k}\right) - v^2 \left(\vb{k}\right) \right)  \sin \theta_0 \left(\vb{k}\right)}{\hbar \omega + \Delta \lambda \left(\vb{k}\right) + i \gamma} \right] \nonumber
\end{IEEEeqnarray*}
where $\Delta \lambda \left(\vb{k}\right) \equiv \lambda_{+} \left(\vb{k}\right) - \lambda_{-} \left(\vb{k}\right)$. $f^{(1)}_{\alpha}, \alpha = e , h$ are independent of spin, and in a reflection-symmetric system, $f^{(1)}_{\alpha} \left(-\vb{k}, -\vb{k}, \omega\right) = - f^{(1)}_{\alpha} \left(\vb{k}, \vb{k}, \omega\right)$.

Eq.\ \eqref{current-def.equ} relates the occupation fluctuation $f^{(1)}_{\alpha}$ to the THz-induced paramagnetic current $J_{\alpha\nu}^{p(1)}$, which gives the paramagnetic conductivity $\sigma_{T\nu}^{p}$ by Eq.\ \eqref{eq:thz-cond-def}. The result is
\begin{IEEEeqnarray*}{l}
\sigma_{T\nu}^{p} \left(\vb{q} = \vb{0}, \omega\right) =\frac{\sum_{\alpha = e , h} J_{\alpha\nu}^{p(1)}  \left(\vb{0} , \omega \right)} {\tilde{E}_{T\nu} \left(\vb{0},\omega\right)} \yesnumber \label{eq:nb25} \\
\phantom{\sigma} =  i c \frac{S_{d}}{\mathcal{A}} \frac{\alpha_{0} }{\hbar \omega} \left(\frac{\hbar^2}{m_r}\right)^2 \sum_{\vb{k}}  \left(\vb{k} \cdot \bm{\epsilon}_{\nu}\right)^2 \left\vert\tilde{p}^{(0)}_{e h} \left(\vb{k}\right)\right\vert u\left(\vb{k}\right) v\left(\vb{k}\right) \\
\phantom{\sigma =}  \cdot \left[ \frac{2 i \left( u^2 \left(\vb{k}\right) - v^2 \left(\vb{k}\right) \right) \sin \theta_0 \left(\vb{k}\right) }{\hbar \omega + i \gamma} \right.  \nonumber \\
\phantom{\sigma = \cdot \left[\right. }
 - \frac{\cos \theta_0 \left(\vb{k}\right) + i \left( u^2 \left(\vb{k}\right) - v^2 \left(\vb{k}\right) \right)  \sin \theta_0 \left(\vb{k}\right) }{\hbar \omega  - \Delta \lambda \left(\vb{k}\right) + i \gamma}  \nonumber \\
\phantom{\sigma = \cdot \left[\right. }
 \left. + \frac{\cos \theta_0 \left(\vb{k}\right) - i \left( u^2 \left(\vb{k}\right) - v^2 \left(\vb{k}\right) \right)  \sin \theta_0 \left(\vb{k}\right) }{\hbar \omega + \Delta \lambda \left(\vb{k}\right) + i \gamma} \right] \nonumber
\end{IEEEeqnarray*}
where $\alpha_0 = q_e^2 / (\hbar c) = 1/137.04$ is the fine structure constant and $S_{d} = 2$ is the spin degeneracy. In the weak absorption limit, the absorption is proportional to $\Re\sigma_{T\nu}^{p} \left(\vb{q} = \vb{0}, \omega\right)$, which is proportional to the sum of $\Re [ f_{e}^{(1)} ( \vb{k} , \vb{k} , \omega) / \tilde{E}_{T\nu} ( \vb{0} , \omega) ]$ over the $\vb{k}$ states.
For each $\vb{k}$, there are 3 peaks in $\Re[ f_{e}^{(1)} ( \vb{k} , \vb{k} , \omega) / \tilde{E}_{T\nu} ( \vb{0} , \omega) ]$, at frequencies $\pm \Delta \lambda$ and 0.
The factor $u \left(\vb{k}\right) v \left(\vb{k}\right)$ limits the contributing $\vb{k}$ region to be around the BCS-like gap.
The peak at $\omega = 0$ is proportional to $\sin \theta_0$, which is non-zero
when the steady-state laser has a non-zero loss rate, corresponding to a non-equilibrium steady state.
For an ideal eh-photon condensate at (quasi-)thermal equilibrium, $\sin \theta_0 = 0$, implying that, under our assumption of $\gamma_p = \gamma_f$, this peak is absent in the THz response of that system; see Appx.\ \ref{appx:bcs-quasi-equil}.
For the peaks at $\hbar \omega = \pm \Delta \lambda$, the THz probe $A_{T\nu}$ perturbs the order parameter $p_{eh}$ directly, not $f_{\alpha}$. The laser field transfers the perturbation from $p_{eh}$ to $f_{\alpha}$, thus creating an intraband current.
We also note that the delta-function in the joint density of states (JDOS), which is derived and Appx.\ \ref{app:jdos} and will be discussed and used in Sec.\ \ref{sec:results},
coincides with the peaks in the integrand of the real part of the  conductivity, which
can be seen in the limit of $\gamma \rightarrow 0$, in which case the real part of Eq.\ \eqref{eq:nb25} gets a delta-function contribution from the second term in the square bracket with the same argument as in the JDOS, cf.\ Eq.\ \eqref{eq:jdos_y_ss_cintra}.

%===========================
% Numerical Results and Discussion
%===========================

\section{Numerical Results and Discussion}

\label{sec:results}

In this section, we present our main results, in particular signatures of light-induced intraband gaps in measurable quantities such as THz transmissivity and absorptivity and the underlying intraband conductivity. In order to build some intuition for the expected results, we first recall a few basic facts of optical response. In a non-excited two-band semiconductor quantum well like (e.g.\ GaAs) without (excitonic) Coulomb effects, a light field in the visible or near-infrared spectrum and
in normal incidence (i.e.\ in-plane wave vector $\vb{q}=0$) can induce vertical (in k-space) transitions between the valence and conduction band.  In contrast, vertical intraband transitions cannot be induced by a THz field, since the limit $\vb{q} \rightarrow 0$ limits optical transitions to zero frequency. For the case of a doped semiconductor at zero temperature, this can be seen from the intraband pair excitation region shown in Fig.\ \ref{fig:pair-excitation}a (this can be obtained from the Lindhard response function, see for example Ref.\ \onlinecite{mahan.00}).

\begin{figure}[t]
\centering
\includegraphics[width=\columnwidth]{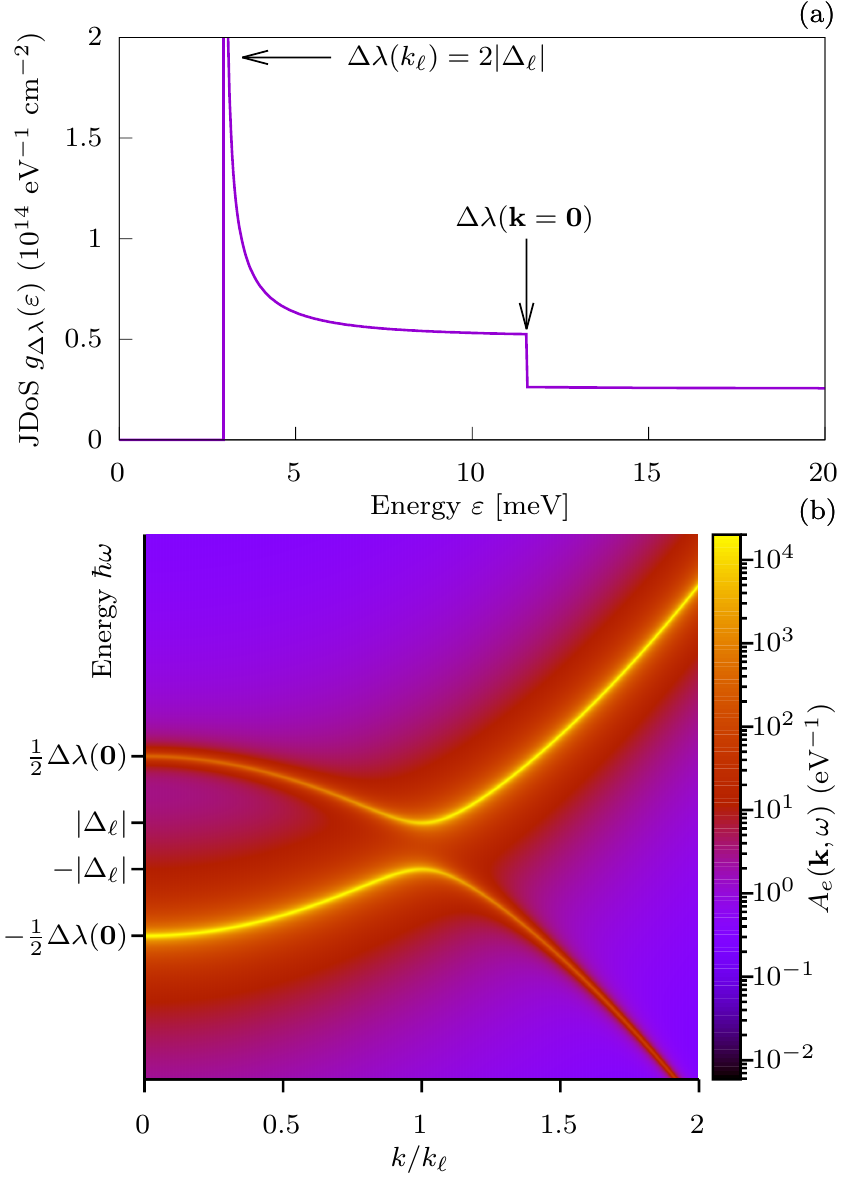}
\caption{(Color online.) (a) Joint density of states for transitions between the lower and upper branches of the renormalized conduction band. (b) Color map of the spectral function from Eq.\ \eqref{eq:A_SS} (cf.\ Eq.\ (30) of Ref.\ \onlinecite{yamaguchi-etal.15}). For this and all subsequent figures, the parameters are give in table \ref{tab:numparams}, unless noted otherwise.
}
\label{fig:JDOS_spect_cl}
\end{figure}

Intraband excitations in normal incidence become possible if one considers intersubband transitions or excitonic effects (e.g.\ 1s to 2p exciton transitions), both of which are not part of our present considerations. Instead, we are interested in vertical (or almost vertical) intraband transitions made possible by the strong coherent light field, which can be either the laser light itself or an external coherent light field. Discussions of single-particle spectral functions of the conduction band (and similarly of the valence band) can be found, for example, in Refs.\ \onlinecite{galitskii-etal.70,jahnke-henneberger.92,kremp-etal.08,yamaguchi-etal.15}, and in Fig.\ \ref{fig:JDOS_spect_cl} we show the conduction band spectral function, Eq.\ \eqref{eq:A_SS}. The light-induced bands are clearly visible, and  transitions between the original branches (i.e.\ the conduction band in the absence of the light field) and the light-induced branches  become possible, see also Fig.\ \ref{fig:bandstruct}.
In this paper, we call these light-induced vertical intraband transitions.

 For the simple case of electrons occupying the states below the light-induced gap, the pair-excitation region is shown in Fig.\ \ref{fig:pair-excitation}b. This includes the possibility of  vertical transitions.  We show the joint density of states (JDOS) for light-induced vertical intraband transitions in Fig.\ \ref{fig:JDOS_spect_cl}, and present mathematical details in Appx.\ \ref{app:jdos}. The JDOS has a lower bound given by the light-induced gap; the (van Hove) singularity is typical for a one-dimensional parabolic band minimum and follows here from the geometry of a ring in two dimensions with non-zero radius $k_{\ell}$, in which case the two-dimensional parabolic band minimum loses its two-dimensional rotation symmetry.
 At higher energies, the JDOS exhibits a step, which is the usual signature of a two-dimensional DOS of parabolic bands and stems here from the vertical transition at $k=0$. In the following, we will focus on measurable signatures due to the light-induced gap and compare features in the measurable quantities to the energetic position of the light-induced gap, or, equivalently, to the lower bound of the JDOS.

\begin{figure}
\centering
\includegraphics{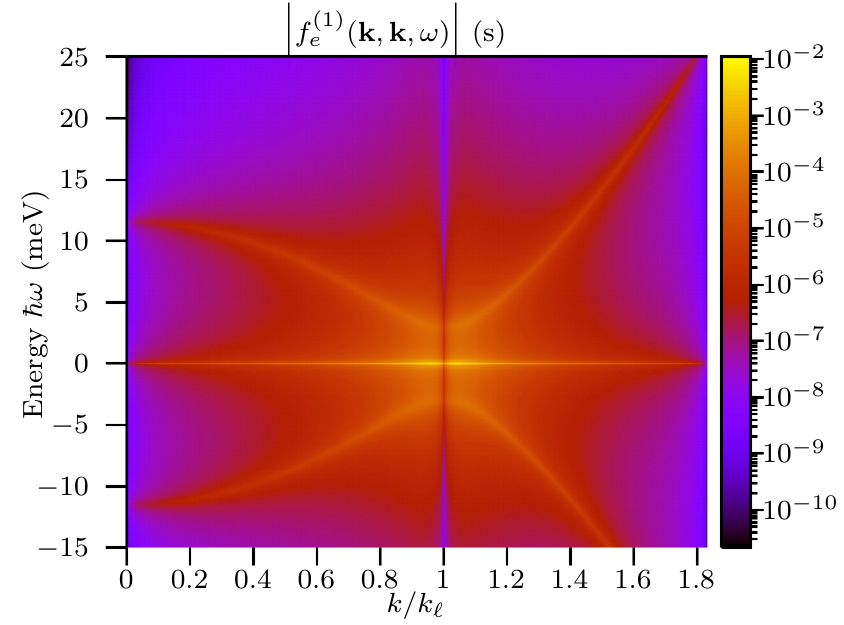}
\caption{(Color online.) Color map of the absolute value of the linear density response function,
$\left|\tilde{f}_{e}^{(1)}(\vb{k},\vb{k},\omega)\right|$ with  $\vb{k} = (0, k)$.
Here, $2\Delta_{\ell} = 2.985$ meV.
}
\label{fig:f1_abs_heatmap}
\end{figure}

In addition to the intraband pair excitations, the low-frequency response of a semiconductor contains a strong component of the Drude response (diamagnetic response), corresponding to a collective excitation of the charge carriers. This response is proportional to $1/(\omega(\omega + i \gamma_D))$, with the phenomenological decay rate $\gamma_D$. In the following, we will first use an (idealized) small value for the Drude decay $\gamma_D$ in order to establish the principal result that the THz measurement can observe the light-induced gaps. For that case, we find that the linear THz response corresponds to THz gain. In a second step, we will look at larger, more realistic values of the Drude decay and show that information on the light-induced gaps can still be extracted, albeit in this case without the presence of THz gain.

In order to relate the intuition, based on vertical light-induced intraband transitions (Fig.\ \ref{fig:bandstruct}) between the original and the light-induced branches to our present theory, we show in Fig.\ \ref{fig:f1_abs_heatmap} the magnitude of the linear response of the carrier density, $\tilde{f}_{e}^{(1)}(\vb{k},\vb{k},\omega)$,
from Eq.\ \eqref{eq:f1eFT}
for the case of $\vb{q}=0$. Note that the vertical scale in Fig.\ \ref{fig:JDOS_spect_cl} is the single-particle energy, while that in Fig.\ \ref{fig:f1_abs_heatmap} is the photon energy (frequency) of the THz field. Our intuition is based on transitions (i.e.\ differences of energies) in Fig.\ \ref{fig:JDOS_spect_cl}, which corresponds to absolute frequencies
in \ref{fig:f1_abs_heatmap}.
 The linear density response exhibits a strong signal at zero frequency, which corresponds to transitions between a certain initial state and the same final state. Light-induced vertical intraband transitions correspond to peaks in $|\tilde{f}_{e}^{(1)}(\vb{k},\vb{k},\omega)|$, not differences as in Fig.\ \ref{fig:JDOS_spect_cl}.  The comparison of Figs.\  \ref{fig:JDOS_spect_cl} and \ref{fig:f1_abs_heatmap} shows that, as mentioned above, our single-time theory does indeed capture the physics usually associated with spectral functions and thus 2-time Green's functions. This correspondence is further established in Sec.\ \ref{sec:nambu-basis}.

\begin{figure*}
\centering
\includegraphics{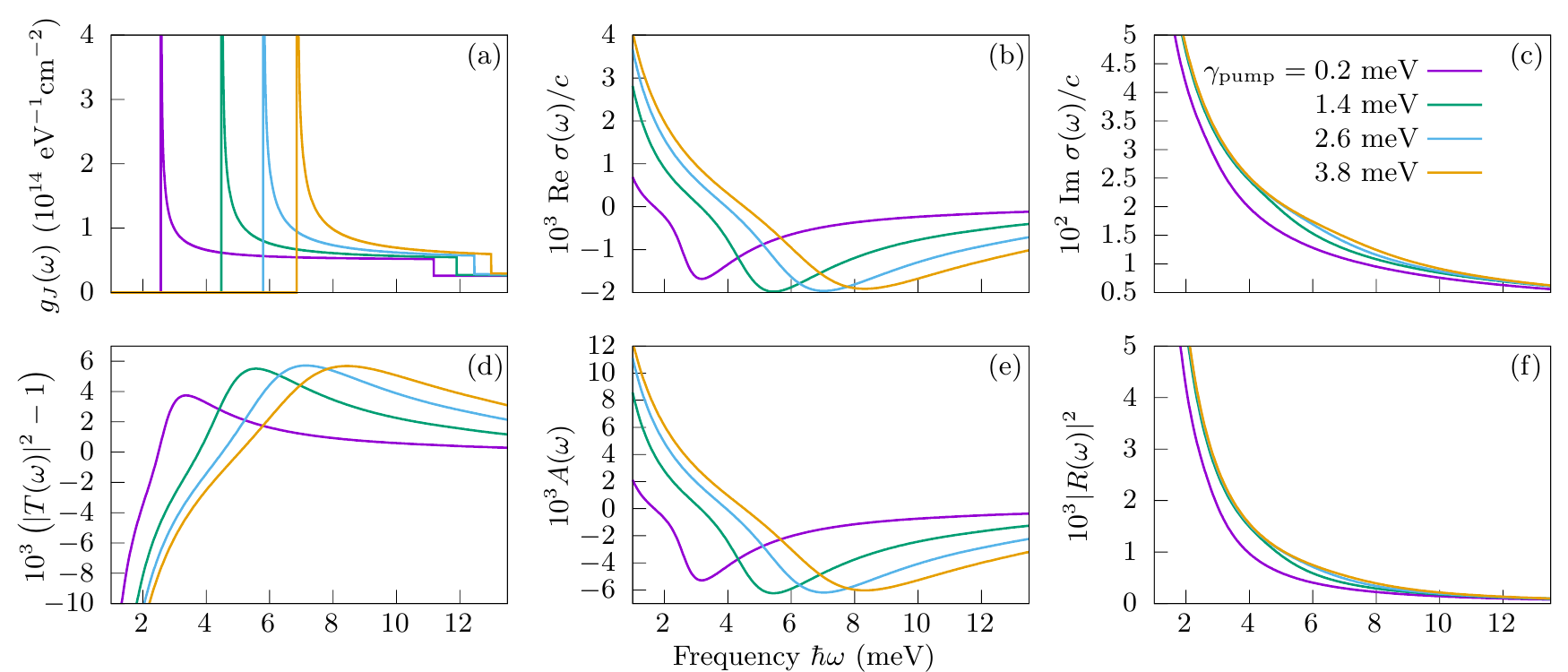}
\caption{(Color online.)
THz probe response for $\gamma_{\mathrm{pump}} = 0.2, 1.4, 2.6, 3.8$ meV, for normal incidence, $\theta_{i}=0$ or equivalently $\vb{q}=0$, and with Drude decay $\gamma_{D} = 0.01$ meV. All parameters except $\gamma_{\mathrm{pump}}$ are as given in table \ref{tab:numparams}. (a) Joint density of states. (cf.\ Appx.\ \ref{app:jdos}.) (b,c) Real and imaginary part of the low-frequency conductivity. Since we are considering the $\vb{q}=0$ case (except in Fig.\ \ref{fig:T_ang}), the conductivity does not depend on the polarization direction, and in this section, we denote it $\sigma(\omega)$. (d) Transmissivity. (e) Absorptivity. (f) Reflectivity.
The singularity in the JDOS comes from transitions across the BCS-like gap, $2|\Delta_{\ell}|$.
The real part of the conductivity, $\Re\sigma_{T\nu}(\omega)$, clearly shows a valley which is tracked by the singularity in JDOS.
For the given (small) value of $\gamma_{D}$, the minima in $\Re\sigma_{T\nu}(\omega)$ then appear as regions of gain, or negative absorptivity, in the absorptivity plot $A\left(\omega\right)$, and also the transmissivity exhibits gain ($|T|^2>1$).
 The Drude response dominates the imaginary part of the conductivity, $\Im\sigma(\omega)$, and the reflectivity $|R|^2$.
 }
\label{fig:art_cl}
\end{figure*}

\begin{figure}
 	\centering
 	\includegraphics{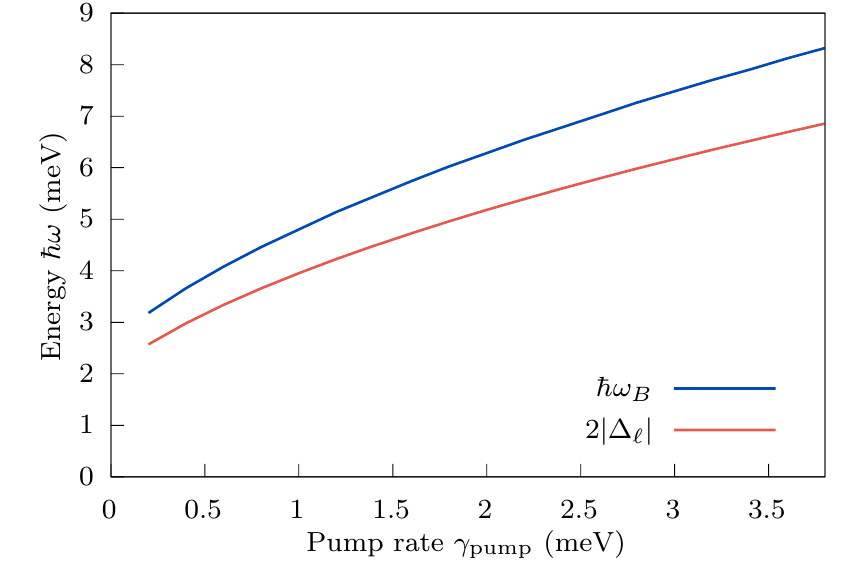}
 	\caption{(Color online.) Comparison of the absorptivity minimum $\hbar\omega_{B}$ and the BCS-like gap $2\Delta_{\ell}$ as a function of the pump rate $\gamma_{\mathrm{pump}}$
 		for the data shown in Fig.\ \protect\ref{fig:art_cl}.
 		The light red line denotes the magnitude of the BCS-like gap, $2\Delta_{\ell}$.
 		The dark blue line shows the frequency $\hbar\omega_{B}$ for which the real part of the conductivity $\Re\sigma_{T\nu}(\omega)$ is a minimum; that is, $\hbar\omega_{B}$ is defined such that $\Re\sigma_{T\nu}(\omega_B)=\min \Re\sigma_{T\nu}(\omega)$.
 		The plot shows that the BCS-like gap is closely tracked by the gain maximum, over a wide range of pumping rates.}
 	\label{fig:bcs_gap_pump}
 \end{figure}

We now show that not only the linear density response, but also the measurable THz transmissivity and absorptivity spectra of a photon laser contain information on the light-induced gaps (which, as noted above, in the case of the photon laser we call BCS-like gaps). Figure \ref{fig:art_cl} shows, in addition to the intraband conductivity, the transmission, reflection and absorption of the THz probe. We see a clear correspondence between the singularity in the JDOS (at the energy of the BCS-like gap) and an extremum in the transmission and absorption. On  the other hand, the reflectivity does not show such an extremum and is dominated by the Drude response. The exact position of the extremum is not exactly at the energy of the BCS-like gap, in part because of line shape effects of the various decay contributions, such as the monotonically decreasing (as a function of $\omega$) Drude contribution,
 but clearly track the BCS-like gap energy as we increase the pump rate. This is seen more clearly in Fig.\ \ref{fig:bcs_gap_pump}, where we show the energetic position of the absorption minimum and the BCS-like gap energy as a function of pump rate. This establishes the main claim of our study, namely that the BCS-like gap can in principle  be observed in THz spectroscopy.

\begin{figure}
\centering
\includegraphics{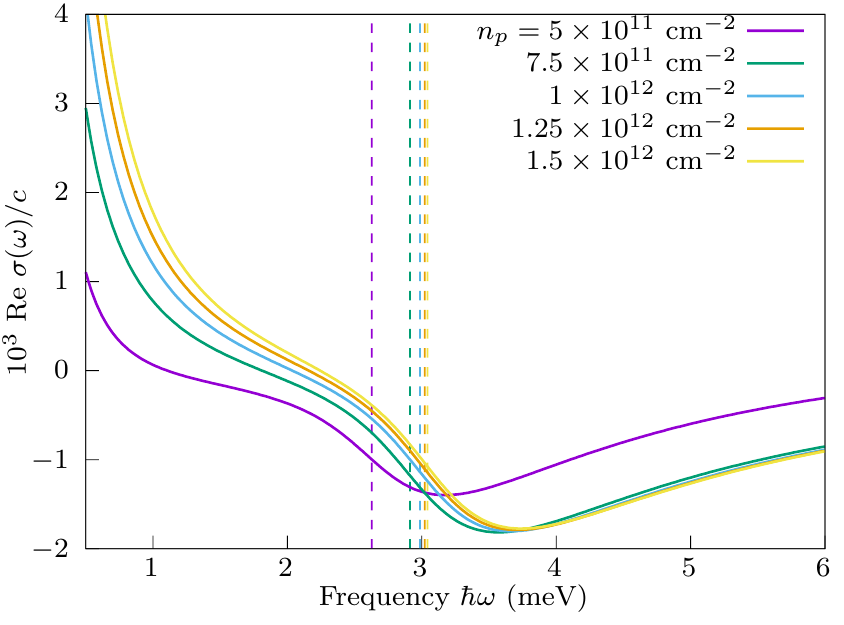}
\caption{(Color online.)
Real part of the conductivity, $\Re\sigma_{T\nu}(\omega)$, which is similar in line shape to the absorptivity  $A\left(\omega\right)$, for various  pump densities $n_{\mathrm{pump}}$.
Here and in the following figures, the vertical dashed lines indicate the frequency of the onset of the JDOS, which is the same as the light-induced gap.}
\label{fig:cond_dens}
\end{figure}

\begin{figure}
\centering
\includegraphics{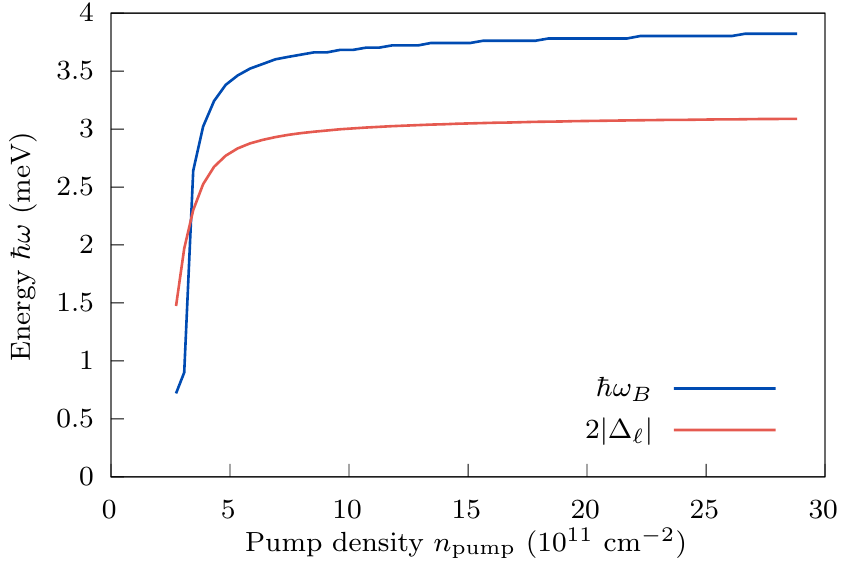}
\caption{(Color online.) Comparison of the absorptivity minimum $\hbar\omega_{B}$ and the BCS-like gap $2\Delta_{\ell}$ as a function of the pump density $n_{\mathrm{pump}}$.
The light red line denotes the magnitude of the BCS-like gap, $2\Delta_{\ell}$.
The dark blue line shows the frequency $\hbar\omega_{B}$ for which the real part of the conductivity $\Re\sigma_{T\nu}(\omega)$ is a minimum; that is, $\hbar\omega_{B}$ is defined such that $\Re\sigma_{T\nu}(\omega_B)=\min \Re\sigma_{T\nu}(\omega)$.
The BCS-like gap is tracked by the minimum in $\Re\sigma_{T\nu}(\omega)$, for changing $n_{\mathrm{pump}}$. $\Delta_{\ell}$ levels off almost completely, while the carrier density $n_{e}$ keeps increasing with increasing $n_{\mathrm{pump}}$.
 Because the Drude conductivity increases for increasing $n_{e}$, the minimum in $\Re\sigma_{T\nu}(\omega)$ does not level off as much as $\Delta_{\ell}$.}
\label{fig:bcs_gap_dens}
\end{figure}

However, the parameters in Figs.\  \ref{fig:art_cl} and \ref{fig:bcs_gap_pump} have been chosen to be idealized in order for the principal effect to be clearly seen. For predictions of measurements under presently realistic conditions, we need to relax the idealized parameter choices. One of the two idealizations in
Figs.\  \ref{fig:art_cl} and \ref{fig:bcs_gap_pump} is the assumption that we can freely vary the pump rate. In typical experiments, not the pump rate but the pump density can readily be varied, since the former is modeling the relaxation of optically or electronically injected charge carriers via electron-phonon and electron-electron interactions, and the latter is determined by the injection power. When we vary the pump density, rather than the pump rate, the BCS-like gaps do not vary strongly once the pump density is sufficiently high. This is because the carrier distribution functions are in the regime of Fermi degenerate functions (basically step-like functions), and additional density is only in k-states that are not involved in the lasing process and thus have no effect on the laser intensity or the BCS-like gap. Nevertheless, Fig.\ \ref{fig:cond_dens} shows that the absorption still has a minimum corresponding to the BCS-like gap, and  Fig.\ \ref{fig:bcs_gap_dens} shows that the minimum still tracks the energy of the BCS-like gap, even though in this case the change of the BCS-like gap  levels off with increasing pump density.

\begin{figure}
\centering
\includegraphics[width=\columnwidth]{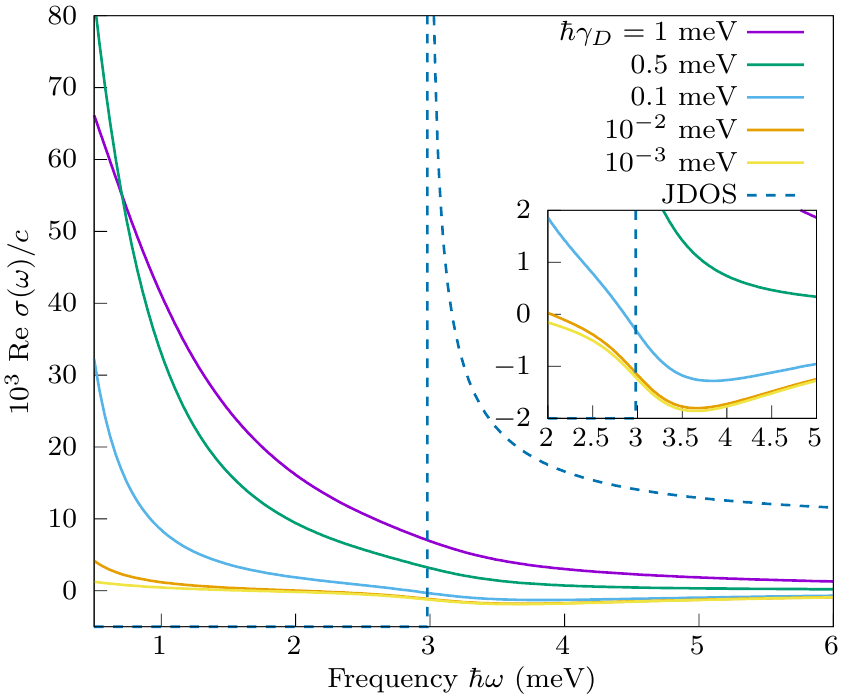}
\caption{(Color online.) Real part of the conductivity vs.\ frequency for various values of the Drude decay rate $\gamma_{D}$.
The inset shows a zoomed-in region for the smaller values of $\gamma_{D}$.
The JDOS is shown as a dashed line. This high-frequency decay of the real part of the conductivity follows approximately that of the JDOS.
}
\label{fig:resig_gm_drude}
\end{figure}

\begin{figure}
\centering
\includegraphics[width=\columnwidth]{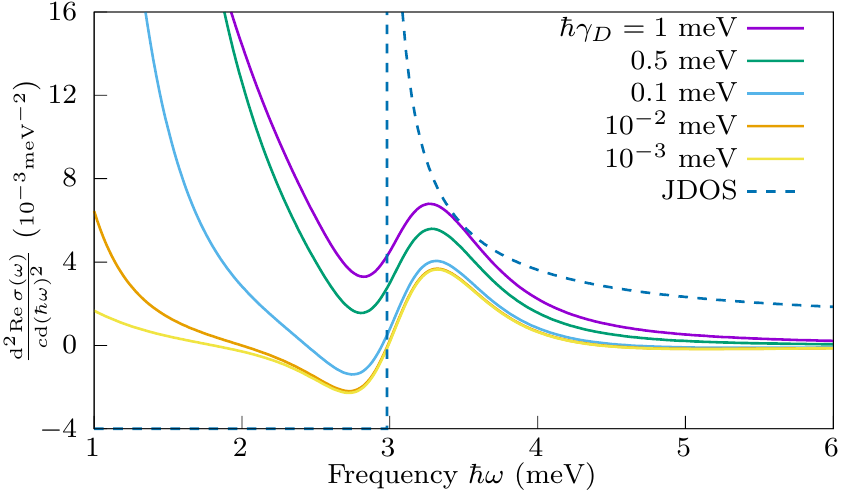}
\caption{(Color online.) Second frequency derivative of the real part of the conductivity vs.\ frequency for various values of the Drude decay  $\gamma_{D}$.
This shows that the BCS-like gap is again visible in $\Re\sigma_{T\nu}(\omega)$ when the second derivative is plotted.
The JDOS is shown as a dashed line.
The high-frequency decay of the second derivative of the real part of the conductivity follows approximately that of the JDOS.
}
\label{fig:resig_dom_gm_drude}
\end{figure}

The second idealization used in Fig.\ \ref{fig:art_cl} is the use of a small Drude decay rate. Increasing the Drude scattering rate so that it is of the same order of magnitude as the BCS gap can make the Drude response dominant. The minima or maxima are then  superposed with a monotonically decreasing Drude contribution and instead of being absolute extrema they are only small variations that cannot be readily identified,
as shown in Fig.\ \ref{fig:resig_gm_drude}.
  In order to still have access to the information from the linear response $\tilde{f}_{e}^{(1)}(\vb{k},\vb{k},\omega)$ and thus the BCS-like gaps, one can take the second frequency derivative, for example of the absorptivity or of the real part of the conductivity.
This is shown in Fig.\
 \ref{fig:resig_dom_gm_drude}. Doing so make the BCS-like gap again readily visible.


\begin{figure}
	\centering
	\includegraphics{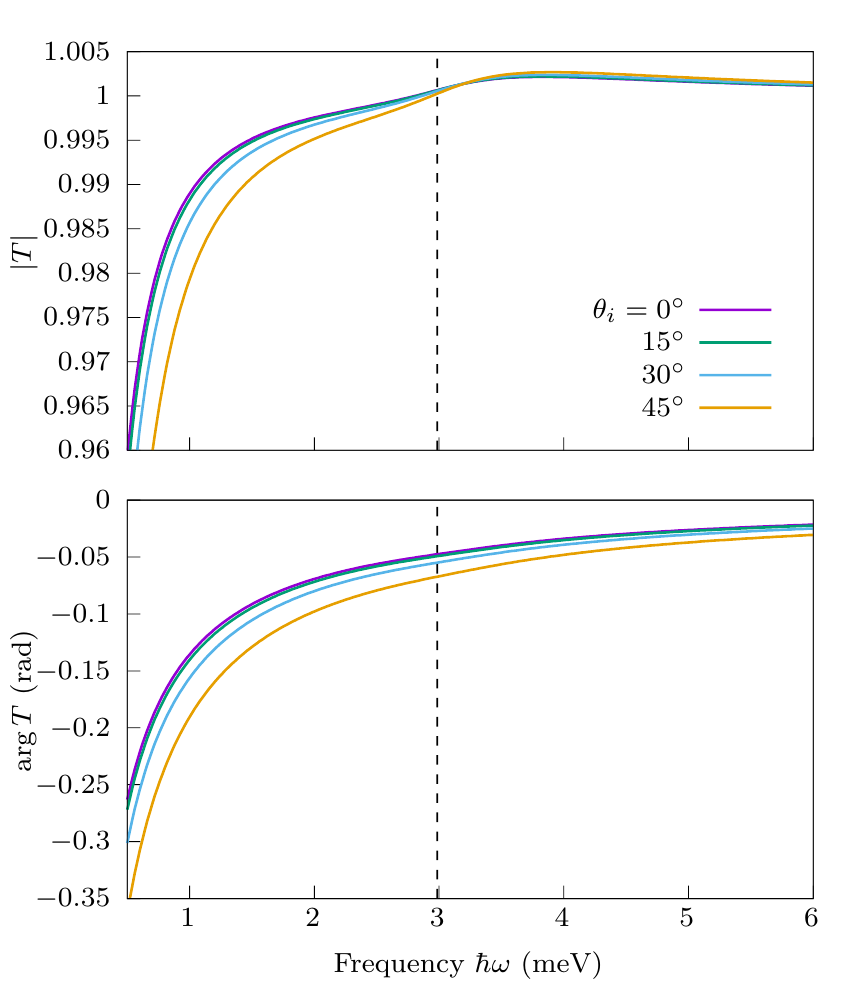}
	\caption{(Color online.)
		The magnitude $|T|$ and phase $\arg T$ of the complex transmission coefficient $T$ of the THz probe wave, as a function of $\omega$  for different angles of incidence $\theta_i$, i.e.\ $\theta_i = 0, 15, 30, 45^{\circ}$, $s$-polarized THz light, and with this paper's default value for the Drude decay,
		$\gamma_{D}=0.01$ meV. $\theta_{i}$ is defined with respect to the background medium, not the angle of incidence from outside the microcavity. The probe geometry is as in Fig.\ \ref{fig:wavevecs}.
		The plot shows that in the region of the BCS-like gap and for higher frequencies, the angle of incidence has only a small effect on the THz transmissivity.}
	\label{fig:T_ang}
\end{figure}

Until now, we have shown results for the case of normal incidence. In experiments, the THz probe might be at oblique incidence, but this has little effect on the findings presented above.
 To show this, Fig.\ \ref{fig:T_ang} has results for various angles of incidence (measured with respect to the surface normal of the quantum well at the position of the quantum well inside the cavity). We see that qualitative features of the transmission depend only weakly on the angle of incidence, with the main variation being the low-frequency Drude response affecting the phase of the complex transmissivity.

\begin{table}
\caption{Material parameters used for the numerical results. Italics denote values different from those used in Ref.\ \onlinecite{hu-etal.21}.
These values are used for the plots, unless otherwise noted.}
\label{tab:numparams}
\centering
\begin{tabular}{|l|l|r|}
\hline Parameter & Symbol & Value \\ \hline
Bohr radius for GaAs & $a_{B}$ & 140 \AA \\
\# of QW's & $N_{\mathrm{QW}}$ & $1$ \\
3D, QW exciton binding energy & $E_{B}^{3\mathrm{D}}$ & 3.2 meV \\
Band gap energy & $E_{g}$ & 1.562 meV\\
\emph{Cavity frequency} & $\omega_{\mathrm{cav}}$ & 1.574 meV \\
$e$-$h$ reduced mass &$m_{r}$&$\frac{\hbar^2}{2 E_{B}^{3\mathrm{D}} a_{B}^{2}}$ \\
$e$ effective mass & $m_{e}$ & $2 m_{r}$ \\
$h$ effective mass & $m_{h}$ & $2 m_{r}$ \\
Relative permittivity & $\epsilon_{b}$ & $\frac{e^2}{2 a_{B}E_{B}^{3\mathrm{D}}}$ \\
Interband coupling strength & $\Gamma(k,q)$ & $6.404$ neV cm \\
Cavity $E$ field loss rate & $\gamma_{E}$ & 0.2 meV \\
\emph{Pump density} & $n_{\mathrm{pump}}$ & $10^{12}$ cm$^{-2}$ \\
\emph{Pump relax.\ rate} & $\gamma_{\mathrm{pump}}$ & 0.4 meV \\
Non-radiative loss rate & $\gamma_{\mathrm{nr}}$& 0.1 $\mu$eV  \\
\emph{Dephasing} & $\gamma_{p}$ & 0.2 meV \\
Fermi dist.\ relax.\ rate & $\gamma_{F}$ & $2\gamma_{p}$  \\
\emph{Effective cond.\ elec.\ temp.} & $T$ & 10 K \\
\emph{Drude scattering rate} & $\gamma_{D}$ & 10 $\mu$eV \\
\emph{THz probe angle of incidence} & $\theta_{i}$ & $0^{\circ}$\\ \hline
\end{tabular}
\end{table}

We now comment further on the issue of THz gain, which can be seen in Fig.\ \ref{fig:art_cl}.
While the THz gain shown in Fig.\ \ref{fig:art_cl} is only obtained for a very small value of the Drude decay $\gamma_{D}$, the fact that it can be obtained as a matter of principle makes some further analysis desirable.

\begin{figure}
	\centering
	\includegraphics{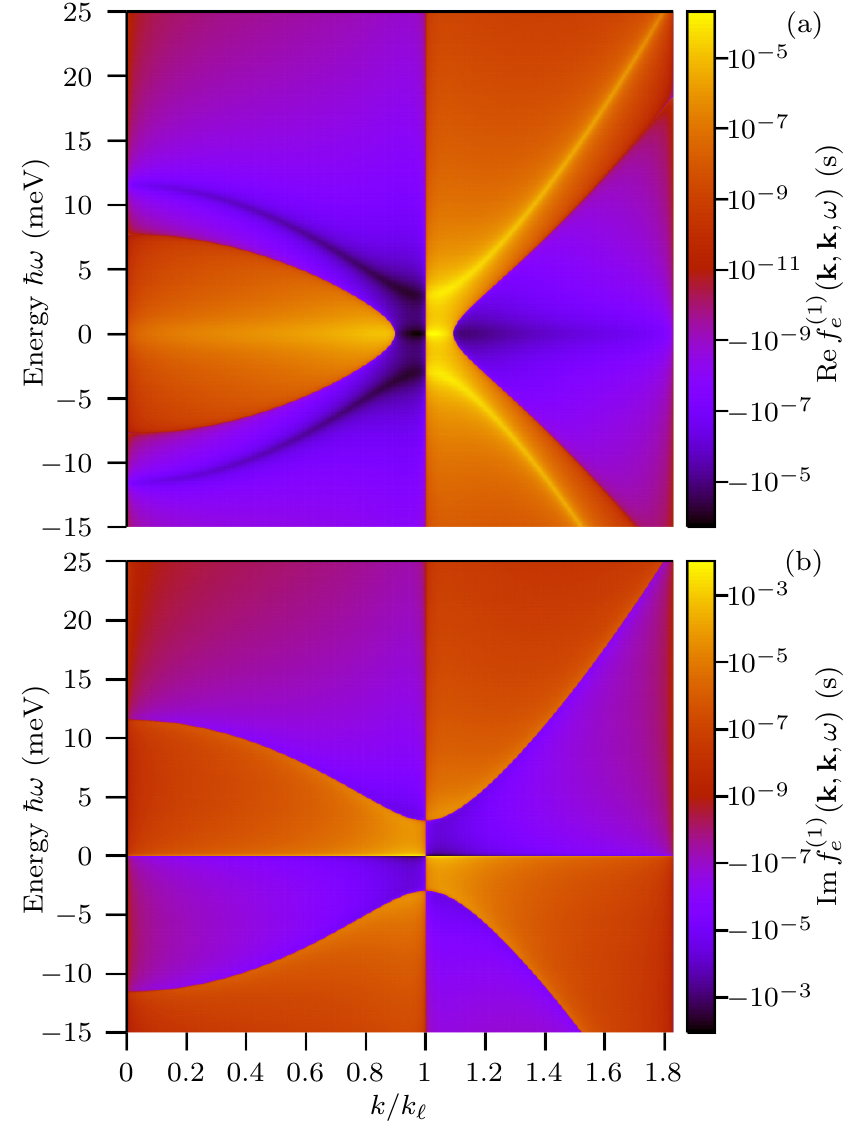}
	\caption{(Color online.) Color map of the linear density response function $\tilde{f}_{e}^{(1)}(\vb{k},\vb{k},\omega)$ with  $\vb{k} = (0, k)$.
		(a) Real part, (b) imaginary part.}
	\label{fig:f1_heatmap}
\end{figure}

\begin{figure}
	\centering
	\includegraphics{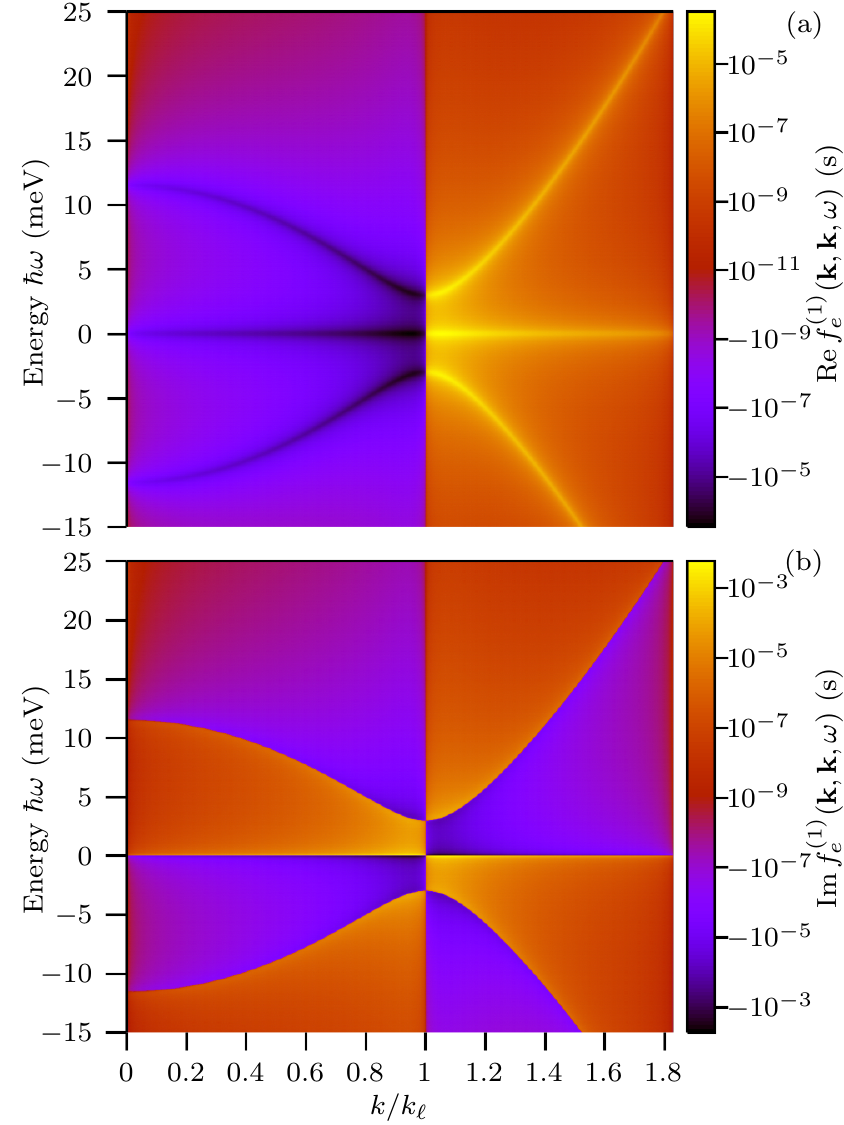}
	\caption{(Color online.) Color map of the linear density response function $\tilde{f}_{e}^{(1)}(\vb{k},\vb{k},\omega)$ with  $\vb{k} = (0, k)$.
		(a) Real part,  (b) imaginary part. In contrast to
		Fig.\ \protect{\ref{fig:f1_heatmap}},
		here $\gamma_p = \gamma_f = 0.2 $ meV,
		so that it can be analyzed using
		the formulas from Sec.\ \protect{\ref{sec:nambu-basis}}.
	}
	\label{fig:f1_heatmap_NB}
\end{figure}

Figure \ref{fig:f1_heatmap} shows again the linear density response, similarly to Fig.\ \ref{fig:f1_abs_heatmap}, but now separately the real and imaginary parts.
We see a sign change across $k_{\ell}$, with $\tilde{f}_{\alpha}^{(1)} (k_{\ell}, \omega ) = 0$.
To further analyze how this sign change is related to the THz gain, it is beneficial to use the theoretical formulation developed in
Sec.\ \ref{sec:nambu-basis}. A plot of the linear density response using $\gamma_p = \gamma_f  $, which obeys Eq.\ \eqref{eq:nb22} of
Sec.\ \ref{sec:nambu-basis}, is shown in Fig.\ \ref{fig:f1_heatmap_NB}. First, we note that a comparison of
Figs.\ \ref{fig:f1_heatmap} and \ref{fig:f1_heatmap_NB} shows that the region close the the BCS-like gap is not sensitive to the choice of the decay constants. Hence, we can analyze the origin of the gain based on  Fig.\ \ref{fig:f1_heatmap_NB} and use the underlying theory,
Eqs.\ \eqref{eq:nb22}, \eqref{eq:nb25} and \eqref{eq:condthz}.

We have THz gain when the real part of the THz conductivity is negative; see Fig.\ \ref{fig:art_cl} or Eqs.\ \eqref{eq:Aqxomega} and \eqref{eq:beta_def}.
Hence the condition for gain is that the real part of the right-hand side (RHS) of Eq.\ \eqref{eq:nb25} is negative. For positive $\omega$, the gain with frequencies close to the BCS-like gap comes from the second term inside the square brackets of Eq.\ \eqref{eq:nb25}. The condition for the summand on the RHS of Eq.\ \eqref{eq:nb25} being negative is the same as the RHS of Eq.\ \eqref{eq:nb22} being positive  if we choose $\tilde{E}_{T\nu}$ to be  positive and real-valued, in order to have a convenient way to relate the signs of
Eqs.\ \eqref{eq:nb22} and \eqref{eq:nb25}. Now our condition for THz gain at a given wave vector is that the real part of
$\tilde{f}_{e}^{(1)} \left(\vb{k}, \vb{k}, \omega\right)$ is positive.
In our case of the photon laser,
the second term inside the square brackets of  Eq.\ \eqref{eq:nb22} changes sign at $k_{\ell}$.
Specifically, both
$\cos \theta_0 \left(\vb{k}\right)$ and
$\left( u^2 \left(\vb{k}\right) - v^2 \left(\vb{k}\right) \right) $ change sign, while
$\sin \theta_0 \left(\vb{k}\right)$ does not. In our case, the $\sin \theta_0 \left(\vb{k}\right)$ term is negligible compared to the
$\cos \theta_0 \left(\vb{k}\right)$ term, which is negative below and positive above $k_{\ell}$, as can be seen from
Fig.\ \ref{fig:0np}, using the fact that with our phase conventions,
$\cos \theta_0 \left(\vb{k}\right) = - \frac{ \Re \tilde{p}_{eh}^{(0)} \left(\vb{k}\right)}{|\tilde{p}_{eh}^{(0)} \left(\vb{k}\right)|}$.
While in our numerical solution we set the arbitrary phase of the laser field $\tilde{E}_{\ell\lambda}^{(0)}$ to zero and use a real-valued coupling constant
$\Gamma_{eh}^{\lambda}$,
one can see from the steady state equations \eqref{eq:ddtP0}--\eqref{eq:ddtE0} that $\theta_{0}$ does not depend on the phase convention of
$\Gamma_{eh}^{\lambda} \tilde{E}_{\ell\lambda}^{(0)} $ .
Hence, we find that in our photon-laser case, in the vicinity of $k_{\ell}$,
$k$-states below $k_{\ell}$ contribute to THz absorption at frequencies in the vicinity of the BCS-like gap,
 while those above $k_{\ell}$ contribute to THz gain.
This is exemplified by the green vertical arrows in Fig.\ \ref{fig:bandstruct}.\

According to Eq.\ \eqref{eq:nb25} or Eq.\ \eqref{eq:condthz}, we have to sum up all $\vb{k}$-contributions to determine whether the absorption or gain contribution is dominant.
In the $\vb{k}$-sum in Eq.\ \eqref{eq:nb25}, the factor of $\left(\vb{k} \cdot \bm{\epsilon}_{\nu}\right)^2$, from the intraband matrix element, favors the $\tilde{f}_{\alpha}^{(1)}$ states with greater $|\vb{k}|$ over those with lower $|\vb{k}|$.
Therefore, from the numerics, we find that the gain overcompensates the absorption, in the frequency region around $2\Delta_{\ell}$; see Fig.\ \ref{fig:art_cl}.

	In the zero-dephasing limit, the dependence of the THz gain on the optical steady state can be shown analytically if the results of Appx. \ref{appx:laserdist} for $|\tilde{p}_{eh}^{(0)}(k)|\cos\theta_{0}(k)$ are used in Eq.\ \eqref{eq:nb25}.
	 Then, the conductivity integrand depends only on the $\cos\theta_{0}(k)$ component of $\tilde{p}_{eh}^{(0)}(k)$, and in the limit of zero dephasing, $\gamma \to 0$,
	\begin{IEEEeqnarray}{r}
		\IEEEeqnarraymulticol{1}{l}{\Re \lim_{\gamma\to 0} \sigma_{T\nu}^{p}(\omega>0) = -\frac{cS_{d}\alpha_{0}\Delta_{\ell}^{2}}{2\hbar^{3}\omega^{3}} \theta\left(\hbar\omega-2\Delta_{\ell}\right) } \yesnumber \label{eq:condtau0gm} \\
		\times \left[
		\left( \sqrt{\hbar^{2}\omega^{2} - 4\Delta_{\ell}^{2}} + \hbar\tilde{\omega}_{\ell}\right) \left(2f_{R}^{+}(\omega)-1\right)
		\right. \nonumber  \\
		\left. +
		\left( \sqrt{\hbar^{2}\omega^{2} - 4\Delta_{\ell}^{2}} - \hbar\tilde{\omega}_{\ell}\right) \left(2f_{R}^{-}(\omega)-1\right) \theta\left(\Delta\lambda(0)-\hbar\omega\right)
		\right] \nonumber
	\end{IEEEeqnarray}
	where $\hbar \tilde{\omega}_{\ell} \equiv \hbar \omega_{\ell} - E_g$, $E_g$ being the band gap, and $f_{R}^{\pm}(\omega) \equiv f_{R}\left(\sqrt{\frac{2m_{r}}{\hbar^{2}}} \sqrt{\hbar\tilde{\omega}_{\ell}\pm\sqrt{\hbar^{2}\omega^{2}-4\Delta_{\ell}^{2}}}\right)$.
	Like the JDOS (cf. Eq.\ \eqref{eq:jdos_y_ss_cintra}),
	Eq.\ \eqref{eq:condtau0gm} has a term that is multiplied by a step function and therefore does not contribute
 for $\hbar\omega>\Delta\lambda(0)$, and a term with no upper-bound step function.
	The bounded term results from the region $k<k_{\ell}$ and is positive, so that it contributes to absorption.
	The other, first term in Eq.\ \eqref{eq:condtau0gm} derives from the integration over $k>k_{\ell}$ and is negative, contributing to gain, until $k \approx k_{F}$, the Fermi wavenumber, where it becomes positive.
	This sign dependence on $k$ is the same as that of $-|\tilde{p}_{eh}^{(0)}(k)|\cos\theta_{0}(k)$.
	Essentially, the THz in-quadrature susceptibility integrates over the optical in-phase polarization, independently of the relative phase of the THz and optical fields.
	In the limit of $2\Delta_{\ell}\ll \hbar\tilde{\omega}_{\ell}$, the THz conductivity is equivalent to a third-order nonlinear susceptibility, $\chi^{(3)}$.%

\begin{figure}
	\centering
	\includegraphics[width=\columnwidth]{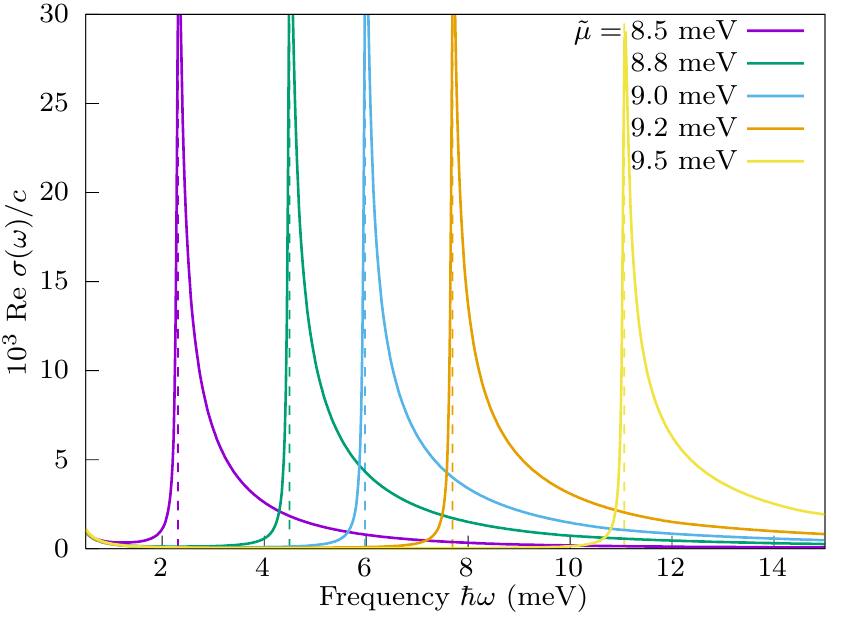}
	\caption{
		(Color online.)
		Real part of the THz conductivity, $\Re \sigma_{Ty}(\vb{q}=0,\omega)$, for a eh-photon system in quasi-equilibrium (cf.\ Appx.\ \ref{appx:bcs-quasi-equil}).}
	\label{fig:quasi-thermal-equi-cond}
\end{figure}

The fact that THz gain is in principle possible results from  the fact that our photon-laser is an open-dissipative and pumped system.
The zeroth-order
intraband polarization and distribution functions given in Appx.\ \ref{appx:laserdist}
 are different from functions in a thermal equilibrium system, or, more precisely, in the case of a microcavity laser, quasi-thermal equilibrium.
 In order to show that the THz gain is indeed absent if the system is in quasi-thermal equilibrium, we use the
 quasi-thermal equilibrium solutions from Ref.\ \onlinecite{yamaguchi-etal.15} and show in Appx.\ \ref{appx:bcs-quasi-equil} that the
 sign of the conductivity is fixed with  the real part being positive, eliminating the possibility of THz gain. We show in
 Fig.\ \ref{fig:quasi-thermal-equi-cond} an example for this case.

\begin{figure}
\centering
\includegraphics{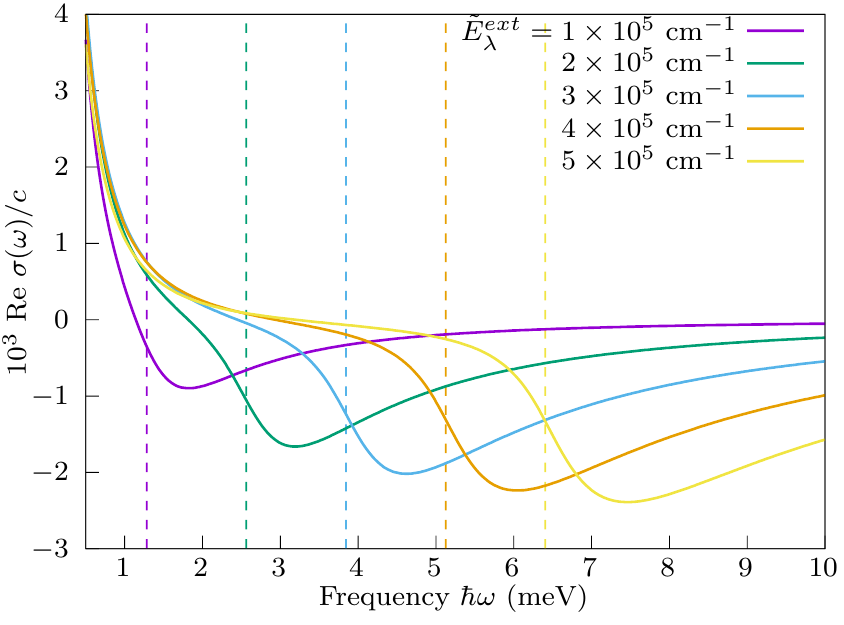}
\caption{(Color online.) Real part of the THz conductivity, $\Re\sigma_{Ty}(\omega)$, for the case of a quantum well irradiated by a strong external coherent light field in normal incidence
with frequency $\hbar \omega_{ext} = E_{g} + 10$ meV.
}
\label{fig:external-field}
\end{figure}

 In addition to the BCS-like gaps in the (non-equilibrium) photon laser, we also analyzed light-induced gaps in the case of a quantum  well irradiated by a strong coherent external light field with frequency in the interband continuum. As mentioned above, we call these gaps Galitskii-Elesin gaps. An example is shown in
 Fig.\ \ref{fig:external-field}. The underlying equations for the zeroth-order solutions
 are obtained from Eqs.\ \eqref{eq:P0motRTAss}
 and \eqref{eq:f0motRTAss} by replacing the cavity field
 $\tilde{E}_{\ell\lambda}^{(0)}$
 with an external coherent field
 $\tilde{E}^{ext}_{\lambda}$,
 and the laser frequency
 $\omega_{\ell}$
 by the frequency of the external field,
 $\omega_{ext}$.
 In this case, Eq.\ \eqref{eq:E0motRTAss} is omitted.
  The electron density is be solved consistently with $n_{e} = n_{F, e}$, and
 the final results follow from Eqs.\ \eqref{eq:f0cauchy} and (\ref{eq:P0lorentz}.
 An example of this case is shown in Fig.\ \ref{fig:external-field}, which shows a result similar to those of the (non-equilibrium) photon-laser. In other words, the Galitskii-Elesin gaps can be related to extrema in the THz spectrum.

Finally, regarding the possibility of observing THz gain in open-dissipative and pumped photon laser as well as the quantum well irradiated by a strong coherent external light field, we note again that the  (paramagnetic) THz gain can be overwhelmed by the absorption due to the (diamagnetic) Drude term if the Drude decay is large.

\section{Conclusion}
\label{sec:conclusion}

To summarize, we have developed a linear response theory for the low-frequency  optical response of a microcavity laser with
 a two-band semiconductor quantum well in the photon-laser approximation (i.e.\ without excitonic Coulomb effects).
 In GaAs (the example treated in this paper) and similar III-V semiconductors, the low-frequency response of interest to this paper is in the THz regime.
 The stationary 2-band system exhibits BCS-like gaps in each of the two bands. We have shown that  these light-induced gaps can give rise to structures (minima or maxima) in the measurable quantities associated with the linear THz response, notably the real part of the (intraband) conductivity and the (THz) absorption. Thus, THz spectroscopy of GaAs or other conventional semiconductors could be a valuable tool to identify BCS-like gaps that cannot be readily seen with light fields whose frequencies are in the vicinity of interband transitions and thus within the stop-band of the microcavity.
  While our numerical results have focused on GaAs, our analysis can in principle also be applied to other semiconductors, where the BCS-like gaps might be larger and the intraband spectroscopy frequencies lie outside the conventional THz band.

 In addition to the correlation between structures in the linear response function and BCS-like gaps, we have found that photon-lasers can in principle exhibit THz amplification. This amplification, however, can only be seen if the Drude decay is sufficiently small. Furthermore, we have shown that the THz amplification is absent if the system is in quasi-thermal equilibrium. While past studies have found the Drude scattering rate under differing conditions,\cite{sernelius.91,leitenstorfer-etal.00,beard-etal.00,shi-etal.08} it would be beneficial if future studies provide a microscopic model for the Drude damping in the photon laser, possibly pointing to pathways toward reducing the magnitude of the Drude damping.

 Finally, we have shown that, as expected,  Galitskii-Elesin gaps  give rise to the same structure in the THz linear response functions as BCS-like gaps.

 In a next step, the role of Coulomb effects, and thus exciton and exciton-polariton effects, in a polariton laser that operates in the polariton-BCS regime, which was studied in Ref.\ \onlinecite{hu-etal.21}, will be investigated.
Furthermore, future studies may investigate the THz emission in the BEC-BCS cross-over regime and in the BEC regime. Such studies could shed light on how the condensation-induced gap transforms during the BEC-BCS cross-over, and could also further elucidate the role of polariton trapping \cite{balili-etal.06,balili-etal.07,berman-etal.08} (which has been found to be an effective tool in the manipulation and creation of polariton BEC states) on the gap in the condensed state.

\begin{acknowledgments}

We gratefully acknowledge useful discussions with Hui Deng, University of Michigan;
financial support from the NSF under grant number DMR 1839570;
and the use of High Performance Computing (HPC) resources supported by the University of Arizona.

\end{acknowledgments}

\appendix

%===========================
%  QW THz Transmission
%===========================
\section{Quantum Well THz Transmission}
\label{appx:THz_transmission}

Eqs.\ \eqref{eq:Tqxomega}--\eqref{eq:beta_def}, relating the THz transmissivity, reflectivity, and absorptivity to the conductivity in the quantum well, are derived in this appendix.
We refer to Fig.\ \ref{fig:wavevecs} for the propagation geometry. The boundary conditions on the THz electric field ${\rm E}_T$ and magnetic field ${\rm B}_T$ at the quantum well ($z=0$) are, for $\eta \downarrow 0$,
\begin{IEEEeqnarray}{rCl}
\epsilon_2 E_{T z} ({\rm r}_\| , \eta) - \epsilon_1 E_{T z} ( {\rm r}_\| , -\eta) &=& 4 \pi \rho_{2\mathrm{D}} ({\rm r}_\| ) \label{pt2eq1} \\
B_{T y} ({\rm r}_\| , \eta) - B_{T y} ({\rm r}_\| , -\eta) &=& - \frac{4 \pi}{c} J_{2\mathrm{D},x} ({\rm r}_\| ) \label{pt2eq2} \\
B_{T x} ({\rm r}_\| , \eta) - B_{T x} ({\rm r}_\| , -\eta) &=& \frac{4 \pi}{c} J_{2\mathrm{D},y} ({\rm r}_\| ) \label{pt2eq3}  \\
E_{T x} ({\rm r}_\| , \eta) &=& E_{T x} ({\rm r}_\| , -\eta) \label{pt2eq4a}   \\
E_{T y} ({\rm r}_\| , \eta) &=& E_{T y} ({\rm r}_\| , -\eta) \label{pt2eq4b} \\
B_{T z} ({\rm r}_\| , \eta) &=& B_{T z} ({\rm r}_\| , -\eta)  \label{pt2eq4c}
\end{IEEEeqnarray}
where ${\rm r}_\| = (x,y)$. The 3D representations of the charge and current densities are
\begin{equation*}
\rho ({\rm r}_\| , z) = \delta(z) \rho_{2\mathrm{D}} ({\rm r}_\|)
\qquad {\rm J} ({\rm r}_\| , z) = \delta(z) {\rm J}_{2\mathrm{D}} ({\rm r}_\|)
\end{equation*}
Expressed in terms of the induced current defined in Eqs.\ \eqref{current-def.equ}, the vector 2D current is
${\rm J}_{2\mathrm{D}} ({\rm r}_\| ) = \sum_{\nu} [ \sum_{e} J_{e\nu}^{(1)} ({\rm r}_\| ) + \sum_{h} J_{h\nu}^{(1)} ({\rm r}_\| ) ] \bm{\epsilon}_{\nu}$.

The incident fields are denoted by ${\rm E}_T^{(i)}$, ${\rm B}_T^{(i)}$, the reflected fields by ${\rm E}_T^{(r)}$, ${\rm B}_T^{(r)}$, and the transmitted fields by ${\rm E}_T^{(t)}$, ${\rm B}_T^{(t)}$. We have ${\rm E}_T ({\rm r}_\| , z) = {\rm E}_T^{(i)} ({\rm r}_\| , z) + {\rm E}_T^{(r)} ({\rm r}_\| , z)$ for $z < 0$ and ${\rm E}_T ({\rm r}_\| , z) = {\rm E}_T^{(t)} ({\rm r}_\| , z)$ for $z > 0$ and similar relations for the magnetic fields. Suppose the incident field is an $s$-polarized (in our coordinate system, $\bm{\epsilon}_{\nu} = \hat{y}$), monochromatic plane wave with frequency $\omega$ and
wave vector ${\rm q}_i = q_{i x} \hat{x} + q_{i z} \hat{z}$, where $q_i \equiv | {\rm q}_i | = \sqrt{\epsilon_1} \omega / c$. The wave's electric and magnetic fields are written as ($z < 0$)
\begin{IEEEeqnarray*}{rCl}
{\rm E}_T^{(i)} ( {\rm x} , t ) &=& \tilde{E}^{(i)}_{T} e^{i ({\rm q}_i \cdot {\rm x} - \omega t)} \hat{y} , \\
{\rm B}_T^{(i)} ( {\rm x} , t ) &=& (q_{ix} \hat{z} - q_{iz} \hat{x}) \frac{c}{\omega} \tilde{E}^{(i)}_{T} e^{i ({\rm q}_i \cdot {\rm x} - \omega t)}
\end{IEEEeqnarray*}
where ${\rm x} = (x,y,z)$.
Similarly, the reflected fields are ($z < 0$)
\begin{IEEEeqnarray*}{rCl}
{\rm E}_T^{(r)} ( {\rm x} , t ) &=& \tilde{E}^{(r)}_{T} e^{i ({\rm q}_r \cdot {\rm x} - \omega t)} \hat{y} , \\
{\rm B}_T^{(r)} ( {\rm x} , t ) &=& (q_{rx} \hat{z} - q_{rz} \hat{x}) \frac{c}{\omega} \tilde{E}^{(r)}_{T} e^{i ({\rm q}_r \cdot {\rm x} - \omega t)}
\end{IEEEeqnarray*}
where ${\rm q}_r = q_{r x} \hat{x} + q_{r z} \hat{z}$, $q_r = \sqrt{\epsilon_1} \omega / c$, and the transmitted fields are ($z > 0$)
\begin{IEEEeqnarray*}{rCl}
{\rm E}_T^{(t)} ( {\rm x} , t ) &=& \tilde{E}^{(t)}_{T} e^{i ({\rm q}_t \cdot {\rm x} - \omega t)} \hat{y} , \\
{\rm B}_T^{(t)} ( {\rm x} , t ) &=& (q_{tx} \hat{z} - q_{tz} \hat{x}) \frac{c}{\omega} \tilde{E}^{(t)}_{T} e^{i ({\rm q}_t \cdot {\rm x} - \omega t)}
\end{IEEEeqnarray*}
where ${\rm q}_t = q_{t x} \hat{x} + q_{t z} \hat{z}$, $q_t = \sqrt{\epsilon_2} \omega / c$.
Substituting these expressions into the condition Eq.\ \eqref{pt2eq4b} gives
\begin{IEEEeqnarray}{rCl}
\tilde{E}^{(i)}_{T} e^{i q_{i x} x} + \tilde{E}^{(r)}_{T} e^{i q_{r x} x} &=& \tilde{E}^{(t)}_{T} e^{i q_{t x} x}, \quad \forall x \nonumber
\end{IEEEeqnarray}
which implies
\begin{IEEEeqnarray}{rCl}
q_{i x} = q_{r x} = q_{t x} &\equiv& q_{x}  \label{pt2eq5} \\
\tilde{E}^{(i)}_{T\nu} + \tilde{E}^{(r)}_{T\nu} &=& \tilde{E}^{(t)}_{T\nu} . \label{pt2eq6}
\end{IEEEeqnarray}
Eq.\ \eqref{pt2eq5}, together with the dispersion relations, lead to $q_{r z} = - q_{i z}$ and $q_t^2 = ({\epsilon_2} / {\epsilon_1}) q_i^2$, which gives
\begin{equation}
q^2_{t z} = \left( \frac {\epsilon_2} {\epsilon_1} - 1 \right) q^2_x + \frac {\epsilon_2} {\epsilon_1} q^2_{i z} \nonumber
\end{equation}
The 2D current is a plane wave inside the quantum well, propagating in the $x$ direction:
\begin{IEEEeqnarray}{rCl}
{\rm J}_{2\mathrm{D}} (x, y, t) &=& \tilde{J}_{2\mathrm{D}} e^{i (q_x x - \omega t)} \hat{y} \label{pt2eq10}
\end{IEEEeqnarray}
Eq.\ \eqref{pt2eq3} becomes
\begin{IEEEeqnarray}{rCl}
q_{iz} \left( \tilde{E}^{(i)}_{T} - \tilde{E}^{(r)}_{T} \right) - q_{t z} \tilde{E}^{(t)}_{T} &=& \frac{4\pi \omega}{c^2} \tilde{J}_{2\mathrm{D}} \label{pt2eq9}
\end{IEEEeqnarray}
The conductivity at frequency $\omega$ for this $y$-polarized probe is given by
\begin{equation}
\sigma_{Ty} (q_x, \omega) = \frac {\tilde{J}_{2\mathrm{D}}} {\tilde{E}^{(t)}_{T}}.
\label{pt2eq11}
\end{equation}
The transmissivity $| T |^2$ and reflectivity $| R |^2$ are the ratios, respectively, of the transmitted and reflected energy fluxes to the incident energy flux.
Equations \eqref{pt2eq6}, \eqref{pt2eq9}, and \eqref{pt2eq11} give these quantities as
\begin{IEEEeqnarray}{rCl}
| T \left( q_x, \omega \right) |^2 &\equiv& \sqrt{\frac {\epsilon_2} {\epsilon_1}} \left| \frac{\tilde{E}^{(t)}_{T}}{\tilde{E}^{(i)}_{T}} \right|^2 = \sqrt{\frac {\epsilon_2} {\epsilon_1}} \left| \frac{2}{1 + \beta \left( q_x, \omega \right)} \right|^2 \nonumber \\ \label{pt2eq12} \\
| R \left( q_x, \omega \right) |^2 &\equiv& \left| \frac{\tilde{E}^{(r)}_{T}}{\tilde{E}^{(i)}_{T}} \right|^2 = \left| \frac{1 - \beta \left( q_x, \omega \right)}{1 + \beta \left( q_x, \omega \right)} \right|^2 \yesnumber \label{pt2e13} \\
A \left( q_x, \omega \right) &=& 1 - |T \left( q_x, \omega \right) |^2 - |R \left( q_x, \omega \right)|^2  \yesnumber \label{eq:Aqxom} \\
&=& \frac{4 \left( {\rm Re} [\beta \left( q_x, \omega \right)] - \sqrt{\epsilon_2 / \epsilon_1} \right)}{\left\vert 1 + \beta \left( q_x, \omega \right) \right\vert^2} \nonumber
\end{IEEEeqnarray}
where $A \left( q_x, \omega \right)$ is the absorptivity and
\begin{IEEEeqnarray}{rCl}
\beta \left( q_x, \omega \right) &=& \frac{1}{q_{ i z}} \left( q_{t z} + \frac{4 \pi \omega}{c^2} \sigma_{Ty} \left( q_x, \omega \right) \right)
\label{pt2eq14} \yesnumber
\end{IEEEeqnarray}
When $\epsilon_1 = \epsilon_2$, Eq.\ \eqref{pt2eq14} reduces to
\begin{equation}
 \beta \left( q_x, \omega \right)  = 1 + \frac{4 \pi \omega}{q_z c^2} \sigma_{Ty} \left( q_x, \omega \right) .
\label{eq:betaqxomsym}
\end{equation}

%===========================
%   Zeroth order equations/ Laser-induced distribution/ Optical steady state
%===========================
\section{Continuous-Wave Laser Distributions}
\label{appx:laserdist}

Our model of a microcavity quantum-well laser is based on Eqs.\ \eqref{p_eh-1.equ}--\eqref{E_ell-1.equ} in the absence of a THz probe.

As explained at the end of Section \ref{sec:theoretical-basis}, the unperturbed (by the THz probe) fields are treated as of zeroth order in $E_{T\nu}$ and carry a superscript $(0)$. Their momentum dependence is simplified to: $E_{\ell \lambda}^{(0)} (\vb{q} , t) = \delta_{\vb{q} \vb{0}} E_{\ell \lambda}^{(0)} ( t )$, $f_{\alpha}^{(0)} (\vb{k}_1, \vb{k}_2, t) = \delta_{\vb{k}_2 , \vb{k}_1} f_{\alpha}^{(0)} (\vb{k}_1, t) , \alpha = e , h$, and $p_{eh}^{(0)} (\vb{k}_1, \vb{k}_2, t) = \delta_{\vb{k}_2 , - \vb{k}_1} p_{eh}^{(0)} (\vb{k}_1, t)$.
Scatterings, e.g.\ carrier-carrier and phonon-carrier, tend to relax the intraband carrier distributions to thermal distributions while pumping keeps the system in a non-equilibrium state. These effects are modeled by adding the following  incoherent terms to the equation for $f_{\alpha}^{(0)} (\vb{k}, t) , \alpha = e,h$, where we distinguish incoherent relaxation and pump terms:
\begin{align}
\hbar \left. \frac {\partial f_{\alpha}^{(0)} (\vb{k}, t)} {\partial t} \right|_{\rm relax} &= - \gamma_F \left( f_{\alpha}^{(0)} (\vb{k}, t) - f_F (\vb{k}) \right) \label{relax-1.equ}\\
& \quad - \gamma_{\mathrm{nr}} f_{\alpha}^{(0)} (\vb{k}, t) \nonumber \\
\hbar \left. \frac {\partial f_{\alpha}^{(0)} (\vb{k}, t)} {\partial t} \right|_{\rm pump} &= - \gamma_{\mathrm{pump}} \left( f_{\alpha}^{(0)} (\vb{k}, t) - f_p (\vb{k}) \right) \label{relax-2.equ}
\end{align}
$f_F (\vb{k})$ is the thermal distribution to which the carriers relax via incoherent intraband scattering, i.e.\ if summed over $\vb{k}$ and spin it yields
the same density as the actual density, i.e.\ the density obtained when
 $f_{\alpha}^{(0)} (\vb{k}, t)$  is summed over $\vb{k}$ and spin,
 and $f_p (\vb{k})$ is a distribution to which the carriers are driven by the pump, and which, if summed over $\vb{k}$ and spin gives the pump density $n_p$,
 which we use as an input parameter.
 We model $f_p (\vb{k})$ also by a Fermi distribution. We have, explicitly,
\begin{equation}
f_{x} \left(\vb{k}; \mu_{x}\right) = \frac{1}{e^{\left( \varepsilon_{\vb{k}} - \mu_{x} \right)/k_B T} + 1} , \label{fermidisteq}
\end{equation}
where $x \in \{F, p\}$.
$T$ is an effective temperature which can generally be different from the lattice temperature, since it accounts for the dynamical equilibrium between the creation of carriers high in the bands and the electron-hole recombination (and other loss) processes.\cite{hu-etal.21}
The thermal ($\mu_{F}$) and pump ($\mu_{p}$) chemical potentials are constrained to give the actual and pump densities, respectively.
Furthermore,
$\gamma_{F}$ is the effective intraband thermalization rate, $\gamma_{\mathrm{pump}}$ is the pump rate, i.e.\ the relaxation rate to the pump distribution, and $\gamma_{\mathrm{nr}}$ is the non-radiative decay rate. For simplicity, we assume in the numerical evaluation the electron and hole
masses to be equal, and thus the electron and hole
populations to share the same distributions $f_F (\vb{k})$ and $f_p (\vb{k})$ and parameters $\gamma_{F}$, $\gamma_{\mathrm{pump}}$, and $\gamma_{\mathrm{nr}}$. We  model the dephasing of the interband polarization and the cavity decay of the cavity field as
\begin{align}
\hbar \left. \frac {\partial p_{eh}^{(0)} (\vb{k}, t)} {\partial t} \right|_{\rm dephasing} &= - \gamma_p p_{\alpha}^{(0)} (\vb{k}, t)
 \label{relax-peh.equ}\\
\hbar \left. \frac {\partial  E_{\ell \lambda}^{(0)} ( t )    } {\partial t} \right|_{\rm decay} &= - \gamma_{E} E_{\ell \lambda}^{(0)} ( t )
  \label{relax-Eell.equ}
\end{align}
This phenomenological model of pump/loss and relaxation is also used in Ref.\ \onlinecite{hu-etal.21} and is more fully discussed there.

The sum of the two terms in Eqs.\ \eqref{relax-1.equ} and \eqref{relax-2.equ} can be simplified to
\begin{multline}
\hbar \left. \frac {\partial f_{\alpha}^{(0)} (\vb{k}, t)} {\partial t} \right|_{\rm relax} + \hbar \left. \frac {\partial f_{\alpha}^{(0)} (\vb{k}, t)} {\partial t} \right|_{\rm pump} \label{relax-12.equ} \\
= - \gamma_f \left( f_{\alpha}^{(0)} (\vb{k}, t) - f_R (\vb{k}) \right)
\end{multline}
where $\gamma_f$ may be interpreted as an effective carrier relaxation rate
\begin{IEEEeqnarray}{rCl}
\gamma_f &=& \gamma_{F} + \gamma_{\mathrm{pump}} + \gamma_{\mathrm{nr}} \label{eq:gmf}
\end{IEEEeqnarray}
and $f_R (\vb{k})$ is an effective target distribution
\begin{IEEEeqnarray}{rCl}
f_R \left(\vb{k}\right) &=& \frac{1}{\gamma_f} \left( \gamma_{F} f_{F} \left(\vb{k}\right) +  \gamma_{\mathrm{pump}} f_{p} \left(\vb{k}\right)\right). \label{eq:f0k}
\end{IEEEeqnarray}
With the incoherent pump, loss and relaxation terms included, Eqs.\ \eqref{p_eh-1.equ}--\eqref{E_ell-1.equ} for the zeroth order fields become
\begin{IEEEeqnarray*}{l}
\IEEEeqnarraymulticol{1}{l}{\left[ i \hbar \frac{\partial}{\partial t} - \left( \varepsilon_{e\vb{k}} + \varepsilon_{h(-\vb{k})} - i \gamma_{p} \right) \right] p^{(0)}_{eh} (\vb{k}, t ) =} \yesnumber \label{eq:ddtP0}  \\
\quad \quad \quad \sum_{\lambda} \left[1 -  f_{e}^{(0)} \left(\vb{k},t\right) - f_{h}^{(0)} \left(-\vb{k},t\right) \right] \Gamma_{eh}^{\lambda} (\vb{k} ,\vb{0}) E^{(0)}_{\ell \lambda} \left(t\right) \nonumber \\
\IEEEeqnarraymulticol{1}{l}{i \hbar \frac{\partial}{\partial t} f_{\alpha}^{(0)} \left(\vb{k}, t\right) = \sum_{\lambda} 2 i \Im \left[ p^{(0) \ast}_{eh} (\pm \vb{k}, t ) \Gamma_{eh}^{\lambda} ( \pm \vb{k} ,\vb{0}) E^{(0)}_{\ell \lambda} ( t ) \right]  }  \nonumber \\
\phantom{i \hbar \frac{\partial}{\partial t} f_{\alpha}^{(0)} \left(\vb{k}, t\right) = \sum_{\lambda}}
- i \gamma_{f} \left( f_{\alpha}^{(0)} \left(\vb{k}, t\right) - f_R \left(\vb{k} \right) \right) \yesnumber \label{eq:ddtf0} \\
\IEEEeqnarraymulticol{1}{l}{\left[ i \hbar \frac{\partial}{\partial t} - \hbar \omega_{\lambda\vb{0}} + i \gamma_E \right] E^{(0)}_{\ell \lambda} (t)
= \sum_{eh \vb{k}} \Gamma_{eh}^{\lambda \ast} (\vb{k}, \vb{0}) p^{(0)}_{eh} (\vb{k}, t)} \nonumber\\
\yesnumber \label{eq:ddtE0}
\end{IEEEeqnarray*}
In Eq.\ \eqref{eq:ddtf0}, the upper sign in $\pm$ is for $\alpha = e$ and the lower sign is for $\alpha = h$.
In the following it is assumed that $f_{e}^{(0)} \left(\vb{k}, t\right) = f_{h}^{(0)} \left(-\vb{k}, t\right) \equiv f^{(0)} \left(\vb{k}, t\right)$, and $f^{(0)} \left(- \vb{k}, t\right) = f^{(0)} \left(\vb{k}, t\right)$.

We seek steady state solutions to Eqs.\ \eqref{eq:ddtP0}-\eqref{eq:ddtE0}, in which the positive frequency parts of $p^{(0)}_{eh}$ and $E^{(0)}_{\ell \lambda}$ oscillate at the lasing frequency $\omega_{\ell}$:
\begin{equation}
E_{\ell \lambda}^{(0)} \left(t\right) = \tilde{E}_{\ell\lambda}^{(0)} e^{-i \omega_{\ell} t}, \
p^{(0)}_{eh} \left(\vb{k}, t\right) = \tilde{p}^{(0)}_{eh} \left(\vb{k}\right) e^{-i \omega_{\ell} t} ,  \label{eq:laserosc}
\end{equation}
and the density distributions are constant in time. We consider only isotropic solutions so that $f^{(0)} (\vb{k})$ and $p^{(0)}_{eh} ( \vb{k} )$ depend only on the momentum's magnitude $k \equiv | \vb{k} |$. Specializing to the case of heavy-hole bands, we have only one electron-hole spin configuration coupled to each photon circular polarization: $\Gamma_{eh}^{\lambda}$ vanish except for $\Gamma_{-1/2,3/2}^{+}$ and $\Gamma^{-}_{1/2,-3/2}$. Since only one term contributes to the sum over $(e,h)$ or $\lambda$ in the equations, we remove the summation symbols over these variables. With these simplifications, we substitute the solution form Eq.\ \eqref{eq:laserosc} into Eqs.\ \eqref{eq:ddtP0}--\eqref{eq:ddtE0} and obtain
\begin{align}
\left[ \hbar\omega_{\ell} -  \Delta\varepsilon \left(k\right)+ i \gamma_{p} \right] \left\vert\tilde{p}^{(0)}_{eh} \left(k\right)\right\vert e^{i \theta_0 ( k )} &= \yesnumber\label{eq:P0motRTAss} \\
[1 - 2 f^{(0)} \left(k\right) ]& | \Gamma^{\lambda}_{eh} (k , 0) | | \tilde{E}_{\ell\lambda}^{(0)} | \nonumber \\
- 2 | \tilde{p}^{(0)}_{eh} \left(k\right) | | \Gamma^{\lambda}_{eh} (k , 0) | | \tilde{E}_{\ell\lambda}^{(0)} | \sin \theta_0 (k) &=
\yesnumber\label{eq:f0motRTAss} \\
\gamma_{f} [f^{(0)} & \left(k\right) - f_{R} \left(k\right)] \nonumber \\
\left[ \hbar\omega_{\ell} -  \hbar \omega_{\lambda\vb{0}} + i \gamma_{E} \right ] | \tilde{E}_{\ell \lambda}^{(0)} | = \sum_{\vb{k}} | \Gamma_{eh}^{\lambda} &(k , 0) | | \tilde{p}^{(0)}_{eh} \left(k\right) | e^{i \theta_0 ( k )}
\label{eq:E0motRTAss}
\end{align}
where $\Delta\varepsilon \left(\vb{k}\right) = \varepsilon_{e \vb{k}} + \varepsilon_{h (-\vb{k})}$. As mentioned above,  we consider parabolic bands $\varepsilon_{\alpha} = \frac{\hbar^2 k^2}{2m_{\alpha}}$. The phase $\theta_0 (k)$ is essentially the relative phase between the interband polarization and the photon field: $\theta_0 (k) = \theta_p (k) - \theta_{\Gamma} (k,0) - \theta_E$ where $\theta_p$, $\theta_{\Gamma}$, and $\theta_E$ are the phases of
$\tilde{p}^{(0)}_{eh}$ , $\Gamma^{\lambda}_{eh}$, and $\tilde{E}_{\ell\lambda}^{(0)}$ respectively.
The above equations are augmented by the density constraints mentioned above, namely that the steady state distribution $f^{(0)} (k)$ and the thermal distribution $f_R (k)$ sum to the same density:
\begin{equation}
n_F = n_{e} = n_{h} \label{eq:nFconstraint}
\end{equation}
where
\begin{equation}
n_{F} = 2 \int \frac{\mathrm{d}^{2} k}{(2\pi)^2} f_{F} \left(\vb{k}\right) , \ n_{\alpha} = 2 \int \frac{\mathrm{d}^{2} k}{(2\pi)^2} f_{\alpha}^{(0)} \left(\vb{k}\right),
\label{nFeq}
\end{equation}
and $\alpha = e, h$.

Apart from the material parameters in the Hamiltonian Eq.\ \eqref{modelH}, the input to solving Eqs.\ \eqref{eq:P0motRTAss}--\eqref{eq:nFconstraint} includes the gain/loss parameters $\{ \gamma_F, \gamma_{\mathrm{pump}}, \gamma_{\mathrm{nr}}, \gamma_p, \gamma_E \}$, the effective temperature $T$
and the pump chemical potential $\mu_p$,
which specifies a target carrier density $n_p$ the pump drives the electron-hole population towards and thus is a measure of the pump strength. The output result includes the field components $| \tilde{E}_{\ell\lambda}^{(0)} |$, $| \tilde{p}^{(0)}_{eh} (k) |$, $\theta_0 (k)$, and $f^{(0)} (k)$, the lasing frequency $\omega_{\ell}$, and the chemical potential $\mu_F$ of the thermal distribution.

\begin{figure}
	\centering
	\includegraphics{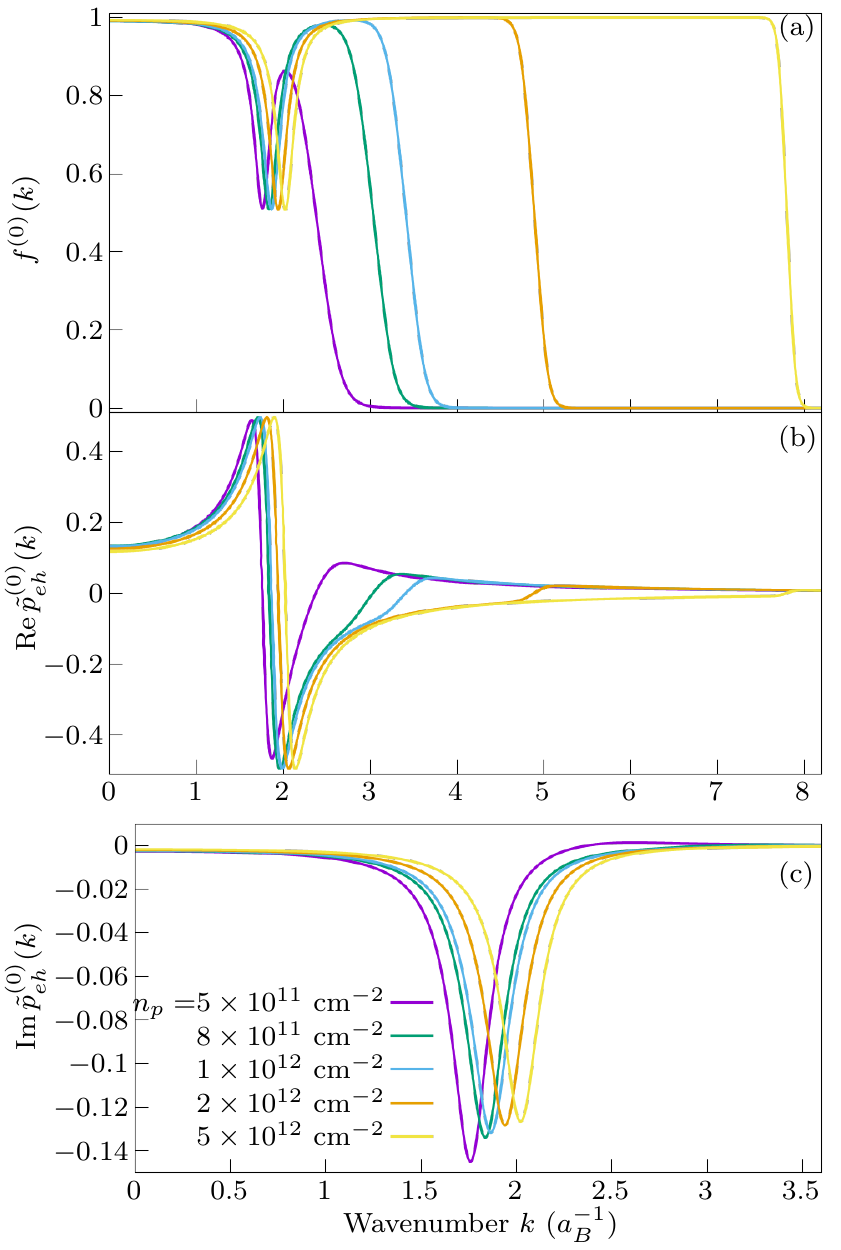}
\caption{(Color online.) (a) The electron density functions $f^{(0)}(k)$ (Eq.\ \eqref{eq:f0cauchy}) and the (b) real and (c) imaginary parts of the  steady-state polarization function $\tilde{p}_{eh}^{(0)}(k)$ (Eq.\ \eqref{eq:P0lorentz}) for different pump densities $n_{p}$.}
	\label{fig:0np}
\end{figure}

Because Eq.\ \eqref{eq:f0motRTAss} is an integral equation, the number of equations to be solved, after discretization, is roughly equal to the number of $k$-points. The numerical task of solution can be made much simpler by some algebraic manipulation of the equations. As we show below, Eqs.\ \eqref{eq:P0motRTAss}--\eqref{eq:nFconstraint} can be reduced to three algebraic equations relating $| \tilde{E}_{\ell\lambda}^{(0)} |$, $\omega_{\ell}$, and $\mu_F$, which are to be solved numerically, and $| \tilde{p}^{(0)}_{eh} (k) |$, $\theta_0 (k)$, and $f^{(0)} (k)$ are expressed explicitly in terms of these three quantities.

The real and imaginary parts of Eq.\ \eqref{eq:P0motRTAss} are
\begin{align}
|\tilde{p}^{(0)}_{e h} (k)| [ \left( \hbar\omega_{\ell} -  \Delta\varepsilon (k) \right) \cos &\theta_0 (k) - \gamma_{p} \sin \theta_0 (k) ] \label{eq:LR211}\\
 &= \left[ 1 - 2f (k) \right] |\Gamma_{eh}^{\lambda} (k,0)| |\tilde{E}_{\ell\lambda}^{(0)}| \nonumber
\end{align}
\begin{equation}
\cos\theta_0 (k)
= - \frac{\hbar\omega_{\ell} -  \Delta\varepsilon (k)}{\gamma_{p}} \sin \theta_0 (k), \ | \tilde{p}_{eh}^{(0)} (k) | \neq 0
\label{eq:LR210}
\end{equation}
Substituting $\cos \theta_0 (k)$ from Eq.\ \eqref{eq:LR210} into Eq.\ \eqref{eq:LR211}, we obtain
\begin{equation}
|\tilde{p}^{(0)}_{e h} (k)| \sin \theta_0 (k)
= -  \frac{\gamma_{p} \left[ 1 - 2f^{(0)} (k) \right] | \Gamma^{\lambda}_{eh} (k , 0) | | \tilde{E}_{\ell\lambda}^{(0)} | }{\left[  \hbar\omega_{\ell} -  \Delta\varepsilon (k) \right]^2 + \gamma_{p}^2  }
\label{eq:LR212}
\end{equation}
Independently, Eq.\ \eqref{eq:f0motRTAss} leads to
\begin{equation}
|\tilde{p}^{(0)}_{e h} (k)| \sin \theta_0 (k)
= -\frac{ \gamma_{f} \left[ f^{(0)} (k) - f_{R} (k) \right]}{2 | \Gamma^{\lambda}_{eh} (k , 0) | | \tilde{E}_{\ell\lambda}^{(0)} | }
\label{eq:LR213}
\end{equation}
We equate the right hand sides of Eqs.\ \eqref{eq:LR212} and \eqref{eq:LR213} to obtain $f^{(0)} (k)$ as a function of $| \tilde{E}_{\ell\lambda}^{(0)} |$, $\mu_F$ (through $f_R (k)$), and $\omega_{\ell}$:
\begin{equation}
	f^{(0)} \left(k\right) =  f_{R} \left(k\right) +  \frac{\left(\tfrac{1}{2} - f_{R} \left(k\right)\right) \gamma_{p}^{2} I}{\left(\hbar\omega_{\ell}- \Delta\varepsilon\left(k\right) \right)^2 + \gamma_{p}^{2}(1+I)} \label{eq:f0cauchy}
\end{equation}
where the dimensionless intensity is $I\equiv\frac{4 \left\vert \Gamma^{\lambda}_{eh}(k,0)\right\vert^{2} \left\vert\tilde{E}_{\ell\lambda}^{(0)}\right\vert^{2}}{\gamma_{p}\gamma_{f}}$. Substituting the expression for $f^{(0)} (k)$ in Eq.\ \eqref{eq:f0cauchy} back into Eq.\ \eqref{eq:P0motRTAss} gives
\begin{multline}
	|\tilde{p}^{(0)}_{eh} (k)| e^{i \theta_0 (k)} =  \left(1 - 2 f_R (k)\right) |\Gamma^{\lambda}_{eh}(k,0)| |\tilde{E}_{\ell\lambda}^{(0)}| \label{eq:P0lorentz} \\
	\times \frac{\hbar \omega_{\ell}  - \Delta\varepsilon\left(k\right) - i \gamma_p}{\left(\hbar\omega_{\ell}- \Delta\varepsilon\left(k\right) \right)^2 + \gamma_{p}^{2}(1+I)}
\end{multline}
Substituting Eq.\ \eqref{eq:P0lorentz} into Eq.\ \eqref{eq:P0motRTAss} and canceling $\tilde{E}_{\ell\lambda}^{(0)}$ gives
\begin{multline}
\hbar \omega_{\ell} - \hbar \omega_{\lambda\vb{0}} + i \gamma_{E} = \sum_{\vb{k}} |\Gamma^{\lambda}_{eh} (k,0)|^2  \left(1 - 2 f_R (k)\right) \\
\times \frac{\hbar \omega_{\ell}  - \Delta\varepsilon\left(k\right) - i \gamma_p}{\left(\hbar\omega_{\ell}- \Delta\varepsilon\left(k\right) \right)^2 + \gamma_{p}^{2}(1+I)} \label{eq:omE0cond}
\end{multline}
Eq.\ \eqref{eq:nFconstraint} and the real and imaginary parts of Eq.\ \eqref{eq:omE0cond} form a set of three independent, nonlinear equations for the three unknown, dependent variables $\omega_{\ell}$, $\tilde{E}_{\ell\lambda}^{(0)}$, and $\mu_{F}$. The system can be solved numerically using e.g.\ Newton's method.\cite{press-etal.92}%\S9.7
Once these dependent variables are obtained, $f^{(0)} \left(k\right)$ can be calculated via Eq.\ \eqref{eq:f0cauchy} and $\tilde{p}^{(0)}_{eh} (k)$ by Eq.\ \eqref{eq:P0lorentz}.

In Eq.\ \eqref{eq:f0cauchy}, the coefficient of $\tfrac{1}{2} - f_{R} \left(k\right)$ is an unnormalized Lorentzian centered at $k=k_{\ell} \equiv \tfrac{1}{\hbar}\sqrt{2m_{r}\hbar\tilde{\omega}_{\ell}}$ and with width $\gamma_{p}^{2}(1+I)$.
$|\tilde{p}^{(0)}_{eh} \left(k\right)| e^{i \theta_0 (k)}$ in Eq.\ \eqref{eq:P0lorentz} is, for a given $k$, formally equivalent to the polarization of a two-level medium under continuous-wave excitation in the rate equation approximation, which shows power-broadening.\cite{meystre-sargent.07}

%===========================
% eh Basis Matrices
%===========================
\section{Linear THz Response Matrices}
\label{appx:ehbasismatrices}

In this appendix, explicit expressions are provided for the linear THz response matrices used in Sec.\ \ref{sec:linear-terahertz-response}. Their behavior for $\vb{q}=0$ is also discussed.

The matrices that appear in Eq.\ \eqref{eq:jecp}, $J$ and $C$, are given by
\begin{IEEEeqnarray}{rCl}
	J(\omega) &=&
	\begin{pmatrix}
		\xi_{\gamma} (\omega) & 0 \\
		0 & \xi_{\gamma}^{\ast} (-\omega)
	\end{pmatrix},
	\yesnumber\label{eq:Jdef}\\
	C\left(\vb{k},\vb{q}\right)
	&=&
	\begin{pmatrix}
		\Gamma_{eh}^{\lambda} (\vb{k}, \vb{q}) & 0 \\
		0 & \Gamma_{eh}^{\lambda\ast} (\vb{k} - \vb{q}, - \vb{q} )
	\end{pmatrix},
	\yesnumber \label{eq:Cdef}
\end{IEEEeqnarray}
where $\xi_{\gamma} (\omega)= \hbar \omega - \left( \hbar \omega_{\lambda\vb{0}} - \hbar \omega_{\ell} \right) + i \gamma_{E}$, and $\hbar\omega_{\lambda\vb{0}}$ is the cavity resonance frequency with zero transverse wavevector.

The matrices $M$, $N$, and $Q$ first appear in Eq.\ \eqref{eq:PEtmat}.
These matrices are written as
\begin{IEEEeqnarray}{rCl}
	M\left(\vb{k},\vb{q},\omega\right) &=&
	\begin{pmatrix}
		a\left(\vb{k},\vb{q},\omega\right) & b\left(\vb{k},\vb{q},\omega\right) \\
		c\left(\vb{k},\vb{q},\omega\right) & d\left(\vb{k},\vb{q},\omega\right)
	\end{pmatrix}, \yesnumber\label{eq:Mmat}\\
N\left(\vb{k},\vb{q},\omega\right) &=&
\begin{pmatrix}
	r\left(\vb{k},\vb{q},\omega\right) & s\left(\vb{k},\vb{q},\omega\right) \\
	t\left(\vb{k},\vb{q},\omega\right) & u\left(\vb{k},\vb{q},\omega\right)
\end{pmatrix}, \yesnumber\label{eq:Nmat}\\
Q\left(\vb{k},\vb{q},\omega\right) &=& \frac{c}{2i\omega}
\begin{pmatrix}
	v\left(\vb{k},\vb{q},\omega\right) & x\left(\vb{k},\vb{q},\omega\right) \\
	y\left(\vb{k},\vb{q},\omega\right) & z\left(\vb{k},\vb{q},\omega\right)
\end{pmatrix}. \label{eq:Qmat}
\end{IEEEeqnarray}
	The energy differences which appear in $M$, $N$, and $Q$ are expressed as
\begin{IEEEeqnarray*}{rClR}
	\xi_{\alpha} \left(\vb{k}_1, \vb{k}_2, \omega\right) &\equiv&  \hbar \omega -  \left( \varepsilon_{\alpha \vb{k}_1 } - \varepsilon_{\alpha \vb{k}_2} \right)  + i \gamma_{f}, & \alpha = e, h, \nonumber \\
	\xi_{e h} (\vb{k}_1, \vb{k}_2, \omega) &\equiv& \hbar \omega - \left( \varepsilon_{e \vb{k}_1 } + \varepsilon_{h \vb{k}_2 } - \hbar \omega_{\ell} \right) + i \gamma_{p} . & \yesnumber
\end{IEEEeqnarray*}
The elements of the matrices $M$, $N$, and $Q$ are given by
\begin{widetext}
{\allowdisplaybreaks
\begin{IEEEeqnarray}{rCl}
 a \left(\vb{k},\vb{q},\omega\right) &=& \xi_{e h} (\vb{k}, \vb{q} - \vb{k}, \omega)
              -\left| \tilde{E}_{\ell\lambda}^{(0)} \right|^2  \left(  \frac{ \left| \Gamma_{eh}^{\lambda}  \left( \vb{k} - \vb{q}, 0 \right) \right|^2  }
                  {\xi_{e} \left( \vb{k}, \vb{k} - \vb{q}, \omega \right) }
 + \frac{ \left| \Gamma_{eh}^{\lambda} \left(\vb{k}, \vb{0} \right) \right|^2 }{\xi_{h} (\vb{q} - \vb{k}, -\vb{k}, \omega)} \right), \yesnumber \\[3pt]
b \left(\vb{k},\vb{q},\omega\right) &=&
 \left(\tilde{E}_{\ell\lambda}^{(0)}\right)^{2} \Gamma_{eh}^{\lambda} (\vb{k} - \vb{q}, 0) \Gamma_{eh}^{\lambda}  (\vb{k}, 0) \left( \frac{
          1 } {  \xi_{e} (\vb{k}, \vb{k} - \vb{q}, \omega) }
         + \frac{1} {  \xi_{h} (\vb{q} - \vb{k}, -\vb{k}, \omega) }\right), \nonumber\\[3pt]
c \left(\vb{k},\vb{q},\omega\right)  &=&
\left(\tilde{E}_{\ell\lambda}^{(0)\ast}\right)^{2}   \Gamma_{eh}^{\lambda\ast} (\vb{k} - \vb{q}, 0) \Gamma_{eh}^{\lambda\ast}  (\vb{k}, 0) \left(
          \frac{ 1 }
                  { \xi_{e}^{\ast} \left( \vb{k} - \vb{q}, \vb{k},-\omega \right) }
         + \frac{1  }
                  { \xi_{h}^{\ast} \left( -\vb{k}, \vb{q} - \vb{k},-\omega \right) } \right), \nonumber\\[3pt]
d  \left(\vb{k},\vb{q},\omega\right) &=& \xi_{e h}^{\ast} \left( \vb{k} - \vb{q}, -\vb{k}, -\omega \right)
                - \left | \tilde{E}_{\ell\lambda}^{(0)} \right|^2 \left(\frac{ \left| \Gamma_{eh}^{\lambda}  \left( \vb{k}, 0 \right)\right|^2    }
                  {  \xi_{e}^{\ast} (\vb{k} - \vb{q}, \vb{k},-\omega)  }
                + \frac{ \left| \Gamma_{eh}^{\lambda} \left( \vb{k} - \vb{q}, 0 \right) \right|^2 }
                  { \xi_{h}^{\ast} (-\vb{k}, \vb{q} - \vb{k},-\omega) }\right); \nonumber
\end{IEEEeqnarray}
\begin{IEEEeqnarray}{rCl}
      r \left(\vb{k},\vb{q},\omega\right) &=&
 \Gamma_{eh}^{\lambda} (\vb{k}, \vb{q}) \left\{ \left(1 - f_{e}^{(0)} \left(\vb{k} \right) - f_{h}^{(0)} \left( \vb{q} - \vb{k} \right) \right)
                - \frac{ \tilde{E}_{\ell\lambda}^{(0)} \tilde{p}_{eh}^{(0)\ast} (\vb{k} - \vb{q})  \Gamma_{eh}^{\lambda} (\vb{k} - \vb{q}, 0)}
                  { \xi_{e} (\vb{k}, \vb{k} - \vb{q}, \omega) } \right. \yesnumber\\
                  &&
                \left. - \frac{ \tilde{E}_{\ell\lambda}^{(0)} \tilde{p}_{eh}^{(0)\ast} (\vb{k})  \Gamma_{eh}^{\lambda}  (\vb{k}, 0) }
                  { \xi_{h} (\vb{q} - \vb{k}, -\vb{k}, \omega) } \right\}, \nonumber\\[3pt]
     s  \left(\vb{k},\vb{q},\omega\right) &=& \Gamma_{eh}^{\lambda\ast} (\vb{k} - \vb{q}, - \vb{q} ) \tilde{E}_{\ell\lambda}^{(0)}   \left(
       \frac{  \tilde{p}_{eh}^{(0)} (\vb{k})   \Gamma_{eh}^{\lambda} (\vb{k} - \vb{q}, 0)    }
                  { \xi_{e} (\vb{k}, \vb{k} - \vb{q}, \omega) }
         + \frac{  \tilde{p}_{eh}^{(0)} (\vb{k} - \vb{q})   \Gamma_{eh}^{\lambda}  (\vb{k}, 0)    }
                  { \xi_{h} (\vb{q} - \vb{k}, -\vb{k}, \omega) }\right), \nonumber\\[3pt]
      t  \left(\vb{k},\vb{q},\omega\right) &=& \Gamma_{eh}^{\lambda} (\vb{k}, \vb{q}) \tilde{E}_{\ell\lambda}^{(0)\ast} \left(
        \frac{     \tilde{p}_{eh}^{(0)\ast} (\vb{k} - \vb{q})   \Gamma_{eh}^{\lambda\ast}  (\vb{k}, 0)    }
                 {  \xi_{e}^{\ast} (\vb{k} - \vb{q}, \vb{k},-\omega) }
         + \frac{  \tilde{p}_{eh}^{(0)\ast} (\vb{k})   \Gamma_{eh}^{\lambda\ast} (\vb{k} - \vb{q}, 0)     }
                  { \xi_{h}^{\ast} (-\vb{k}, \vb{q} - \vb{k},-\omega)  }\right), \nonumber\\[3pt]
      u \left(\vb{k},\vb{q},\omega\right) &=&
 \Gamma_{eh}^{\lambda\ast} (\vb{k} - \vb{q}, - \vb{q} ) \left\{   \left(1 - f_{e}^{(0)} \left(\vb{k} -  \vb{q} \right) - f_{h}^{(0)} \left(-\vb{k} \right) \right)
           \right. \nonumber \\
   &&   \left. -  \tilde{E}_{\ell\lambda}^{(0)\ast} \left[  \frac{  \Gamma_{eh}^{\lambda\ast} (\vb{k}, 0)   \tilde{p}_{eh}^{(0)} (\vb{k}) }
   { \xi_{e}^{\ast} (\vb{k} - \vb{q}, \vb{k},-\omega) } + \frac{    \Gamma_{eh}^{\lambda\ast} (\vb{k} - \vb{q}, 0)   \tilde{p}_{eh}^{(0)} (\vb{k} - \vb{q}) }
                 {  \xi_{h}^{\ast} (-\vb{k}, \vb{q} - \vb{k},-\omega)  } \right] \right\}; \nonumber
\end{IEEEeqnarray}
and
\begin{IEEEeqnarray*}{rCl}
      v \left(\vb{k},\vb{q},\omega\right) &=&
 - \tilde{E}_{\ell\lambda}^{(0)}  \left\{ \frac{\Gamma_{eh}^{\lambda} (\vb{k} - \vb{q}, 0)}{\xi_{e} (\vb{k}, \vb{k} - \vb{q}, \omega)}
            \left[ \tilde{f}_e^{(0)} (\vb{k} - \vb{q}) - \tilde{f}_e^{(0)} (\vb{k}) \right] g_{e}^{\nu}\left(\vb{k} - \tfrac{1}{2} \vb{q}\right) \right.  \yesnumber \label{eq:Qelem} \\
      &&  \phantom{- \tilde{E}_{\ell\lambda}^{(0)}\left(\right.} \left. + \frac{ \Gamma_{eh}^{\lambda}  (\vb{k}, 0)}{\xi_{h} (\vb{q} - \vb{k}, -\vb{k}, \omega)} \left[  \tilde{f}_h^{(0)} (-\vb{k}) - \tilde{f}_h^{(0)} (\vb{q} - \vb{k}) \right]  g_{h}^{\nu}\left(\tfrac{1}{2}\vb{q}-\vb{k}\right)  \right\}  \nonumber\\
        && +     \tilde{p}_{eh}^{(0)} (\vb{k} - \vb{q}) g_{e}^{\nu}\left(\vb{k} - \tfrac{1}{2} \vb{q}\right)
         +     \tilde{p}_{eh}^{(0)} (\vb{k}) g_{h}^{\nu}\left(\tfrac{1}{2}\vb{q}-\vb{k}\right), \nonumber\\[4pt]
      x \left(\vb{k},\vb{q},\omega\right) &=&
 - \tilde{E}_{\ell\lambda}^{(0)}  \left\lbrace
          \frac{  \Gamma_{eh}^{\lambda} (\vb{k} - \vb{q}, 0) }{ \xi_{e} (\vb{k}, \vb{k} - \vb{q}, \omega)  } \left[  \tilde{f}_e^{(0)} (\vb{k} - \vb{q}) - \tilde{f}_e^{(0)} (\vb{k})\right]  g_{e}^{\nu}\left(\vb{k} - \tfrac{1}{2} \vb{q} \right)   \right. \nonumber\\
      &&  \phantom{- \tilde{E}_{\ell\lambda}^{(0)}\left(\right.}
       \left. +  \frac{ \Gamma_{eh}^{\lambda}  (\vb{k}, 0) }{ \xi_{h} (\vb{q} - \vb{k}, -\vb{k}, \omega) } \left[  \tilde{f}_h^{(0)} (-\vb{k}) - \tilde{f}_h^{(0)} (\vb{q} - \vb{k}) \right]  g_{h}^{\nu}\left(\tfrac{1}{2}\vb{q}-\vb{k}\right) \right\rbrace  \nonumber\\
  &&       +     \tilde{p}_{eh}^{(0)} (\vb{k} - \vb{q}) g_{e}^{\nu}\left(\vb{k} - \tfrac{1}{2} \vb{q}\right)
         +     \tilde{p}_{eh}^{(0)} (\vb{k}) g_{h}^{\nu}\left(\tfrac{1}{2}\vb{q}-\vb{k}\right) , \nonumber\\[4pt]
     y \left(\vb{k},\vb{q},\omega\right) &=&
 - \tilde{E}_{\ell\lambda}^{(0)\ast} \left\lbrace
            \frac{\Gamma_{eh}^{\lambda\ast}  (\vb{k}, 0)  }{\xi_{e}^{\ast} (\vb{k} - \vb{q}, \vb{k},-\omega)  }  \left[  \tilde{f}_e^{(0)} (\vb{k}) - \tilde{f}_e^{(0)} (\vb{k} - \vb{q}) \right]  g_{e}^{\nu}\left(\vb{k} - \tfrac{1}{2} \vb{q} \right)   \right. \nonumber\\
   &&   \phantom{- \tilde{E}_{\ell\lambda}^{(0)\ast}\left(\right.} \left.  +  \frac{ \Gamma_{eh}^{\lambda\ast} (\vb{k} - \vb{q}, 0) }{ \xi_{h}^{\ast} (-\vb{k}, \vb{q} - \vb{k},-\omega)  }   \left[  \tilde{f}_h^{(0)} (\vb{q} - \vb{k}) - \tilde{f}_h^{(0)} (-\vb{k}) \right]  g_{h}^{\nu}\left(\tfrac{1}{2}\vb{q}-\vb{k}\right)  \right\rbrace  \nonumber\\
 &&        +     \tilde{p}_{eh}^{(0)\ast} (\vb{k})  g_{e}^{\nu}\left(\vb{k} - \tfrac{1}{2} \vb{q} \right)
         +     \tilde{p}_{eh}^{(0)\ast} (\vb{k} - \vb{q})  g_{h}^{\nu}\left(\tfrac{1}{2}\vb{q}-\vb{k}\right), \nonumber\\[4pt]
      z \left(\vb{k},\vb{q},\omega\right) &=&
 - \tilde{E}_{\ell\lambda}^{(0)\ast} \left\lbrace
            \frac{\Gamma_{eh}^{\lambda\ast}  (\vb{k}, 0)}{ \xi_{e}^{\ast} (\vb{k} - \vb{q}, \vb{k},-\omega) }  \left[  \tilde{f}_e^{(0)} (\vb{k}) - \tilde{f}_e^{(0)} (\vb{k} - \vb{q})  \right]  g_{e}^{\nu}\left(\vb{k} - \tfrac{1}{2} \vb{q}\right)  \right. \nonumber\\
  &&  \phantom{- \tilde{E}_{\ell\lambda}^{(0)\ast}\left(\right.}   \left.   +  \frac{ \Gamma_{eh}^{\lambda\ast} (\vb{k} - \vb{q}, 0) }{ \xi_{h}^{\ast} (-\vb{k}, \vb{q} - \vb{k},-\omega)  }  \left[  \tilde{f}_h^{(0)} (\vb{q} - \vb{k}) - \tilde{f}_h^{(0)} (-\vb{k})  \right]  g_{h}^{\nu}\left(\tfrac{1}{2}\vb{q}-\vb{k}\right)  \right\rbrace  \nonumber\\
   &&      +     \tilde{p}_{eh}^{(0)\ast} (\vb{k})   g_{e}^{\nu}\left(\vb{k}-\tfrac{1}{2} \vb{q}\right)
         +     \tilde{p}_{eh}^{(0)\ast} (\vb{k} - \vb{q})  g_{h}^{\nu}\left(\tfrac{1}{2}\vb{q}-\vb{k}\right) . \nonumber
\end{IEEEeqnarray*}

The matrices in Eq.\ \eqref{eq:f1ETmatform} are defined as
\begin{IEEEeqnarray*}{rCl}
G\left(\vb{k},\vb{q}\right) &=& \begin{pmatrix}
\tilde{p}_{eh}^{(0)\ast} (\vb{k} - \vb{q}) \Gamma_{eh}^{\lambda} (\vb{k}, \vb{q})
& - \tilde{p}_{eh}^{(0)} (\vb{k}) \Gamma_{eh}^{\lambda\ast} (\vb{k} - \vb{q}, - \vb{q} )
\\
\tilde{p}_{eh}^{(0)\ast} (\vb{k}) \Gamma_{eh}^{\lambda} (\vb{k}, \vb{q})
&  - \tilde{p}_{eh}^{(0)} (\vb{k} - \vb{q}) \Gamma_{eh}^{\lambda\ast} (\vb{k} - \vb{q}, - \vb{q} )
\end{pmatrix} \yesnumber\label{eq:Gmat} \\[4pt]
H\left(\vb{k},\vb{q}\right)  &=& \begin{pmatrix}
- \tilde{E}_{\ell\lambda}^{(0)\ast} \Gamma_{eh}^{\lambda\ast}  (\vb{k} - \vb{q}, \vb{0})
& \tilde{E}_{\ell\lambda}^{(0)} \Gamma_{eh}^{\lambda} (\vb{k}, \vb{0})   \\
 - \tilde{E}_{\ell\lambda}^{(0)\ast} \Gamma_{eh}^{\lambda\ast}  (\vb{k}, \vb{0})
&  \tilde{E}_{\ell\lambda}^{(0)} \Gamma_{eh}^{\lambda} (\vb{k} - \vb{q}, \vb{0})
\end{pmatrix} \yesnumber\label{eq:Hmat} \\[4pt]
L\left(\vb{k},\vb{q},\omega\right)  &=& \frac{c}{2i\omega}
\begin{pmatrix}
f_{e}^{(0)} (\vb{k} - \vb{q}) - f_{e}^{(0)} (\vb{k})  & 0 \\
0 & f_{h}^{(0)} (- \vb{k} ) - f_{h}^{(0)} ( \vb{q} - \vb{k})
\end{pmatrix}
\begin{pmatrix}
g_{e}^{\nu}\left(\vb{k}-\tfrac{1}{2}\vb{q}\right)
& g_{e}^{\nu}\left(\vb{k}-\tfrac{1}{2}\vb{q}\right) \\
g_{h}^{\nu}\left(\tfrac{1}{2}\vb{q}-\vb{k}\right)
& g_{h}^{\nu}\left(\tfrac{1}{2}\vb{q}-\vb{k}\right)
\end{pmatrix} \yesnumber\label{eq:lindmat} \\[4pt]
\Xi\left(\vb{k},\vb{q},\omega\right) &=&
\begin{pmatrix}
\frac{1}{\xi_{e} (\vb{k}, \vb{k} - \vb{q}, \omega)} & 0 \\
0 &  \frac{1}{\xi_{h} (\vb{q} - \vb{k}, - \vb{k}, \omega)}
\end{pmatrix}. \yesnumber\label{eq:Ximat}
\end{IEEEeqnarray*}}
\end{widetext}

The matrix
\begin{equation}
	\hat{f}_{\mathrm{L}}^{(1)}
	\equiv
	\begin{pmatrix}
		\tilde{f}_{e,\mathrm{L}}^{(1)} (\vb{k}, \vb{k} - \vb{q}, \omega)
		\\
		\tilde{f}_{h,\mathrm{L}}^{(1)} (\vb{q} - \vb{k}, - \vb{k}, \omega)
	\end{pmatrix}
	= \Xi L \hat{E}_{T\nu}
	\label{eq:lindrespmatdef}
\end{equation}
gives the Lindhard-like linear response of the carrier densities. This is evaluated analytically in the limits of zero temperature $T \to 0$ and zero pump and decay $\gamma_{f} \to 0$ in Appx.\ \ref{appx:Lindhard}. However, for normal THz wavelengths, $\vb{q} \ll \vb{k}$, the matrices $D^{\nu}$ and $L$ are quite small compared to $X^{\nu}$.
When $\vb{q} = 0$, $D^{\nu} = 0$ and $L =0$ exactly.
That $L(\vb{q}=0)=0$ can be seen directly from the first matrix in the definition of $L$, Eq.\ \eqref{eq:lindmat}.

It is also true that $D^{\nu}(\vb{q}=0,\omega)=0$, because $D^{\nu}$ is a sum of terms linear in the factors $g_{\alpha}^{\nu}(\vb{k})$. To see this, note that of the matrices that appear in the definition of $D$ in Eq.\ \eqref{eq:Dmatdef}, only $Q$ contains any factors of $g_{\alpha}^{\nu}$. As is shown in Eqs.\ \eqref{eq:Qmat} and \eqref{eq:Qelem}, $Q$ is linear in the factors $g_{\alpha}^{\nu}$. All of the other (zeroth order) functions that enter the linear response matrices are taken to be isotropic in the QW's plane, $f(\vb{k})=f(k)$. As given by Eq.\ \eqref{eq:gnuform}, $g_{\alpha}^{\nu} \propto \vb{k}$. Therefore the elements of $D^{\nu}(q=0)$ have the form $\sum_{\vb{k}} \vb{k} f(k)$, where $f(k)$ is isotropic in $k$. As this is an integral with even limits over an odd function, it is zero.

More generally, any of the quantities that are $\vb{k}$-sums of terms linear in $g_{\alpha}^{\nu}$ will be proportional to $\vb{q}$. Because at THz frequencies the EM field has $\vb{q}\ll\vb{k}$, those quantities that contain sums over odd powers of $g_{\alpha}^{\nu}$ are negligible compared to those that contain even powers of $g_{\alpha}^{\nu}$. Fundamentally, the intraband processes induced by the THz field break the rotational symmetry of the quantum well.

%===========================
% Joint Density of States
%===========================
\section{Joint Density of States}
\label{app:jdos}

In this appendix we provide brief remarks on the joint density of states governing vertical transitions between the original and the light-induced branches, see Fig.\ \ref{fig:JDOS_spect_cl}.

Under the influence of the optical field, the band structure is effectively modified to have two branches with energies $\lambda_{\pm}(\vb{k})$, given by Eq.\ \eqref{eq:nb4}. The transition energy between the two band branches is
\begin{equation}
\Delta\lambda(\vb{k})=\lambda_{+}(\vb{k})-\lambda_{-}(\vb{k}) = \sqrt{4\Delta_{\ell}^{2} + \left(\frac{\hbar^2 k^2}{2m_{r}} - \hbar\tilde{\omega}_{\ell}\right)^2},
\end{equation}
where the Rabi frequency is $\Delta_{\ell} \equiv \left\vert\Gamma_{eh}^{\lambda}(0) \tilde{E}_{\ell\lambda}^{(0)}\right\vert$, the reduced mass is given by $\frac{1}{m_r} = \frac{1}{m_e} + \frac{1}{m_h}$, and here, $\hbar\tilde{\omega}_{\ell} = \hbar\omega_{\ell}-E_g$, $E_{g}$ being the band gap. The minimum value of the transition energy is $\min \Delta\lambda(\vb{k}) = \Delta\lambda(k_{\ell}) = 2\Delta_{\ell}$, which occurs at $k_{\ell} = \sqrt{\frac{2m_{r}\tilde{\omega}_{\ell}}{\hbar}}$. $k_{\ell}$ is the quasi-momentum at which the electron and hole energies in the undressed bands are resonant with the laser frequency $\hbar\omega_{\ell}$.

The density of states $g(\varepsilon)$ is defined by
\begin{equation*}
\frac{S_d}{\mathcal{A}} \sum_{\vb{k}} f(\vb{k}) = \int \mathrm{d} \varepsilon \, g(\varepsilon) f(\varepsilon(\vb{k})) ,
\end{equation*}
where $S_{d}$ is the spin degeneracy factor.
The joint density of states (JDOS)\cite{kalt-klingshirn.19} for a transition from the $\alpha$ band to the $\beta$ band is
\begin{equation*}
g_{J} (\varepsilon) = \frac{S_d}{\mathcal{A}}  \sum_{\vb{k}} \delta \left( \varepsilon - \left[ \varepsilon_{\beta \vb{k}} - \varepsilon_{\alpha \vb{k}} \right] \right).
\label{eq:JDOSdef}
\end{equation*}

The JDOS for transitions from the lower branch to the upper branch of the dressed conduction band is given by
\begin{IEEEeqnarray}{rCl}
g_{J}(\varepsilon) &=& \frac{ S_d }{ 2 \pi}  \, \int_{0}^{\infty} k \, \mathrm{d}k \,   \delta \left( \varepsilon - \Delta\lambda(\vb{k})  \right) \\
 &=&  \frac{ S_d }{ 2 \pi} \sum_{i}\frac{k_i}{ \left|{ \frac{\mathrm{d}\Delta\lambda(k_i)}{\mathrm{d}k} }\right| }, \nonumber \\
\text{where} \quad \Delta \lambda (k_i) &=& \varepsilon, \
k_i \geq 0  \nonumber \\
\Rightarrow k_i &=&  \frac{\sqrt{2 m_{r}}}{\hbar} \sqrt{\hbar\tilde{\omega}_{\ell} \pm \sqrt{\varepsilon^{2}-4 |\Delta_{\ell}|^{2}}}
; \nonumber
\end{IEEEeqnarray}
and where the spin degeneracy is $S_{d}=2$.
This is evaluated to give the JDOS as
\begin{equation}
g_{J} (\varepsilon) = \frac{m_{r}}{\pi\hbar^2} \frac{\varepsilon}{{\sqrt{\varepsilon^{2}-4\Delta_{\ell}^{2}}}} \theta \left(\varepsilon - 2 \Delta_{\ell} \right) \yesnumber \label{eq:jdos_y_ss_cintra}  \left[1+\theta\left(\Delta\lambda(0)-\varepsilon\right)\right].
\end{equation}
where $\Delta\lambda(k=0)=\sqrt{\hbar^{2}\tilde{\omega}_{\ell}^{2}+4\Delta_{\ell}^{2}}$ is the maximum transition energy for $k<k_{\ell}$.

The joint density of states, Eq.\ \eqref{eq:jdos_y_ss_cintra}, is shown in figure \ref{fig:JDOS_spect_cl}(a).
At the BCS gap $\varepsilon = 2\Delta_{\ell}$, there is a Van Hove singularity in the JDOS.\cite{ashcroft-mermin.76}
The limit as $\varepsilon$ approaches $2\Delta_{\ell}$ is
\begin{equation*}
\lim_{\delta\equiv\varepsilon-2|\Delta_{\ell}| \to 0^{+}} g_{J} (\varepsilon) = \lim_{\delta \to 0^{+}} \frac{2m_{r}}{\pi\hbar^2} \sqrt{\frac{\Delta_{\ell}}{\delta}} \theta \left(\delta\right).
\end{equation*}
The plot in figure \ref{fig:JDOS_spect_cl}(a) shows that the BCS gap Van Hove singularity is the dominant feature in the JDOS. Semiconductor absorption spectra are often functions of the JDOS.\cite{kalt-klingshirn.19} Therefore the BCS gap is expected to be a noticeable feature in the THz spectra. In many of the plots in this paper, the BCS gap is indicated with a dashed, vertical line.

For $\varepsilon<\Delta\lambda(0)$, there are two equal contributions to the JDOS.
One contribution is from transitions between the two dressed band branches for $k \leq k_{\ell}$, and the other is from transitions with $k \geq k_{\ell}$, as shown in figure \ref{fig:JDOS_spect_cl}(b). When $\varepsilon>\Delta\lambda(0) $, transitions with $k<k_{\ell}$ no longer contribute to the JDOS.
This gives rise to the step at high energies in Fig.\ \ref{fig:JDOS_spect_cl}(a).

%===========================
% Pair Excitation Region
%===========================
\section{Pair Excitation Region}
\label{sec:app:per}

In this appendix, the calculations of the pair excitation regions shown in Fig.\ \ref{fig:pair-excitation} are explained.

By an extension of an argument in Ref.\ \onlinecite{mahan.00}, the rate of pair excitations from one band $\alpha$ to another $\alpha^{\prime}$ for a given $\vb{q},\omega$ is
\begin{multline}
	R_{+,\alpha\to\alpha^{\prime}}(\vb{q},\omega) = \sum_{\vb{k}} \delta\left(\hbar\omega-\varepsilon_{\alpha^{\prime}}(\vb{k})+\varepsilon_{\alpha} (\vb{k}-\vb{q})\right) \\
	\times f_{\alpha}(\vb{k}-\vb{q}) \left[1-f_{\alpha^{\prime}}(\vb{k})\right],
	\label{eq:PERdef}
\end{multline}
up to a proportionality factor.
The rate of pair de-excitations or relaxations back from $\alpha^{\prime}$ to $\alpha$ for the same $\vb{q},\omega$ is $R_{-,\alpha\to\alpha^{\prime}}(\vb{q},\omega)$, which is equal to $R_{+,\alpha^{\prime}\to\alpha}(-\vb{q},-\omega)$.
The net rate of pair excitations is then $R_{\alpha\to\alpha^{\prime}} = R_{+,\alpha\to\alpha^{\prime}}-R_{-,\alpha\to\alpha^{\prime}}$.

We define the pair excitation region (PER) as that $\vb{q},\omega$ domain for which $R_{+,\alpha\to\alpha^{\prime}}(\vb{q},\omega) \neq 0$.
In the following, we consider the PER for $\vb{q},\omega> 0$.
Analytical formulae are provided for two limiting cases of transitions accessible in the band structure depicted in Fig.\ \ref{fig:bandstruct}.
In general, pair excitation regions can be calculated using Eq.\ \eqref{eq:PERdef}.

For an e-h plasma at zero temperature and not coupled to a cavity, the intraband pair excitation region is given in Ref.\ \onlinecite{mahan.00} and shown in Fig.\ \ref{fig:pair-excitation}(a). For a given $q$, the maximum and minimum $\omega >0$ transitions begin from the Fermi wavenumber, $k_{F}$. The boundaries of the PER are parabolic, like the original band structure.

In a photon laser, the band structure is modified and the eigenenergies are $\lambda_{\pm}(k)$, given in Eq.\ \eqref{eq:nb4}.
The BCS-like gap is clearly revealed in the PER if the $\lambda_{-}(k)$ light-induced branch is occupied for approximately $k \leq k_{\ell}$, (and so here we define $k_{F}\equiv k_{\ell}$), and the $\lambda_{+}(k)$ light-induced branch is unoccupied. This is the case for an electron-hole-photon system in quasi-thermal equilibrium, in the $T\to0$ limit (cf.\ Appx.\ \ref{appx:bcs-quasi-equil}).

If only transitions to the $\lambda_{+}(k)$ branch, and to $k\geq k_{\ell}$, are allowed, then the pair excitation region is given by
\begin{IEEEeqnarray}{rCCClCrCl}
	2\Delta_{\ell} & \leq & \hbar\omega &\leq& \Delta_{\ell} + \lambda_{+}(k_{\ell}+q)   & \ \  \text{if} \ \  & q &<& 2 k_{\ell}; \qquad \yesnumber\label{eq:kfklper}\\
	\Delta_{\ell} + \lambda_{+}(k_{\ell}-q)  & \leq & \hbar\omega &\leq& \Delta_{\ell} + \lambda_{+}(k_{\ell}+q)  & \ \ \text{if} \ \  & q &>& 2 k_{\ell}. \nonumber
\end{IEEEeqnarray}
If transitions from the lower light-induced branch $\lambda_{-} (k)$ to the upper light-induced branch $\lambda_{+}(k)$ are allowed for all $k$,
then the PER is modified in the domain $q < 2k_{\ell}$ to be
\begin{equation}
	2\Delta_{\ell} \leq \hbar\omega \leq
	\max \begin{Bmatrix} \lambda_{+}(q) - \lambda_{-} (0) \\
	\lambda_{+}(k_{\ell}+q) + \Delta_{\ell}  \end{Bmatrix} \label{eq:om1dckf}
\end{equation}
This case is plotted in Fig.\ \ref{fig:pair-excitation}(b).

%===========================
%Quasi-Equilibrium Model of the Optical Steady State
%===========================
\section{Quasi-thermal-equilibrium model}
\label{appx:bcs-quasi-equil}

A quasi-thermal equilibrium model assumes that the electrons, holes and photons are in thermal equilibrium with each other at temperature $T$. The populations are not determined by pump and decay processes, as in the open system, but by the chemical potential $\mu$, which is used as a parameter. Within the Hartree-Fock (or self-consistent field) approximation, the expectation values for the interband polarization, carrier distributions and coherent light field amplitude can be obtained from the Hamiltonian and a density operator for the grand-canonical ensemble. This has been done in Ref.\ \onlinecite{yamaguchi-etal.15}.
Our Hamiltonian, Eq.\ \eqref{modelH}, coincides with that used in Ref.\ \onlinecite{yamaguchi-etal.15} if the Coulomb contributions in Ref.\ \onlinecite{yamaguchi-etal.15} are neglected. Hence, in this case we can use the quasi-thermal equilibrium solutions given as Eqs.\ (16)--(18) in Ref.\ \onlinecite{yamaguchi-etal.15}. Written in the notation of this paper, the solutions are
\begin{IEEEeqnarray}{rCl}
\tilde{E}_{\ell\lambda}^{(0)} &=& \sum_{\vb{k}^{\prime}} \frac{|\Gamma_{eh}^{\lambda}|}{\hbar\omega_{\vb{0}\lambda} -\mu} \tilde{p}_{eh}^{(0)}(\vb{k}^{\prime}), \yesnumber\label{eq:bcsEl0}\\
\tilde{p}_{eh}^{(0)}(\vb{k}) &=& \frac{\Delta_{\ell}}{2E_{eh}(\vb{k})}\tanh\left(\frac{E_{eh}(\vb{k})}{2k_{B}T}\right), \yesnumber\label{eq:BCSpeh0}\\
f_{e/h}^{(0)}(\pm\vb{k}) &=& \frac{1}{2} \left[ 1 - \frac{\xi_{eh}^{+}(\vb{k})}{E_{eh}(\vb{k})}\tanh\left(\frac{E_{eh}(\vb{k})}{2k_{B}T}\right)\right].
\yesnumber\label{eq:BCSfeh0}
\end{IEEEeqnarray}
The distribution function in Eq.\ \eqref{eq:BCSfeh0} denotes $f_{e}^{(0)}(\vb{k})$ for electrons and $f_{h}^{(0)}(-\vb{k})$ for holes.
$\xi_{eh}^{+}(\vb{k})$ and $E_{eh}(\vb{k})$ are defined in Eq.\ \eqref{eq:Eehk}, and $\hbar\omega_{\vb{0}\lambda}$ is the fundamental cavity mode.
In the absence of Coulomb interactions, and with $\Gamma_{eh}^{\lambda}$ taken as independent of $\vb{k}$, the Rabi energy $\Delta_{\ell} = \left\vert\Gamma_{eh}^{\lambda} \tilde{E}_{\ell\lambda}^{(0)}\right\vert$ is also $\vb{k}$-independent.
The laser frequency is given by $\hbar\omega_{\ell} = \mu$.
By substituting Eq.\ \eqref{eq:BCSpeh0} into Eq.\ \eqref{eq:bcsEl0}, a single nonlinear equation is found which determines the Rabi frequency:
\begin{equation}
1 = \sum_{\vb{k}} \frac{|\Gamma_{eh}^{\lambda}|^2}{\hbar\omega_{\vb{0}\lambda}-\mu}\frac{1}{2E_{eh}(\vb{k})}\tanh\left(\frac{E_{eh}(\vb{k})}{2k_{B}T}\right) . \label{eq:BCSDeltaeq}
\end{equation}
It is necessary for $\hbar\omega_{\vb{0}\lambda} \geq \mu$ for Eq.\ \eqref{eq:BCSDeltaeq} to have a solution.
Once Eq.\ \eqref{eq:BCSDeltaeq} is solved for $\Delta_{\ell}$ for a given set of input parameters, the expectation values given by Eqs.\ \eqref{eq:bcsEl0}--\eqref{eq:BCSfeh0} can be calculated.

Eqs.\ \eqref{eq:bcsEl0}--\eqref{eq:BCSfeh0} satisfy the monochromatic semiconductor Bloch equations (SBEs) Eqs.\ \eqref{eq:P0motRTAss}--\eqref{eq:E0motRTAss} when $\hbar\omega_{\ell}=\mu$ and the phenomenological reservoir interaction terms are zero, i.e.\ $\gamma_{p}$, $\gamma_{f}$, and $\gamma_{E} =0$.
Therefore, as the THz spectrum is calculated as a perturbation on the SBEs, the formalism of this paper can be used to calculate the THz spectrum for the quasi-equilibrium case.
Using these conditions and Eq.\ \eqref{eq:nb25}, the $\omega>0$-resonant paramagnetic THz conductivity is
\newcommand{\PV}{{\cal{P}\,}}
\begin{IEEEeqnarray}{rCl}
\sigma_{T\nu}^{p}(\omega)&=& \frac{c}{2\pi} \frac{\alpha_{0}}{\hbar\omega}\left(\frac{\hbar^2}{m_{r}}\right)^{2}\int_{0}^{\infty} k^{3}\mathrm{d}k \,  u(k) v(k)  \yesnumber\\
&& \times \frac{\Delta_{\ell}}{2E_{eh}(k)} \tanh\left(\frac{E_{eh}(k)}{2k_{B}T}\right) \nonumber\\
&& \times \left[ \PV \frac{i}{\hbar\omega -\Delta \lambda (k)} + \pi \delta(\hbar\omega-\Delta\lambda(k)) \right] \nonumber
\end{IEEEeqnarray}
The delta function in the last line gives the real part of the conductivity, while the Cauchy principal value ($\cal{P}$) integral gives the imaginary part. This delta function is the same as that which appears in the formula for the JDOS.
Evaluating the delta function and simplifying gives the real part of the conductivity,
\begin{IEEEeqnarray}{rCl}
\Re \sigma_{T\nu}^{p}(\omega)&=&  \frac{c \alpha_{0} \Delta_{\ell}^{2}}{\hbar^{2}\omega^{2}} \frac{\theta(\hbar\omega-2\Delta_{\ell})}{\sqrt{\hbar^2\omega^2-4\Delta_{\ell}^{2}}} \tanh\left(\frac{\hbar\omega}{4k_{B}T}\right) \yesnumber\\
&& \times \left[  2\hbar\tilde{\omega}_{\ell} \theta\left(\Delta\lambda(0)-\hbar\omega\right)
 \right. \nonumber\\
&& \left. + \left(\hbar\tilde{\omega}_{\ell} + \sqrt{\hbar^2\omega^2-4\Delta_{\ell}^{2}}\right)\theta\left(\hbar\omega-\Delta\lambda(0)\right) \right]. \nonumber
\end{IEEEeqnarray}
For $\omega$, $T$, and $\hbar\tilde{\omega}_{\ell} > 0$, the real part of the paramagnetic THz conductivity is nonnegative, $\Re \sigma_{T\nu}^{p}(\omega) \geq 0$.
In turn, %
the absorptivity is nonnegative, i.e.\ there is no THz gain in the quasi-equilibrium case. This is verified numerically. The absence of THz gain means that a system in chemical and thermal quasi-equilibrium would not impart energy to a probe beam.
The THz spectrum is dominated by a single, sharp peak at the BCS gap $2\Delta_{\ell}$.

\section{Analytical Solution for the Lindhard Response}
\label{appx:Lindhard}

In this appendix, we summarize the derivation of the analytical expression of the THz paramagnetic conductivity in a non-interacting gas of electrons and holes in a quantum well that is not coupled to a laser field. The result is related to the 2D Lindhard function
(cf.\ \onlinecite{mihaila.11}).

According to Eqs.\ \eqref{current-def.equ} and \eqref{eq:thz-cond-def}, the paramagnetic conductivity is given by the THz-induced current $J_{\alpha}^{p\nu(1)}$ which depends on the induced density fluctuations $\tilde{f}_{\alpha}^{(1)}$, $\alpha = e , h$. Setting $\tilde{E}_{\ell\lambda}^{(0)}$ and $\tilde{p}^{(0)}_{e h}$ to zero in Eqs.\ \eqref{eq:f1eFT} and \eqref{eq:f1hFT} gives the linear density fluctuation as
\begin{multline}
\tilde{f}_{\alpha\mathrm{L}}^{(1)} ( \vb{k} +\vb{q} , \vb{k} , \omega ) = \frac{c} {i \omega} \left( \frac{f_{\alpha}^{(0)} (\vb{k})-f_{\alpha}^{(0)} (\vb{k} + \vb{q})}{\hbar\omega - \varepsilon_{\alpha,\vb{k}+\vb{q}} + \varepsilon_{\alpha,\vb{k}} + i\gamma_{f}}\right) \\ \times g_{\alpha}^{\nu} ( \vb{k} + \vb{q} / 2 ) \tilde{E}_{T\nu} ( \vb{q} , \omega ) , \quad \alpha = e , h .
\end{multline}

We substitute this into Eqs.\ \eqref{current-def.equ} and \eqref{eq:thz-cond-def} and, after some algebraic simplification, obtain the paramagnetic conductivity as
\begin{equation}
\sigma_{T\nu}^{p} (\vb{q},\omega) = c \frac{4\alpha_{0}}{ i}  \sum_{\alpha \in \{e,h\}} \frac {1} {\hbar \omega_{\alpha}^{\prime}}
L_{\alpha\nu} \left(\vb{q}^{\prime},\omega_{\alpha}^{\prime}\right)
\label{eq:ldhd15}
\end{equation}
where $\alpha_{0} = \frac{q_e^2}{\hbar c} = \frac{1}{137.04}$ is the fine structure constant and
\begin{multline}
L_{\alpha\nu} (\vb{q}^{\prime},\omega^{\prime}_{\alpha}) \equiv  \int \frac{\mathrm{d}k_{x}^{\prime}\mathrm{d}k_{y}^{\prime}}{(2\pi)^2}
 \left[(\vb{k}^{\prime}+\tfrac{1}{2}\vb{q}^{\prime})\cdot\bm{\epsilon}_{\nu}\right]^2 \\
 \times \left(\frac{f_{\alpha}^{(0)} (\vb{k}+\vb{q})-f_{\alpha}^{(0)}(\vb{k})}{\hbar\omega_{\alpha}^{\prime}-(2\vb{k}^{\prime}\cdot\vb{q}^{\prime}+\vb{q}^{\prime 2})+i\gamma_{\alpha}^{\prime}}\right)
\label{eq:ldhd16}
\end{multline}
We have scaled the momenta by the Fermi momentum $k_{F}$ and the energies by the Fermi energy $\varepsilon_{\alpha F} = \frac{\hbar^2 k_{F}^{2}}{2 m_{\alpha}}$:
\begin{multline}
\vb{k}^{\prime} =\frac{\vb{k}}{k_{F}} , \,
\vb{q}^{\prime} =\frac{\vb{q}}{k_{F}} , \, \\
\hbar \omega_{\alpha}^{\prime} = \frac{\hbar\omega}{\varepsilon_{\alpha F}} , \,
\varepsilon_{\alpha,\vb{k}}^{\prime} = \frac{\varepsilon_{\alpha,\vb{k}}}{\varepsilon_{\alpha F}} , \,
\gamma_{\alpha}^{\prime} = \frac{\gamma_{f}}{\varepsilon_{\alpha F}} , \quad
\alpha = e,h.
\end{multline}
The Fermi momentum is defined as the radius of the (zero-temperature) Fermi sphere corresponding to a given density, $\frac{S_d k_{F}^{2}}{4\pi} = n_e = n_h$, where $S_d =$ spin degeneracy, assumed to be the same for $e$ and $h$.

The steady-state distribution $f^{(0)}_{\alpha}$ has so far been left as arbitrary. To gain analytic insight, we consider the case of a quasi-equilibrium electron-hole plasma at zero temperature ($T = 0$) and vanishing damping losses ($\gamma_{\alpha}^{\prime} \to 0$). In this case, the steady-state occupation is $f_{\alpha}^{(0)}(\vb{k})=\theta(k_{F}-k)$, and
\begin{multline}
	f_{\alpha}^{(0)}(\vb{k}+\vb{q})-f_{\alpha}^{(0)}(\vb{k}) = -\theta(\vert\vb{k}+\vb{q}\vert-k_F)\theta(k_{F}-k) \\ +\theta(k_{F}-\vert\vb{k}+\vb{q}\vert)\theta(k-k_{F}) \label{eq:ldhd17}
\end{multline}
In the limit $\gamma_{\alpha}^{\prime} \to 0$, the Sochocki-Plemelj theorem yields
\begin{multline}
	\lim_{\gamma_{\alpha}^{\prime} \to 0} \frac{1}{\hbar\omega_{\alpha}^{\prime}-(2\vb{k}^{\prime}\cdot\vb{q}^{\prime}+\vb{q}^{\prime 2})+i\gamma_{\alpha}^{\prime}} = \\  \mathcal{P}\frac{1}{\hbar\omega_{\alpha}^{\prime}-(2\vb{k}^{\prime}\cdot\vb{q}^{\prime}+\vb{q}^{\prime 2})}
-i\pi\delta(\hbar\omega_{\alpha}^{\prime}-(2\vb{k}^{\prime}\cdot\vb{q}^{\prime}+\vb{q}^{\prime 2})) \label{eq:ldhd18}
\end{multline}
With Eqs.\ \eqref{eq:ldhd17} and \eqref{eq:ldhd18}, the imaginary part of Eq.\ \eqref{eq:ldhd16} becomes
\begin{IEEEeqnarray}{l}
\Im L_{\alpha\nu} (\vb{q}^{\prime}, \omega_{\alpha}^{\prime}) = \int \frac{\mathrm{d}^{2}\vb{k}^{\prime}}{4\pi} \left[(\vb{k}^{\prime}+\tfrac{1}{2}\vb{q}^{\prime})\cdot\bm{\epsilon}_{\nu}\right]^2 \yesnumber\label{eq:ldhd19} \\
 \times \delta \left(\hbar\omega_{\alpha}^{\prime}-(2\vb{k}^{\prime}\cdot\vb{q}^{\prime}+\vb{q}^{\prime 2})\right)  \left[\theta(\vert\vb{k}+\vb{q}\vert-k_F)\theta(k_{F}-k) \right. \nonumber \\
 \IEEEeqnarraymulticol{1}{r}{
 \left. -\theta(k_{F}-\vert\vb{k}+\vb{q}\vert)\theta(k-k_{F}) \right]} \nonumber
\end{IEEEeqnarray}
We orient our coordinate system so that $\hat{x}$ points in the direction of $\vb{q}$. Then the factor in the integrand  in Eq.\ \eqref{eq:ldhd19} that depends on the THz polarization vector becomes, for the two linear polarization directions,
\begin{IEEEeqnarray}{srCl}
s-polarized: & \left[(\vb{k}^{\prime}+\tfrac{1}{2}\vb{q}^{\prime})\cdot\bm{\epsilon}_{\nu}\right]^2 &=& k_{y}^{\prime 2} \nonumber\\
p-polarized: & \left[(\vb{k}^{\prime}+\tfrac{1}{2}\vb{q}^{\prime})\cdot\bm{\epsilon}_{\nu}\right]^2 &=& (k_{x}^{\prime} + \tfrac{1}{2}q^{\prime})^2 \cos^2\phi  \nonumber
\end{IEEEeqnarray}
where $\phi$ is the angle between $\bm{\epsilon}_{\nu}$ and the $\hat{x}$-axis.
The result of the integration in Eq.\ \eqref{eq:ldhd19} is as follows.\\
For $0 \leq q' \leq 2$, $0 \leq \hbar \omega_{\alpha}' \leq 2 q' - q'^2$,
\begin{multline*}
\Im L_{\alpha\nu} (\vb{q}^{\prime}, \omega_{\alpha}^{\prime}) = \frac{1}{4 \pi q'} \times \\ \begin{cases}
\frac{1}{3} \left[ (1-a_{-}^2)^{3/2} - (1-a_+^2)^{3/2} \right], & \text{s-pol.} \\
\left( \frac{\hbar \omega_{\alpha}'}{2q'} \right)^2 \cos^2 \phi  \left[ \sqrt{1- a_-^2} - \sqrt{1 + a_+^2} \right] & \text{p-pol.}
\end{cases}
\end{multline*}
For $0 \leq q' \leq 2$, $2 q' - q'^2 \leq \hbar \omega_{\alpha}' \leq 2 q' + q'^2$,
\begin{equation*}
\Im L_{\alpha\nu} (\vb{q}^{\prime}, \omega_{\alpha}^{\prime}) = \frac{1}{4 \pi q'} \begin{cases}
\frac{1}{3}(1-a_{-}^2)^{\frac{3}{2}}  , & \text{s-pol.} \\
\left( \frac{\hbar \omega_{\alpha}'}{2q'} \right)^2 \cos^2 \phi  \sqrt{1- a_-^2}  & \text{p-pol.}
\end{cases}
\end{equation*}
For $ q' \geq 2$, $ q'^2 - 2q' \leq \hbar \omega_{\alpha}' \leq q'^2 + 2q'$,
\begin{equation*}
\Im L_{\alpha\nu} (\vb{q}^{\prime}, \omega_{\alpha}^{\prime}) = \frac{1}{4 \pi q'} \begin{cases}
\frac{1}{3}(1-a_{-}^2)^{\frac{3}{2}}  , & \text{s-pol.} \\
\left( \frac{\hbar \omega_{\alpha}'}{2q'} \right)^2 \cos^2 \phi  \sqrt{1- a_-^2}  & \text{p-pol.}
\end{cases}
\end{equation*}
$\Im L_{\alpha\nu} (\vb{q}^{\prime}, \omega_{\alpha}^{\prime})  = 0$ for elsewhere in the range $\omega_{\alpha}' \geq 0$. In the above,
\begin{equation*}
a_{\pm} = \frac{\hbar \omega_{\alpha}'}{2 q'} \pm \frac{q'}{2}.
\end{equation*}
The value of $\Im L_{\alpha\nu} (\vb{q}^{\prime}, \omega^{\prime}_{\alpha})$ for negative $\omega^{\prime}_{\alpha}$ is obtained via the symmetry relation (for inversion-symmetric systems)
\begin{equation}
L_{\alpha\nu} (\vb{q}^{\prime}, - \omega^{\prime}_{\alpha}) = L_{\alpha\nu}^{\ast} (\vb{q}^{\prime}, \omega^{\prime}_{\alpha}).
\end{equation}

%======================================================
%   References and Notes
%======================================================
%


\begin{thebibliography}{100}%
	\makeatletter
	\providecommand \@ifxundefined [1]{%
		\@ifx{#1\undefined}
	}%
	\providecommand \@ifnum [1]{%
		\ifnum #1\expandafter \@firstoftwo
		\else \expandafter \@secondoftwo
		\fi
	}%
	\providecommand \@ifx [1]{%
		\ifx #1\expandafter \@firstoftwo
		\else \expandafter \@secondoftwo
		\fi
	}%
	\providecommand \natexlab [1]{#1}%
	\providecommand \enquote  [1]{``#1''}%
	\providecommand \bibnamefont  [1]{#1}%
	\providecommand \bibfnamefont [1]{#1}%
	\providecommand \citenamefont [1]{#1}%
	\providecommand \href@noop [0]{\@secondoftwo}%
	\providecommand \href [0]{\begingroup \@sanitize@url \@href}%
	\providecommand \@href[1]{\@@startlink{#1}\@@href}%
	\providecommand \@@href[1]{\endgroup#1\@@endlink}%
	\providecommand \@sanitize@url [0]{\catcode `\\12\catcode `\$12\catcode
		`\&12\catcode `\#12\catcode `\^12\catcode `\_12\catcode `\%12\relax}%
	\providecommand \@@startlink[1]{}%
	\providecommand \@@endlink[0]{}%
	\providecommand \url  [0]{\begingroup\@sanitize@url \@url }%
	\providecommand \@url [1]{\endgroup\@href {#1}{\urlprefix }}%
	\providecommand \urlprefix  [0]{URL }%
	\providecommand \Eprint [0]{\href }%
	\providecommand \doibase [0]{https://doi.org/}%
	\providecommand \selectlanguage [0]{\@gobble}%
	\providecommand \bibinfo  [0]{\@secondoftwo}%
	\providecommand \bibfield  [0]{\@secondoftwo}%
	\providecommand \translation [1]{[#1]}%
	\providecommand \BibitemOpen [0]{}%
	\providecommand \bibitemStop [0]{}%
	\providecommand \bibitemNoStop [0]{.\EOS\space}%
	\providecommand \EOS [0]{\spacefactor3000\relax}%
	\providecommand \BibitemShut  [1]{\csname bibitem#1\endcsname}%
	\let\auto@bib@innerbib\@empty
	%</preamble>
	\bibitem [{\citenamefont {Fan}\ \emph {et~al.}(1997)\citenamefont {Fan},
		\citenamefont {Wang}, \citenamefont {Hou},\ and\ \citenamefont
		{Hammons}}]{fan-etal.97pra}%
	\BibitemOpen
	\bibfield  {author} {\bibinfo {author} {\bibfnamefont {X.}~\bibnamefont
			{Fan}}, \bibinfo {author} {\bibfnamefont {H.}~\bibnamefont {Wang}}, \bibinfo
		{author} {\bibfnamefont {H.~Q.}\ \bibnamefont {Hou}},\ and\ \bibinfo {author}
		{\bibfnamefont {B.~E.}\ \bibnamefont {Hammons}},\ }\bibfield  {title}
	{\bibinfo {title} {Laser emission from semiconductor microcavities: {The}
			role of cavity polaritons},\ }\href
	{https://doi.org/10.1103/PhysRevA.56.3233} {\bibfield  {journal} {\bibinfo
			{journal} {Phys. Rev. A}\ }\textbf {\bibinfo {volume} {56}},\ \bibinfo
		{pages} {3233} (\bibinfo {year} {1997})}\BibitemShut {NoStop}%
	\bibitem [{\citenamefont {Cao}\ \emph {et~al.}(1997)\citenamefont {Cao},
		\citenamefont {Pau}, \citenamefont {Jacobson}, \citenamefont {Bj{\"o}rk},
		\citenamefont {Yamamoto},\ and\ \citenamefont
		{Imamo{\u{g}}lu}}]{cao-etal.97}%
	\BibitemOpen
	\bibfield  {author} {\bibinfo {author} {\bibfnamefont {H.}~\bibnamefont
			{Cao}}, \bibinfo {author} {\bibfnamefont {S.}~\bibnamefont {Pau}}, \bibinfo
		{author} {\bibfnamefont {J.~M.}\ \bibnamefont {Jacobson}}, \bibinfo {author}
		{\bibfnamefont {G.}~\bibnamefont {Bj{\"o}rk}}, \bibinfo {author}
		{\bibfnamefont {Y.}~\bibnamefont {Yamamoto}},\ and\ \bibinfo {author}
		{\bibfnamefont {A.}~\bibnamefont {Imamo{\u{g}}lu}},\ }\bibfield  {title}
	{\bibinfo {title} {Transition from a microcavity exciton polariton to a
			photon laser},\ }\href {https://doi.org/10.1103/PhysRevA.55.4632} {\bibfield
		{journal} {\bibinfo  {journal} {Phys. Rev. A}\ }\textbf {\bibinfo {volume}
			{55}},\ \bibinfo {pages} {4632} (\bibinfo {year} {1997})}\BibitemShut
	{NoStop}%
	\bibitem [{\citenamefont {Kuwata-Gonokami}\ \emph {et~al.}(1997)\citenamefont
		{Kuwata-Gonokami}, \citenamefont {Inouye}, \citenamefont {Suzuura},
		\citenamefont {Shirane}, \citenamefont {Shimano}, \citenamefont {Someya},\
		and\ \citenamefont {Sakaki}}]{kuwata-gonokami-etal.97}%
	\BibitemOpen
	\bibfield  {author} {\bibinfo {author} {\bibfnamefont {M.}~\bibnamefont
			{Kuwata-Gonokami}}, \bibinfo {author} {\bibfnamefont {S.}~\bibnamefont
			{Inouye}}, \bibinfo {author} {\bibfnamefont {H.}~\bibnamefont {Suzuura}},
		\bibinfo {author} {\bibfnamefont {M.}~\bibnamefont {Shirane}}, \bibinfo
		{author} {\bibfnamefont {R.}~\bibnamefont {Shimano}}, \bibinfo {author}
		{\bibfnamefont {T.}~\bibnamefont {Someya}},\ and\ \bibinfo {author}
		{\bibfnamefont {H.}~\bibnamefont {Sakaki}},\ }\bibfield  {title} {\bibinfo
		{title} {Parametric scattering of cavity polaritons},\ }\href
	{https://doi.org/10.1103/PhysRevLett.79.1341} {\bibfield  {journal} {\bibinfo
			{journal} {Phys. Rev. Lett.}\ }\textbf {\bibinfo {volume} {79}},\ \bibinfo
		{pages} {1341} (\bibinfo {year} {1997})}\BibitemShut {NoStop}%
	\bibitem [{\citenamefont {Kira}\ \emph {et~al.}(1999)\citenamefont {Kira},
		\citenamefont {Jahnke}, \citenamefont {Hoyer},\ and\ \citenamefont
		{Koch}}]{kira-etal.99b}%
	\BibitemOpen
	\bibfield  {author} {\bibinfo {author} {\bibfnamefont {M.}~\bibnamefont
			{Kira}}, \bibinfo {author} {\bibfnamefont {F.}~\bibnamefont {Jahnke}},
		\bibinfo {author} {\bibfnamefont {W.}~\bibnamefont {Hoyer}},\ and\ \bibinfo
		{author} {\bibfnamefont {S.}~\bibnamefont {Koch}},\ }\bibfield  {title}
	{\bibinfo {title} {Quantum theory of spontaneous emission and coherent
			effects in semiconductor microstructures},\ }\href
	{https://doi.org/10.1016/s0079-6727(99)00008-7} {\bibfield  {journal}
		{\bibinfo  {journal} {Prog Quant Electron}\ }\textbf {\bibinfo {volume}
			{23}},\ \bibinfo {pages} {189} (\bibinfo {year} {1999})}\BibitemShut
	{NoStop}%
	\bibitem [{\citenamefont {Moskalenko}\ and\ \citenamefont
		{Snoke}(2000)}]{moskalenko-snoke.00}%
	\BibitemOpen
	\bibfield  {author} {\bibinfo {author} {\bibfnamefont {S.~A.}\ \bibnamefont
			{Moskalenko}}\ and\ \bibinfo {author} {\bibfnamefont {D.~W.}\ \bibnamefont
			{Snoke}},\ }\href {https://doi.org/10.1017/CBO9780511721687} {\emph {\bibinfo
			{title} {Bose-{Einstein} {Condensation} of {Excitons} and {Biexcitons}: and
				{Coherent} {Nonlinear} {Optics} with {Excitons}}}}\ (\bibinfo  {publisher}
	{Cambridge University Press},\ \bibinfo {address} {Cambridge},\ \bibinfo
	{year} {2000})\BibitemShut {NoStop}%
	\bibitem [{\citenamefont {Ciuti}\ \emph {et~al.}(2000)\citenamefont {Ciuti},
		\citenamefont {Schwendimann}, \citenamefont {Deveaud},\ and\ \citenamefont
		{Quattropani}}]{ciuti-etal.00}%
	\BibitemOpen
	\bibfield  {author} {\bibinfo {author} {\bibfnamefont {C.}~\bibnamefont
			{Ciuti}}, \bibinfo {author} {\bibfnamefont {P.}~\bibnamefont {Schwendimann}},
		\bibinfo {author} {\bibfnamefont {B.}~\bibnamefont {Deveaud}},\ and\ \bibinfo
		{author} {\bibfnamefont {A.}~\bibnamefont {Quattropani}},\ }\bibfield
	{title} {\bibinfo {title} {Theory of the angle-resonant polariton
			amplifier},\ }\href {https://doi.org/10.1103/PhysRevB.62.R4825} {\bibfield
		{journal} {\bibinfo  {journal} {Phys. Rev. B}\ }\textbf {\bibinfo {volume}
			{62}},\ \bibinfo {pages} {R4825} (\bibinfo {year} {2000})}\BibitemShut
	{NoStop}%
	\bibitem [{\citenamefont {Savvidis}\ \emph {et~al.}(2000)\citenamefont
		{Savvidis}, \citenamefont {Baumberg}, \citenamefont {Stevenson},
		\citenamefont {Skolnick}, \citenamefont {Whittaker},\ and\ \citenamefont
		{Roberts}}]{savvidis-etal.00}%
	\BibitemOpen
	\bibfield  {author} {\bibinfo {author} {\bibfnamefont {P.~G.}\ \bibnamefont
			{Savvidis}}, \bibinfo {author} {\bibfnamefont {J.~J.}\ \bibnamefont
			{Baumberg}}, \bibinfo {author} {\bibfnamefont {R.~M.}\ \bibnamefont
			{Stevenson}}, \bibinfo {author} {\bibfnamefont {M.~S.}\ \bibnamefont
			{Skolnick}}, \bibinfo {author} {\bibfnamefont {D.~M.}\ \bibnamefont
			{Whittaker}},\ and\ \bibinfo {author} {\bibfnamefont {J.~S.}\ \bibnamefont
			{Roberts}},\ }\bibfield  {title} {\bibinfo {title} {Angle-resonant stimulated
			polariton amplifier},\ }\href {https://doi.org/10.1103/physrevlett.84.1547}
	{\bibfield  {journal} {\bibinfo  {journal} {Phys. Rev. Lett.}\ }\textbf
		{\bibinfo {volume} {84}},\ \bibinfo {pages} {1547} (\bibinfo {year}
		{2000})}\BibitemShut {NoStop}%
	\bibitem [{\citenamefont {Kwong}\ \emph {et~al.}(2001)\citenamefont {Kwong},
		\citenamefont {Takayama}, \citenamefont {Rumyantsev}, \citenamefont
		{Kuwata-Gonokami},\ and\ \citenamefont {Binder}}]{kwong-etal.01prl}%
	\BibitemOpen
	\bibfield  {author} {\bibinfo {author} {\bibfnamefont {N.~H.}\ \bibnamefont
			{Kwong}}, \bibinfo {author} {\bibfnamefont {R.}~\bibnamefont {Takayama}},
		\bibinfo {author} {\bibfnamefont {I.}~\bibnamefont {Rumyantsev}}, \bibinfo
		{author} {\bibfnamefont {M.}~\bibnamefont {Kuwata-Gonokami}},\ and\ \bibinfo
		{author} {\bibfnamefont {R.}~\bibnamefont {Binder}},\ }\bibfield  {title}
	{\bibinfo {title} {Evidence of nonperturbative continuum correlations in
			two-dimensional exciton systems in semiconductor microcavities},\ }\href
	{https://doi.org/https://doi.org/10.1103/PhysRevLett.87.027402} {\bibfield
		{journal} {\bibinfo  {journal} {Phys. Rev. Lett.}\ }\textbf {\bibinfo
			{volume} {87}},\ \bibinfo {pages} {027402} (\bibinfo {year}
		{2001})}\BibitemShut {NoStop}%
	\bibitem [{\citenamefont {Baumberg}\ and\ \citenamefont
		{Lagoudakis}(2005)}]{baumberg-lagoudakis.05}%
	\BibitemOpen
	\bibfield  {author} {\bibinfo {author} {\bibfnamefont {J.~J.}\ \bibnamefont
			{Baumberg}}\ and\ \bibinfo {author} {\bibfnamefont {P.~G.}\ \bibnamefont
			{Lagoudakis}},\ }\bibfield  {title} {\bibinfo {title} {Parametric
			amplification and polariton liquids in semiconductor microcavities},\ }\href
	{https://doi.org/https://doi.org/10.1002/pssb.200560960} {\bibfield
		{journal} {\bibinfo  {journal} {Phys. Status Solidi B}\ }\textbf {\bibinfo
			{volume} {242}},\ \bibinfo {pages} {2210} (\bibinfo {year}
		{2005})}\BibitemShut {NoStop}%
	\bibitem [{\citenamefont {Balili}\ \emph {et~al.}(2006)\citenamefont {Balili},
		\citenamefont {Snoke}, \citenamefont {Pfeiffer},\ and\ \citenamefont
		{West}}]{balili-etal.06}%
	\BibitemOpen
	\bibfield  {author} {\bibinfo {author} {\bibfnamefont {R.~B.}\ \bibnamefont
			{Balili}}, \bibinfo {author} {\bibfnamefont {D.~W.}\ \bibnamefont {Snoke}},
		\bibinfo {author} {\bibfnamefont {L.}~\bibnamefont {Pfeiffer}},\ and\
		\bibinfo {author} {\bibfnamefont {K.}~\bibnamefont {West}},\ }\bibfield
	{title} {\bibinfo {title} {Actively tuned and spatially trapped polaritons},\
	}\href {https://doi.org/10.1063/1.2164431} {\bibfield  {journal} {\bibinfo
			{journal} {Appl. Phys. Lett.}\ }\textbf {\bibinfo {volume} {88}},\ \bibinfo
		{pages} {031110} (\bibinfo {year} {2006})}\BibitemShut {NoStop}%
	\bibitem [{\citenamefont {Balili}\ \emph {et~al.}(2007)\citenamefont {Balili},
		\citenamefont {Hartwell}, \citenamefont {Snoke}, \citenamefont {Pfeiffer},\
		and\ \citenamefont {West}}]{balili-etal.07}%
	\BibitemOpen
	\bibfield  {author} {\bibinfo {author} {\bibfnamefont {R.}~\bibnamefont
			{Balili}}, \bibinfo {author} {\bibfnamefont {V.}~\bibnamefont {Hartwell}},
		\bibinfo {author} {\bibfnamefont {D.}~\bibnamefont {Snoke}}, \bibinfo
		{author} {\bibfnamefont {L.}~\bibnamefont {Pfeiffer}},\ and\ \bibinfo
		{author} {\bibfnamefont {K.}~\bibnamefont {West}},\ }\bibfield  {title}
	{\bibinfo {title} {Bose-{Einstein} {Condensation} of {Microcavity}
			{Polaritons} in a {Trap}},\ }\href {https://doi.org/10.1126/science.1140990}
	{\bibfield  {journal} {\bibinfo  {journal} {Science}\ }\textbf {\bibinfo
			{volume} {316}},\ \bibinfo {pages} {1007} (\bibinfo {year}
		{2007})}\BibitemShut {NoStop}%
	\bibitem [{\citenamefont {Keeling}\ \emph {et~al.}(2007)\citenamefont
		{Keeling}, \citenamefont {Marchetti}, \citenamefont {Szymanska},\ and\
		\citenamefont {Littlewood}}]{keeling_collective_2007}%
	\BibitemOpen
	\bibfield  {author} {\bibinfo {author} {\bibfnamefont {J.}~\bibnamefont
			{Keeling}}, \bibinfo {author} {\bibfnamefont {F.~M.}\ \bibnamefont
			{Marchetti}}, \bibinfo {author} {\bibfnamefont {M.~H.}\ \bibnamefont
			{Szymanska}},\ and\ \bibinfo {author} {\bibfnamefont {P.~B.}\ \bibnamefont
			{Littlewood}},\ }\bibfield  {title} {\bibinfo {title} {Collective coherence
			in planar semiconductor microcavities},\ }\href
	{https://doi.org/10.1088/0268-1242/22/5/R01} {\bibfield  {journal} {\bibinfo
			{journal} {Semicond Sci. Technol.}\ }\textbf {\bibinfo {volume} {22}},\
		\bibinfo {pages} {R1} (\bibinfo {year} {2007})}\BibitemShut {NoStop}%
	\bibitem [{\citenamefont {Schumacher}\ \emph {et~al.}(2007)\citenamefont
		{Schumacher}, \citenamefont {Kwong},\ and\ \citenamefont
		{Binder}}]{schumacher-etal.07prb}%
	\BibitemOpen
	\bibfield  {author} {\bibinfo {author} {\bibfnamefont {S.}~\bibnamefont
			{Schumacher}}, \bibinfo {author} {\bibfnamefont {N.~H.}\ \bibnamefont
			{Kwong}},\ and\ \bibinfo {author} {\bibfnamefont {R.}~\bibnamefont
			{Binder}},\ }\bibfield  {title} {\bibinfo {title} {Influence of
			exciton-exciton correlations on the polarization characteristics of polariton
			amplification in semiconductor microcavities},\ }\href
	{https://doi.org/10.1103/PhysRevB.76.245324} {\bibfield  {journal} {\bibinfo
			{journal} {Phys. Rev. B}\ }\textbf {\bibinfo {volume} {76}},\ \bibinfo
		{pages} {245324} (\bibinfo {year} {2007})}\BibitemShut {NoStop}%
	\bibitem [{\citenamefont {Bajoni}\ \emph {et~al.}(2008)\citenamefont {Bajoni},
		\citenamefont {Senellart}, \citenamefont {Wertz}, \citenamefont {Sagnes},
		\citenamefont {Miard}, \citenamefont {Lema{\^\i}tre},\ and\ \citenamefont
		{Bloch}}]{bajoni-etal.08}%
	\BibitemOpen
	\bibfield  {author} {\bibinfo {author} {\bibfnamefont {D.}~\bibnamefont
			{Bajoni}}, \bibinfo {author} {\bibfnamefont {P.}~\bibnamefont {Senellart}},
		\bibinfo {author} {\bibfnamefont {E.}~\bibnamefont {Wertz}}, \bibinfo
		{author} {\bibfnamefont {I.}~\bibnamefont {Sagnes}}, \bibinfo {author}
		{\bibfnamefont {A.}~\bibnamefont {Miard}}, \bibinfo {author} {\bibfnamefont
			{A.}~\bibnamefont {Lema{\^\i}tre}},\ and\ \bibinfo {author} {\bibfnamefont
			{J.}~\bibnamefont {Bloch}},\ }\bibfield  {title} {\bibinfo {title} {Polariton
			laser using single micropillar {GaAs-GaAlAs} semiconductor cavities},\ }\href
	{https://doi.org/10.1103/PhysRevLett.100.047401} {\bibfield  {journal}
		{\bibinfo  {journal} {Phys. Rev. Lett.}\ }\textbf {\bibinfo {volume} {100}},\
		\bibinfo {pages} {047401} (\bibinfo {year} {2008})}\BibitemShut {NoStop}%
	\bibitem [{\citenamefont {Berman}\ \emph {et~al.}(2008)\citenamefont {Berman},
		\citenamefont {Lozovik},\ and\ \citenamefont {Snoke}}]{berman-etal.08}%
	\BibitemOpen
	\bibfield  {author} {\bibinfo {author} {\bibfnamefont {O.~L.}\ \bibnamefont
			{Berman}}, \bibinfo {author} {\bibfnamefont {Y.~E.}\ \bibnamefont
			{Lozovik}},\ and\ \bibinfo {author} {\bibfnamefont {D.~W.}\ \bibnamefont
			{Snoke}},\ }\bibfield  {title} {\bibinfo {title} {Theory of {Bose}-{Einstein}
			condensation and superfluidity of two-dimensional polaritons in an in-plane
			harmonic potential},\ }\href {https://doi.org/10.1103/PhysRevB.77.155317}
	{\bibfield  {journal} {\bibinfo  {journal} {Phys. Rev. B}\ }\textbf {\bibinfo
			{volume} {77}},\ \bibinfo {pages} {155317} (\bibinfo {year}
		{2008})}\BibitemShut {NoStop}%
	\bibitem [{\citenamefont {Berney}\ \emph {et~al.}(2008)\citenamefont {Berney},
		\citenamefont {Portella-Oberli},\ and\ \citenamefont
		{Deveaud}}]{berney-etal.08}%
	\BibitemOpen
	\bibfield  {author} {\bibinfo {author} {\bibfnamefont {J.}~\bibnamefont
			{Berney}}, \bibinfo {author} {\bibfnamefont {M.~T.}\ \bibnamefont
			{Portella-Oberli}},\ and\ \bibinfo {author} {\bibfnamefont {B.}~\bibnamefont
			{Deveaud}},\ }\bibfield  {title} {\bibinfo {title} {Dressed excitons within
			an incoherent electron gas: {Observation} of a {Mollow} triplet and an
			{Autler}-{Townes} doublet},\ }\href
	{https://doi.org/10.1103/PhysRevB.77.121301} {\bibfield  {journal} {\bibinfo
			{journal} {Phys. Rev. B}\ }\textbf {\bibinfo {volume} {77}},\ \bibinfo
		{pages} {121301} (\bibinfo {year} {2008})}\BibitemShut {NoStop}%
	\bibitem [{\citenamefont {Amo}\ \emph {et~al.}(2009)\citenamefont {Amo},
		\citenamefont {Sanvitto}, \citenamefont {Laussy}, \citenamefont {Ballarini},
		\citenamefont {del Valle}, \citenamefont {Martin}, \citenamefont
		{Lema{\^{\i}}tre}, \citenamefont {Bloch}, \citenamefont {Krizhanovskii},
		\citenamefont {Skolnick}, \citenamefont {Tejedor},\ and\ \citenamefont
		{Vi{\~{n}}a}}]{amo-etal.09}%
	\BibitemOpen
	\bibfield  {author} {\bibinfo {author} {\bibfnamefont {A.}~\bibnamefont
			{Amo}}, \bibinfo {author} {\bibfnamefont {D.}~\bibnamefont {Sanvitto}},
		\bibinfo {author} {\bibfnamefont {F.~P.}\ \bibnamefont {Laussy}}, \bibinfo
		{author} {\bibfnamefont {D.}~\bibnamefont {Ballarini}}, \bibinfo {author}
		{\bibfnamefont {E.}~\bibnamefont {del Valle}}, \bibinfo {author}
		{\bibfnamefont {M.~D.}\ \bibnamefont {Martin}}, \bibinfo {author}
		{\bibfnamefont {A.}~\bibnamefont {Lema{\^{\i}}tre}}, \bibinfo {author}
		{\bibfnamefont {J.}~\bibnamefont {Bloch}}, \bibinfo {author} {\bibfnamefont
			{D.~N.}\ \bibnamefont {Krizhanovskii}}, \bibinfo {author} {\bibfnamefont
			{M.~S.}\ \bibnamefont {Skolnick}}, \bibinfo {author} {\bibfnamefont
			{C.}~\bibnamefont {Tejedor}},\ and\ \bibinfo {author} {\bibfnamefont
			{L.}~\bibnamefont {Vi{\~{n}}a}},\ }\bibfield  {title} {\bibinfo {title}
		{Collective fluid dynamics of a polariton condensate in a semiconductor
			microcavity},\ }\href {https://doi.org/10.1038/nature07640} {\bibfield
		{journal} {\bibinfo  {journal} {Nature}\ }\textbf {\bibinfo {volume} {457}},\
		\bibinfo {pages} {291} (\bibinfo {year} {2009})}\BibitemShut {NoStop}%
	\bibitem [{\citenamefont {Timofeev}\ and\ \citenamefont
		{Sanvitto}(2012)}]{timofeev-sanvitto.12}%
	\BibitemOpen
	\bibinfo {editor} {\bibfnamefont {V.}~\bibnamefont {Timofeev}}\ and\ \bibinfo
	{editor} {\bibfnamefont {D.}~\bibnamefont {Sanvitto}},\ eds.,\ \href
	{https://doi.org/10.1007/978-3-642-24186-4} {\emph {\bibinfo {title} {Exciton
				Polaritons in Microcavities}}},\ Vol.\ \bibinfo {volume} {172}\ (\bibinfo
	{publisher} {Springer},\ \bibinfo {address} {Berlin, Heidelberg},\ \bibinfo
	{year} {2012})\BibitemShut {NoStop}%
	\bibitem [{\citenamefont {Semkat}\ \emph {et~al.}(2009)\citenamefont {Semkat},
		\citenamefont {Richter}, \citenamefont {Kremp}, \citenamefont {Manzke},
		\citenamefont {Kraeft},\ and\ \citenamefont {Henneberger}}]{semkat-etal.09}%
	\BibitemOpen
	\bibfield  {author} {\bibinfo {author} {\bibfnamefont {D.}~\bibnamefont
			{Semkat}}, \bibinfo {author} {\bibfnamefont {F.}~\bibnamefont {Richter}},
		\bibinfo {author} {\bibfnamefont {D.}~\bibnamefont {Kremp}}, \bibinfo
		{author} {\bibfnamefont {G.}~\bibnamefont {Manzke}}, \bibinfo {author}
		{\bibfnamefont {W.-D.}\ \bibnamefont {Kraeft}},\ and\ \bibinfo {author}
		{\bibfnamefont {K.}~\bibnamefont {Henneberger}},\ }\bibfield  {title}
	{\bibinfo {title} {Ionization equilibrium in an excited semiconductor: {Mott}
			transition versus {Bose}-{Einstein} condensation},\ }\href
	{https://doi.org/10.1103/PhysRevB.80.155201} {\bibfield  {journal} {\bibinfo
			{journal} {Phys. Rev. B}\ }\textbf {\bibinfo {volume} {80}},\ \bibinfo
		{pages} {155201} (\bibinfo {year} {2009})}\BibitemShut {NoStop}%
	\bibitem [{\citenamefont {Kamide}\ and\ \citenamefont
		{Ogawa}(2010)}]{kamide-ogawa.10}%
	\BibitemOpen
	\bibfield  {author} {\bibinfo {author} {\bibfnamefont {K.}~\bibnamefont
			{Kamide}}\ and\ \bibinfo {author} {\bibfnamefont {T.}~\bibnamefont {Ogawa}},\
	}\bibfield  {title} {\bibinfo {title} {What determines the wave function of
			electron-hole pairs in polariton condensates?},\ }\href
	{https://doi.org/10.1103/physrevlett.105.056401} {\bibfield  {journal}
		{\bibinfo  {journal} {Phys. Rev. Lett.}\ }\textbf {\bibinfo {volume} {105}},\
		\bibinfo {pages} {056401} (\bibinfo {year} {2010})}\BibitemShut {NoStop}%
	\bibitem [{\citenamefont {Deng}\ \emph {et~al.}(2010)\citenamefont {Deng},
		\citenamefont {Haug},\ and\ \citenamefont {Yamamoto}}]{deng-etal.10}%
	\BibitemOpen
	\bibfield  {author} {\bibinfo {author} {\bibfnamefont {H.}~\bibnamefont
			{Deng}}, \bibinfo {author} {\bibfnamefont {H.}~\bibnamefont {Haug}},\ and\
		\bibinfo {author} {\bibfnamefont {Y.}~\bibnamefont {Yamamoto}},\ }\bibfield
	{title} {\bibinfo {title} {Exciton-polariton {Bose}-{Einstein}
			condensation},\ }\href {https://doi.org/10.1103/RevModPhys.82.1489}
	{\bibfield  {journal} {\bibinfo  {journal} {Rev. Mod. Phys.}\ }\textbf
		{\bibinfo {volume} {82}},\ \bibinfo {pages} {1489} (\bibinfo {year}
		{2010})}\BibitemShut {NoStop}%
	\bibitem [{\citenamefont {Snoke}\ and\ \citenamefont
		{Littlewood}(2010)}]{snoke-littlewood.10}%
	\BibitemOpen
	\bibfield  {author} {\bibinfo {author} {\bibfnamefont {D.}~\bibnamefont
			{Snoke}}\ and\ \bibinfo {author} {\bibfnamefont {P.}~\bibnamefont
			{Littlewood}},\ }\bibfield  {title} {\bibinfo {title} {Polariton
			condensates},\ }\href {https://doi.org/10.1063/1.3480075} {\bibfield
		{journal} {\bibinfo  {journal} {Phys. Today}\ }\textbf {\bibinfo {volume}
			{63}},\ \bibinfo {pages} {42} (\bibinfo {year} {2010})}\BibitemShut {NoStop}%
	\bibitem [{\citenamefont {Liu}\ \emph {et~al.}(2015)\citenamefont {Liu},
		\citenamefont {Galfsky}, \citenamefont {Sun}, \citenamefont {Xia},
		\citenamefont {Lin}, \citenamefont {Lee}, \citenamefont {K{\'e}na-Cohen},\
		and\ \citenamefont {Menon}}]{liu-etal.15}%
	\BibitemOpen
	\bibfield  {author} {\bibinfo {author} {\bibfnamefont {X.}~\bibnamefont
			{Liu}}, \bibinfo {author} {\bibfnamefont {T.}~\bibnamefont {Galfsky}},
		\bibinfo {author} {\bibfnamefont {Z.}~\bibnamefont {Sun}}, \bibinfo {author}
		{\bibfnamefont {F.}~\bibnamefont {Xia}}, \bibinfo {author} {\bibfnamefont
			{E.-c.}\ \bibnamefont {Lin}}, \bibinfo {author} {\bibfnamefont {Y.-H.}\
			\bibnamefont {Lee}}, \bibinfo {author} {\bibfnamefont {S.}~\bibnamefont
			{K{\'e}na-Cohen}},\ and\ \bibinfo {author} {\bibfnamefont {V.~M.}\
			\bibnamefont {Menon}},\ }\bibfield  {title} {\bibinfo {title} {Strong
			light-matter coupling in two-dimensional atomic crystals},\ }\href
	{https://doi.org/10.1038/nphoton.2014.304} {\bibfield  {journal} {\bibinfo
			{journal} {Nat. Photonics}\ }\textbf {\bibinfo {volume} {9}},\ \bibinfo
		{pages} {30} (\bibinfo {year} {2015})}\BibitemShut {NoStop}%
	\bibitem [{\citenamefont {Schulze}\ \emph {et~al.}(2014)\citenamefont
		{Schulze}, \citenamefont {Lingnau}, \citenamefont {Hein}, \citenamefont
		{Carmele}, \citenamefont {Sch{\"o}ll}, \citenamefont {L{\"u}dge},\ and\
		\citenamefont {Knorr}}]{schulze-etal.14}%
	\BibitemOpen
	\bibfield  {author} {\bibinfo {author} {\bibfnamefont {F.}~\bibnamefont
			{Schulze}}, \bibinfo {author} {\bibfnamefont {B.}~\bibnamefont {Lingnau}},
		\bibinfo {author} {\bibfnamefont {S.~M.}\ \bibnamefont {Hein}}, \bibinfo
		{author} {\bibfnamefont {A.}~\bibnamefont {Carmele}}, \bibinfo {author}
		{\bibfnamefont {E.}~\bibnamefont {Sch{\"o}ll}}, \bibinfo {author}
		{\bibfnamefont {K.}~\bibnamefont {L{\"u}dge}},\ and\ \bibinfo {author}
		{\bibfnamefont {A.}~\bibnamefont {Knorr}},\ }\bibfield  {title} {\bibinfo
		{title} {Feedback-induced steady-state light bunching above the lasing
			threshold},\ }\href {https://doi.org/10.1103/PhysRevA.89.041801} {\bibfield
		{journal} {\bibinfo  {journal} {Phys. Rev. A}\ }\textbf {\bibinfo {volume}
			{89}},\ \bibinfo {pages} {041801} (\bibinfo {year} {2014})}\BibitemShut
	{NoStop}%
	\bibitem [{\citenamefont {M{\'{e}}nard}\ \emph {et~al.}(2014)\citenamefont
		{M{\'{e}}nard}, \citenamefont {Poellmann}, \citenamefont {Porer},
		\citenamefont {Leierseder}, \citenamefont {Galopin}, \citenamefont
		{Lema{\^{\i}}tre}, \citenamefont {Amo}, \citenamefont {Bloch},\ and\
		\citenamefont {Huber}}]{menard-etal.14}%
	\BibitemOpen
	\bibfield  {author} {\bibinfo {author} {\bibfnamefont {J.~M.}\ \bibnamefont
			{M{\'{e}}nard}}, \bibinfo {author} {\bibfnamefont {C.}~\bibnamefont
			{Poellmann}}, \bibinfo {author} {\bibfnamefont {M.}~\bibnamefont {Porer}},
		\bibinfo {author} {\bibfnamefont {U.}~\bibnamefont {Leierseder}}, \bibinfo
		{author} {\bibfnamefont {E.}~\bibnamefont {Galopin}}, \bibinfo {author}
		{\bibfnamefont {A.}~\bibnamefont {Lema{\^{\i}}tre}}, \bibinfo {author}
		{\bibfnamefont {A.}~\bibnamefont {Amo}}, \bibinfo {author} {\bibfnamefont
			{J.}~\bibnamefont {Bloch}},\ and\ \bibinfo {author} {\bibfnamefont
			{R.}~\bibnamefont {Huber}},\ }\bibfield  {title} {\bibinfo {title} {Revealing
			the dark side of a bright exciton{\textendash}polariton condensate},\ }\href
	{https://doi.org/10.1038/ncomms5648} {\bibfield  {journal} {\bibinfo
			{journal} {Nat Commun}\ }\textbf {\bibinfo {volume} {5}},\ \bibinfo {pages}
		{4648} (\bibinfo {year} {2014})}\BibitemShut {NoStop}%
	\bibitem [{\citenamefont {Kamandar~Dezfouli}\ \emph {et~al.}(2014)\citenamefont
		{Kamandar~Dezfouli}, \citenamefont {Dignam}, \citenamefont {Steel},\ and\
		\citenamefont {Sipe}}]{kamandardezfouli-etal.14}%
	\BibitemOpen
	\bibfield  {author} {\bibinfo {author} {\bibfnamefont {M.}~\bibnamefont
			{Kamandar~Dezfouli}}, \bibinfo {author} {\bibfnamefont {M.~M.}\ \bibnamefont
			{Dignam}}, \bibinfo {author} {\bibfnamefont {M.~J.}\ \bibnamefont {Steel}},\
		and\ \bibinfo {author} {\bibfnamefont {J.~E.}\ \bibnamefont {Sipe}},\
	}\bibfield  {title} {\bibinfo {title} {Heisenberg treatment of pair
			generation in lossy coupled-cavity systems},\ }\href
	{https://doi.org/10.1103/PhysRevA.90.043832} {\bibfield  {journal} {\bibinfo
			{journal} {Phys. Rev. A}\ }\textbf {\bibinfo {volume} {90}},\ \bibinfo
		{pages} {043832} (\bibinfo {year} {2014})}\BibitemShut {NoStop}%
	\bibitem [{\citenamefont {Schmutzler}\ \emph {et~al.}(2015)\citenamefont
		{Schmutzler}, \citenamefont {Lewandowski}, \citenamefont {A{\ss}mann},
		\citenamefont {Niemietz}, \citenamefont {Schumacher}, \citenamefont {Kamp},
		\citenamefont {Schneider}, \citenamefont {H{\"o}fling},\ and\ \citenamefont
		{Bayer}}]{schmutzler-etal.15}%
	\BibitemOpen
	\bibfield  {author} {\bibinfo {author} {\bibfnamefont {J.}~\bibnamefont
			{Schmutzler}}, \bibinfo {author} {\bibfnamefont {P.}~\bibnamefont
			{Lewandowski}}, \bibinfo {author} {\bibfnamefont {M.}~\bibnamefont
			{A{\ss}mann}}, \bibinfo {author} {\bibfnamefont {D.}~\bibnamefont
			{Niemietz}}, \bibinfo {author} {\bibfnamefont {S.}~\bibnamefont
			{Schumacher}}, \bibinfo {author} {\bibfnamefont {M.}~\bibnamefont {Kamp}},
		\bibinfo {author} {\bibfnamefont {C.}~\bibnamefont {Schneider}}, \bibinfo
		{author} {\bibfnamefont {S.}~\bibnamefont {H{\"o}fling}},\ and\ \bibinfo
		{author} {\bibfnamefont {M.}~\bibnamefont {Bayer}},\ }\bibfield  {title}
	{\bibinfo {title} {All-optical flow control of a polariton condensate using
			nonresonant excitation},\ }\href {https://doi.org/10.1103/PhysRevB.91.195308}
	{\bibfield  {journal} {\bibinfo  {journal} {Phys. Rev. B}\ }\textbf {\bibinfo
			{volume} {91}},\ \bibinfo {pages} {195308} (\bibinfo {year}
		{2015})}\BibitemShut {NoStop}%
	\bibitem [{\citenamefont {de~Leeuw}\ \emph {et~al.}(2016)\citenamefont
		{de~Leeuw}, \citenamefont {van~der Wurff}, \citenamefont {Duine},
		\citenamefont {van Oosten},\ and\ \citenamefont {Stoof}}]{leeuw-etal.16}%
	\BibitemOpen
	\bibfield  {author} {\bibinfo {author} {\bibfnamefont {A.-W.}\ \bibnamefont
			{de~Leeuw}}, \bibinfo {author} {\bibfnamefont {E.~C.~I.}\ \bibnamefont
			{van~der Wurff}}, \bibinfo {author} {\bibfnamefont {R.~A.}\ \bibnamefont
			{Duine}}, \bibinfo {author} {\bibfnamefont {D.}~\bibnamefont {van Oosten}},\
		and\ \bibinfo {author} {\bibfnamefont {H.~T.~C.}\ \bibnamefont {Stoof}},\
	}\bibfield  {title} {\bibinfo {title} {Theory for {Bose}-{Einstein}
			condensation of light in nanofabricated semiconductor microcavities},\ }\href
	{https://doi.org/10.1103/PhysRevA.94.013615} {\bibfield  {journal} {\bibinfo
			{journal} {Phys. Rev. A}\ }\textbf {\bibinfo {volume} {94}},\ \bibinfo
		{pages} {013615} (\bibinfo {year} {2016})}\BibitemShut {NoStop}%
	\bibitem [{\citenamefont {Hayenga}\ and\ \citenamefont
		{Khajavikhan}(2017)}]{hayenga-khajavikhan.17}%
	\BibitemOpen
	\bibfield  {author} {\bibinfo {author} {\bibfnamefont {W.~E.}\ \bibnamefont
			{Hayenga}}\ and\ \bibinfo {author} {\bibfnamefont {M.}~\bibnamefont
			{Khajavikhan}},\ }\bibfield  {title} {\bibinfo {title} {Unveiling the physics
			of microcavity lasers},\ }\href {https://doi.org/10.1038/lsa.2017.91}
	{\bibfield  {journal} {\bibinfo  {journal} {Light: Science \& Applications}\
		}\textbf {\bibinfo {volume} {6}},\ \bibinfo {pages} {e17091} (\bibinfo {year}
		{2017})}\BibitemShut {NoStop}%
	\bibitem [{\citenamefont {Kavokin}\ \emph {et~al.}(2017)\citenamefont
		{Kavokin}, \citenamefont {Baumberg}, \citenamefont {Malpuech},\ and\
		\citenamefont {Laussy}}]{kavokin-etal.17}%
	\BibitemOpen
	\bibfield  {author} {\bibinfo {author} {\bibfnamefont {A.}~\bibnamefont
			{Kavokin}}, \bibinfo {author} {\bibfnamefont {J.}~\bibnamefont {Baumberg}},
		\bibinfo {author} {\bibfnamefont {G.}~\bibnamefont {Malpuech}},\ and\
		\bibinfo {author} {\bibfnamefont {F.}~\bibnamefont {Laussy}},\ }\href
	{https://books.google.com/books?id=cUi1DgAAQBAJ} {\emph {\bibinfo {title}
			{Microcavities}}},\ Oxford science publications\ (\bibinfo  {publisher}
	{Oxford University Press},\ \bibinfo {address} {Oxford, England},\ \bibinfo
	{year} {2017})\BibitemShut {NoStop}%
	\bibitem [{\citenamefont {Bao}\ \emph {et~al.}(2019)\citenamefont {Bao},
		\citenamefont {Liu}, \citenamefont {Xue}, \citenamefont {Zheng},
		\citenamefont {Tao}, \citenamefont {Wang}, \citenamefont {Xia}, \citenamefont
		{Zhao}, \citenamefont {Kim}, \citenamefont {Yang}, \citenamefont {Li},
		\citenamefont {Wang}, \citenamefont {Wang}, \citenamefont {Wang},
		\citenamefont {MacDonald},\ and\ \citenamefont {Zhang}}]{bao-etal.19}%
	\BibitemOpen
	\bibfield  {author} {\bibinfo {author} {\bibfnamefont {W.}~\bibnamefont
			{Bao}}, \bibinfo {author} {\bibfnamefont {X.}~\bibnamefont {Liu}}, \bibinfo
		{author} {\bibfnamefont {F.}~\bibnamefont {Xue}}, \bibinfo {author}
		{\bibfnamefont {F.}~\bibnamefont {Zheng}}, \bibinfo {author} {\bibfnamefont
			{R.}~\bibnamefont {Tao}}, \bibinfo {author} {\bibfnamefont {S.}~\bibnamefont
			{Wang}}, \bibinfo {author} {\bibfnamefont {Y.}~\bibnamefont {Xia}}, \bibinfo
		{author} {\bibfnamefont {M.}~\bibnamefont {Zhao}}, \bibinfo {author}
		{\bibfnamefont {J.}~\bibnamefont {Kim}}, \bibinfo {author} {\bibfnamefont
			{S.}~\bibnamefont {Yang}}, \bibinfo {author} {\bibfnamefont {Q.}~\bibnamefont
			{Li}}, \bibinfo {author} {\bibfnamefont {Y.}~\bibnamefont {Wang}}, \bibinfo
		{author} {\bibfnamefont {Y.}~\bibnamefont {Wang}}, \bibinfo {author}
		{\bibfnamefont {L.-W.}\ \bibnamefont {Wang}}, \bibinfo {author}
		{\bibfnamefont {A.~H.}\ \bibnamefont {MacDonald}},\ and\ \bibinfo {author}
		{\bibfnamefont {X.}~\bibnamefont {Zhang}},\ }\bibfield  {title} {\bibinfo
		{title} {Observation of {Rydberg} exciton polaritons and their condensate in
			a perovskite cavity},\ }\href {https://doi.org/10.1073/pnas.1909948116}
	{\bibfield  {journal} {\bibinfo  {journal} {Proc. Natl. Acad. Sci.}\ }\textbf
		{\bibinfo {volume} {116}},\ \bibinfo {pages} {20274} (\bibinfo {year}
		{2019})}\BibitemShut {NoStop}%
	\bibitem [{\citenamefont {Carcamo}\ \emph {et~al.}(2020)\citenamefont
		{Carcamo}, \citenamefont {Schumacher}, \citenamefont {Schumacher},\ and\
		\citenamefont {Binder}}]{carcamo-etal.20}%
	\BibitemOpen
	\bibfield  {author} {\bibinfo {author} {\bibfnamefont {M.}~\bibnamefont
			{Carcamo}}, \bibinfo {author} {\bibfnamefont {S.}~\bibnamefont {Schumacher}},
		\bibinfo {author} {\bibfnamefont {S.}~\bibnamefont {Schumacher}},\ and\
		\bibinfo {author} {\bibfnamefont {R.}~\bibnamefont {Binder}},\ }\bibfield
	{title} {\bibinfo {title} {Transfer function replacement of phenomenological
			single-mode equations in semiconductor microcavity modeling},\ }\href
	{https://doi.org/10.1364/AO.392014} {\bibfield  {journal} {\bibinfo
			{journal} {Appl. Opt.}\ }\textbf {\bibinfo {volume} {59}},\ \bibinfo {pages}
		{G112} (\bibinfo {year} {2020})}\BibitemShut {NoStop}%
	\bibitem [{\citenamefont {Comte}\ and\ \citenamefont
		{Nozi{\`e}res}(1982)}]{comte-nozieres.82}%
	\BibitemOpen
	\bibfield  {author} {\bibinfo {author} {\bibfnamefont {C.}~\bibnamefont
			{Comte}}\ and\ \bibinfo {author} {\bibfnamefont {P.}~\bibnamefont
			{Nozi{\`e}res}},\ }\bibfield  {title} {\bibinfo {title} {Exciton {Bose}
			{Condensation}: the ground state of an electrion-hole gas {I}. {Mean} field
			description of a simplified model},\ }\href@noop {} {\bibfield  {journal}
		{\bibinfo  {journal} {J. Physique}\ }\textbf {\bibinfo {volume} {43}},\
		\bibinfo {pages} {1069} (\bibinfo {year} {1982})}\BibitemShut {NoStop}%
	\bibitem [{\citenamefont {Keeling}\ \emph {et~al.}(2005)\citenamefont
		{Keeling}, \citenamefont {Eastham}, \citenamefont {Szymanska},\ and\
		\citenamefont {Littlewood}}]{keeling-etal.05}%
	\BibitemOpen
	\bibfield  {author} {\bibinfo {author} {\bibfnamefont {J.}~\bibnamefont
			{Keeling}}, \bibinfo {author} {\bibfnamefont {P.~R.}\ \bibnamefont
			{Eastham}}, \bibinfo {author} {\bibfnamefont {M.~H.}\ \bibnamefont
			{Szymanska}},\ and\ \bibinfo {author} {\bibfnamefont {P.~B.}\ \bibnamefont
			{Littlewood}},\ }\bibfield  {title} {\bibinfo {title} {{BCS}-{BEC} crossover
			in a system of microcavity polaritons},\ }\href
	{https://doi.org/10.1103/PhysRevB.72.115320} {\bibfield  {journal} {\bibinfo
			{journal} {Phys. Rev. B}\ }\textbf {\bibinfo {volume} {72}},\ \bibinfo
		{pages} {115320} (\bibinfo {year} {2005})}\BibitemShut {NoStop}%
	\bibitem [{\citenamefont {Kremp}\ \emph {et~al.}(2008)\citenamefont {Kremp},
		\citenamefont {Semkat},\ and\ \citenamefont {Henneberger}}]{kremp-etal.08}%
	\BibitemOpen
	\bibfield  {author} {\bibinfo {author} {\bibfnamefont {D.}~\bibnamefont
			{Kremp}}, \bibinfo {author} {\bibfnamefont {D.}~\bibnamefont {Semkat}},\ and\
		\bibinfo {author} {\bibfnamefont {K.}~\bibnamefont {Henneberger}},\
	}\bibfield  {title} {\bibinfo {title} {Quantum condensation in electron-hole
			plasmas},\ }\href {https://doi.org/10.1103/physrevb.78.125315} {\bibfield
		{journal} {\bibinfo  {journal} {Phys. Rev. B}\ }\textbf {\bibinfo {volume}
			{78}},\ \bibinfo {pages} {125315} (\bibinfo {year} {2008})}\BibitemShut
	{NoStop}%
	\bibitem [{\citenamefont {Byrnes}\ \emph {et~al.}(2010)\citenamefont {Byrnes},
		\citenamefont {Horikiri}, \citenamefont {Ishida},\ and\ \citenamefont
		{Yamamoto}}]{byrnes-etal.10}%
	\BibitemOpen
	\bibfield  {author} {\bibinfo {author} {\bibfnamefont {T.}~\bibnamefont
			{Byrnes}}, \bibinfo {author} {\bibfnamefont {T.}~\bibnamefont {Horikiri}},
		\bibinfo {author} {\bibfnamefont {N.}~\bibnamefont {Ishida}},\ and\ \bibinfo
		{author} {\bibfnamefont {Y.}~\bibnamefont {Yamamoto}},\ }\bibfield  {title}
	{\bibinfo {title} {{BCS} wave-function approach to the {BEC}-{BCS} crossover
			of exciton-polariton condensates},\ }\href
	{https://doi.org/10.1103/PhysRevLett.105.186402} {\bibfield  {journal}
		{\bibinfo  {journal} {Phys. Rev. Lett.}\ }\textbf {\bibinfo {volume} {105}},\
		\bibinfo {pages} {186402} (\bibinfo {year} {2010})}\BibitemShut {NoStop}%
	\bibitem [{\citenamefont {Combescot}\ and\ \citenamefont
		{Shiau}(2015)}]{combescot-shiau.15}%
	\BibitemOpen
	\bibfield  {author} {\bibinfo {author} {\bibfnamefont {M.}~\bibnamefont
			{Combescot}}\ and\ \bibinfo {author} {\bibfnamefont {S.-Y.}\ \bibnamefont
			{Shiau}},\ }\href {https://doi.org/10.1093/acprof:oso/9780198753735.001.0001}
	{\emph {\bibinfo {title} {Excitons and Cooper Pairs}}}\ (\bibinfo
	{publisher} {Oxford University Press},\ \bibinfo {address} {Oxford, UK},\
	\bibinfo {year} {2015})\BibitemShut {NoStop}%
	\bibitem [{\citenamefont {Hu}\ and\ \citenamefont {Liu}(2020)}]{hu-liu.20}%
	\BibitemOpen
	\bibfield  {author} {\bibinfo {author} {\bibfnamefont {H.}~\bibnamefont
			{Hu}}\ and\ \bibinfo {author} {\bibfnamefont {X.-J.}\ \bibnamefont {Liu}},\
	}\bibfield  {title} {\bibinfo {title} {Quantum fluctuations in a strongly
			interacting {Bardeen-Cooper-Schrieffer} polariton condensate at thermal
			equilibrium},\ }\href {https://doi.org/10.1103/physreva.101.011602}
	{\bibfield  {journal} {\bibinfo  {journal} {Phys. Rev. A}\ }\textbf {\bibinfo
			{volume} {101}},\ \bibinfo {pages} {011602} (\bibinfo {year}
		{2020})}\BibitemShut {NoStop}%
	\bibitem [{\citenamefont {Hu}\ \emph {et~al.}(2021)\citenamefont {Hu},
		\citenamefont {Wang}, \citenamefont {Kim}, \citenamefont {Deng},
		\citenamefont {Brodbeck}, \citenamefont {Schneider}, \citenamefont
		{H{\"o}fling}, \citenamefont {Kwong},\ and\ \citenamefont
		{Binder}}]{hu-etal.21}%
	\BibitemOpen
	\bibfield  {author} {\bibinfo {author} {\bibfnamefont {J.}~\bibnamefont
			{Hu}}, \bibinfo {author} {\bibfnamefont {Z.}~\bibnamefont {Wang}}, \bibinfo
		{author} {\bibfnamefont {S.}~\bibnamefont {Kim}}, \bibinfo {author}
		{\bibfnamefont {H.}~\bibnamefont {Deng}}, \bibinfo {author} {\bibfnamefont
			{S.}~\bibnamefont {Brodbeck}}, \bibinfo {author} {\bibfnamefont
			{C.}~\bibnamefont {Schneider}}, \bibinfo {author} {\bibfnamefont
			{S.}~\bibnamefont {H{\"o}fling}}, \bibinfo {author} {\bibfnamefont {N.~H.}\
			\bibnamefont {Kwong}},\ and\ \bibinfo {author} {\bibfnamefont
			{R.}~\bibnamefont {Binder}},\ }\bibfield  {title} {\bibinfo {title}
		{Polariton {Laser} in the {Bardeen}-{Cooper}-{Schrieffer} {Regime}},\ }\href
	{https://doi.org/10.1103/PhysRevX.11.011018} {\bibfield  {journal} {\bibinfo
			{journal} {Phys. Rev. X}\ }\textbf {\bibinfo {volume} {11}},\ \bibinfo
		{pages} {011018} (\bibinfo {year} {2021})},\ \Eprint
	{https://arxiv.org/abs/1902.00142} {arXiv:1902.00142 [cond-mat.mes-hall]}
	\BibitemShut {NoStop}%
	\bibitem [{\citenamefont {Binder}\ and\ \citenamefont
		{Kwong}(2021)}]{binder-kwong.21}%
	\BibitemOpen
	\bibfield  {author} {\bibinfo {author} {\bibfnamefont {R.}~\bibnamefont
			{Binder}}\ and\ \bibinfo {author} {\bibfnamefont {N.~H.}\ \bibnamefont
			{Kwong}},\ }\bibfield  {title} {\bibinfo {title} {Metamorphosis of
			{Goldstone} and soft fluctuation modes in polariton lasers},\ }\href
	{https://doi.org/10.1103/PhysRevB.103.085304} {\bibfield  {journal} {\bibinfo
			{journal} {Phys. Rev. B}\ }\textbf {\bibinfo {volume} {103}},\ \bibinfo
		{pages} {085304} (\bibinfo {year} {2021})},\ \Eprint
	{https://arxiv.org/abs/2007.13253} {arXiv:2007.13253 [cond-mat.mes-hall]}
	\BibitemShut {NoStop}%
	\bibitem [{\citenamefont {Galitskii}\ \emph {et~al.}(1970)\citenamefont
		{Galitskii}, \citenamefont {Goreslavskii},\ and\ \citenamefont
		{Elesin}}]{galitskii-etal.70}%
	\BibitemOpen
	\bibfield  {author} {\bibinfo {author} {\bibfnamefont {V.~M.}\ \bibnamefont
			{Galitskii}}, \bibinfo {author} {\bibfnamefont {S.~P.}\ \bibnamefont
			{Goreslavskii}},\ and\ \bibinfo {author} {\bibfnamefont {V.~F.}\ \bibnamefont
			{Elesin}},\ }\bibfield  {title} {\bibinfo {title} {Electric and magnetic
			properties of a semiconductor in the field of a strong electromagnetic
			wave},\ }\href {http://jetp.ac.ru/cgi-bin/dn/e_030_01_0117.pdf} {\bibfield
		{journal} {\bibinfo  {journal} {Sov. Phys. JETP}\ }\textbf {\bibinfo {volume}
			{30}},\ \bibinfo {pages} {117} (\bibinfo {year} {1970})}\BibitemShut
	{NoStop}%
	\bibitem [{\citenamefont {Nishimura}\ and\ \citenamefont
		{Nishimura}(1973)}]{nishimura-nishimura.73}%
	\BibitemOpen
	\bibfield  {author} {\bibinfo {author} {\bibfnamefont {Y.}~\bibnamefont
			{Nishimura}}\ and\ \bibinfo {author} {\bibfnamefont {Y.}~\bibnamefont
			{Nishimura}},\ }\bibfield  {title} {\bibinfo {title} {Spectral hole-burning
			and nonlinear-gain decrease in a band-to-level transition semiconductor
			laser},\ }\href {https://doi.org/10.1109/jqe.1973.1077406} {\bibfield
		{journal} {\bibinfo  {journal} {{IEEE} Journal of Quantum Electronics}\
		}\textbf {\bibinfo {volume} {9}},\ \bibinfo {pages} {1011} (\bibinfo {year}
		{1973})}\BibitemShut {NoStop}%
	\bibitem [{\citenamefont {Keldysh}(1995)}]{keldysh.95}%
	\BibitemOpen
	\bibfield  {author} {\bibinfo {author} {\bibfnamefont {L.~V.}\ \bibnamefont
			{Keldysh}},\ }\bibfield  {title} {\bibinfo {title} {Correlations in the
			coherent transient electron-hole system},\ }\href
	{https://doi.org/10.1002/pssb.2221880102} {\bibfield  {journal} {\bibinfo
			{journal} {Phys. Status Solidi B}\ }\textbf {\bibinfo {volume} {188}},\
		\bibinfo {pages} {11} (\bibinfo {year} {1995})}\BibitemShut {NoStop}%
	\bibitem [{\citenamefont {Patel}\ \emph {et~al.}(1979)\citenamefont {Patel},
		\citenamefont {Brosson},\ and\ \citenamefont {Ripper}}]{patel-etal.79}%
	\BibitemOpen
	\bibfield  {author} {\bibinfo {author} {\bibfnamefont {N.~B.}\ \bibnamefont
			{Patel}}, \bibinfo {author} {\bibfnamefont {P.}~\bibnamefont {Brosson}},\
		and\ \bibinfo {author} {\bibfnamefont {J.~E.}\ \bibnamefont {Ripper}},\
	}\bibfield  {title} {\bibinfo {title} {Spectral hole burning in {GaAs}
			junction lasers},\ }\href {https://doi.org/10.1063/1.90776} {\bibfield
		{journal} {\bibinfo  {journal} {Appl. Phys. Lett.}\ }\textbf {\bibinfo
			{volume} {34}},\ \bibinfo {pages} {330} (\bibinfo {year} {1979})}\BibitemShut
	{NoStop}%
	\bibitem [{\citenamefont {Schmitt-Rink}\ \emph {et~al.}(1988)\citenamefont
		{Schmitt-Rink}, \citenamefont {Chemla},\ and\ \citenamefont
		{Haug}}]{schmitt-rink-etal.88}%
	\BibitemOpen
	\bibfield  {author} {\bibinfo {author} {\bibfnamefont {S.}~\bibnamefont
			{Schmitt-Rink}}, \bibinfo {author} {\bibfnamefont {D.~S.}\ \bibnamefont
			{Chemla}},\ and\ \bibinfo {author} {\bibfnamefont {H.}~\bibnamefont {Haug}},\
	}\bibfield  {title} {\bibinfo {title} {Nonequilibrium theory of the optical
			stark effect and spectral hole burning in semiconductors},\ }\href
	{https://doi.org/10.1103/physrevb.37.941} {\bibfield  {journal} {\bibinfo
			{journal} {Phys. Rev. B}\ }\textbf {\bibinfo {volume} {37}},\ \bibinfo
		{pages} {941} (\bibinfo {year} {1988})}\BibitemShut {NoStop}%
	\bibitem [{\citenamefont {Paul}\ \emph {et~al.}(1992)\citenamefont {Paul},
		\citenamefont {Binder},\ and\ \citenamefont {Koch}}]{paul-etal.92}%
	\BibitemOpen
	\bibfield  {author} {\bibinfo {author} {\bibfnamefont {A.~E.}\ \bibnamefont
			{Paul}}, \bibinfo {author} {\bibfnamefont {R.}~\bibnamefont {Binder}},\ and\
		\bibinfo {author} {\bibfnamefont {S.~W.}\ \bibnamefont {Koch}},\ }\bibfield
	{title} {\bibinfo {title} {Spectral hole burning and light-induced band
			splitting in the gain region of highly excited semiconductors},\ }\href
	{https://doi.org/10.1103/physrevb.45.5879} {\bibfield  {journal} {\bibinfo
			{journal} {Phys. Rev. B}\ }\textbf {\bibinfo {volume} {45}},\ \bibinfo
		{pages} {5879} (\bibinfo {year} {1992})}\BibitemShut {NoStop}%
	\bibitem [{\citenamefont {Henneberger}\ \emph {et~al.}(1992)\citenamefont
		{Henneberger}, \citenamefont {Herzel}, \citenamefont {Koch}, \citenamefont
		{Binder}, \citenamefont {Paul},\ and\ \citenamefont
		{Scott}}]{henneberger-etal.92}%
	\BibitemOpen
	\bibfield  {author} {\bibinfo {author} {\bibfnamefont {K.}~\bibnamefont
			{Henneberger}}, \bibinfo {author} {\bibfnamefont {F.}~\bibnamefont {Herzel}},
		\bibinfo {author} {\bibfnamefont {S.~W.}\ \bibnamefont {Koch}}, \bibinfo
		{author} {\bibfnamefont {R.}~\bibnamefont {Binder}}, \bibinfo {author}
		{\bibfnamefont {A.~E.}\ \bibnamefont {Paul}},\ and\ \bibinfo {author}
		{\bibfnamefont {D.}~\bibnamefont {Scott}},\ }\bibfield  {title} {\bibinfo
		{title} {Spectral hole burning and gain saturation in short-cavity
			semiconductor lasers},\ }\href {https://doi.org/10.1103/physreva.45.1853}
	{\bibfield  {journal} {\bibinfo  {journal} {Phys. Rev. A}\ }\textbf {\bibinfo
			{volume} {45}},\ \bibinfo {pages} {1853} (\bibinfo {year}
		{1992})}\BibitemShut {NoStop}%
	\bibitem [{\citenamefont {Meissner}\ \emph {et~al.}(1993)\citenamefont
		{Meissner}, \citenamefont {Fluegel}, \citenamefont {Giessen}, \citenamefont
		{McGinnis}, \citenamefont {Paul}, \citenamefont {Binder}, \citenamefont
		{Koch}, \citenamefont {Peyghambarian}, \citenamefont {Gr{\"{u}}n},\ and\
		\citenamefont {Klingshirn}}]{meissner-etal.93}%
	\BibitemOpen
	\bibfield  {author} {\bibinfo {author} {\bibfnamefont {K.}~\bibnamefont
			{Meissner}}, \bibinfo {author} {\bibfnamefont {B.}~\bibnamefont {Fluegel}},
		\bibinfo {author} {\bibfnamefont {H.}~\bibnamefont {Giessen}}, \bibinfo
		{author} {\bibfnamefont {B.~P.}\ \bibnamefont {McGinnis}}, \bibinfo {author}
		{\bibfnamefont {A.}~\bibnamefont {Paul}}, \bibinfo {author} {\bibfnamefont
			{R.}~\bibnamefont {Binder}}, \bibinfo {author} {\bibfnamefont {S.~W.}\
			\bibnamefont {Koch}}, \bibinfo {author} {\bibfnamefont {N.}~\bibnamefont
			{Peyghambarian}}, \bibinfo {author} {\bibfnamefont {M.}~\bibnamefont
			{Gr{\"{u}}n}},\ and\ \bibinfo {author} {\bibfnamefont {C.}~\bibnamefont
			{Klingshirn}},\ }\bibfield  {title} {\bibinfo {title} {Spectral hole burning
			in the gain region of an inverted semiconductor},\ }\href
	{https://doi.org/10.1103/physrevb.48.15472} {\bibfield  {journal} {\bibinfo
			{journal} {Phys. Rev. B}\ }\textbf {\bibinfo {volume} {48}},\ \bibinfo
		{pages} {15472} (\bibinfo {year} {1993})}\BibitemShut {NoStop}%
	\bibitem [{\citenamefont {Yamaguchi}\ \emph {et~al.}(2015)\citenamefont
		{Yamaguchi}, \citenamefont {Nii}, \citenamefont {Kamide}, \citenamefont
		{Ogawa},\ and\ \citenamefont {Yamamoto}}]{yamaguchi-etal.15}%
	\BibitemOpen
	\bibfield  {author} {\bibinfo {author} {\bibfnamefont {M.}~\bibnamefont
			{Yamaguchi}}, \bibinfo {author} {\bibfnamefont {R.}~\bibnamefont {Nii}},
		\bibinfo {author} {\bibfnamefont {K.}~\bibnamefont {Kamide}}, \bibinfo
		{author} {\bibfnamefont {T.}~\bibnamefont {Ogawa}},\ and\ \bibinfo {author}
		{\bibfnamefont {Y.}~\bibnamefont {Yamamoto}},\ }\bibfield  {title} {\bibinfo
		{title} {Generating functional approach for spontaneous coherence in
			semiconductor electron-hole-photon systems},\ }\href
	{https://doi.org/10.1103/physrevb.91.115129} {\bibfield  {journal} {\bibinfo
			{journal} {Phys. Rev. B}\ }\textbf {\bibinfo {volume} {91}},\ \bibinfo
		{pages} {115129} (\bibinfo {year} {2015})}\BibitemShut {NoStop}%
	\bibitem [{\citenamefont {Murotani}\ \emph {et~al.}(2019)\citenamefont
		{Murotani}, \citenamefont {Kim}, \citenamefont {Akiyama}, \citenamefont
		{Pfeiffer}, \citenamefont {West},\ and\ \citenamefont
		{Shimano}}]{murotani-etal.19}%
	\BibitemOpen
	\bibfield  {author} {\bibinfo {author} {\bibfnamefont {Y.}~\bibnamefont
			{Murotani}}, \bibinfo {author} {\bibfnamefont {C.}~\bibnamefont {Kim}},
		\bibinfo {author} {\bibfnamefont {H.}~\bibnamefont {Akiyama}}, \bibinfo
		{author} {\bibfnamefont {L.~N.}\ \bibnamefont {Pfeiffer}}, \bibinfo {author}
		{\bibfnamefont {K.~W.}\ \bibnamefont {West}},\ and\ \bibinfo {author}
		{\bibfnamefont {R.}~\bibnamefont {Shimano}},\ }\bibfield  {title} {\bibinfo
		{title} {Light-driven electron-hole {Bardeen-Cooper-Schrieffer}-like state in
			bulk {GaAs}},\ }\href {https://doi.org/10.1103/physrevlett.123.197401}
	{\bibfield  {journal} {\bibinfo  {journal} {Phys. Rev. Lett.}\ }\textbf
		{\bibinfo {volume} {123}},\ \bibinfo {pages} {197401} (\bibinfo {year}
		{2019})}\BibitemShut {NoStop}%
	\bibitem [{\citenamefont {Quochi}\ \emph {et~al.}(1998)\citenamefont {Quochi},
		\citenamefont {Bongiovanni}, \citenamefont {Mura}, \citenamefont {Staehli},
		\citenamefont {Deveaud}, \citenamefont {Stanley}, \citenamefont {Oesterle},\
		and\ \citenamefont {Houdr{\'e}}}]{quochi-etal.98}%
	\BibitemOpen
	\bibfield  {author} {\bibinfo {author} {\bibfnamefont {F.}~\bibnamefont
			{Quochi}}, \bibinfo {author} {\bibfnamefont {G.}~\bibnamefont {Bongiovanni}},
		\bibinfo {author} {\bibfnamefont {A.}~\bibnamefont {Mura}}, \bibinfo {author}
		{\bibfnamefont {J.~L.}\ \bibnamefont {Staehli}}, \bibinfo {author}
		{\bibfnamefont {B.}~\bibnamefont {Deveaud}}, \bibinfo {author} {\bibfnamefont
			{R.~P.}\ \bibnamefont {Stanley}}, \bibinfo {author} {\bibfnamefont
			{U.}~\bibnamefont {Oesterle}},\ and\ \bibinfo {author} {\bibfnamefont
			{R.}~\bibnamefont {Houdr{\'e}}},\ }\bibfield  {title} {\bibinfo {title}
		{Strongly {Driven} {Semiconductor} {Microcavities}: {From} the {Polariton}
			{Doublet} to an ac {Stark} {Triplet}},\ }\href
	{https://doi.org/10.1103/PhysRevLett.80.4733} {\bibfield  {journal} {\bibinfo
			{journal} {Phys. Rev. Lett.}\ }\textbf {\bibinfo {volume} {80}},\ \bibinfo
		{pages} {4733} (\bibinfo {year} {1998})}\BibitemShut {NoStop}%
	\bibitem [{\citenamefont {Horikiri}\ \emph {et~al.}(2016)\citenamefont
		{Horikiri}, \citenamefont {Yamaguchi}, \citenamefont {Kamide}, \citenamefont
		{Matsuo}, \citenamefont {Byrnes}, \citenamefont {Ishida}, \citenamefont
		{L{\"o}ffler}, \citenamefont {H{\"o}fling}, \citenamefont {Shikano},
		\citenamefont {Ogawa}, \citenamefont {Forchel},\ and\ \citenamefont
		{Yamamoto}}]{horikiri-etal.16}%
	\BibitemOpen
	\bibfield  {author} {\bibinfo {author} {\bibfnamefont {T.}~\bibnamefont
			{Horikiri}}, \bibinfo {author} {\bibfnamefont {M.}~\bibnamefont {Yamaguchi}},
		\bibinfo {author} {\bibfnamefont {K.}~\bibnamefont {Kamide}}, \bibinfo
		{author} {\bibfnamefont {Y.}~\bibnamefont {Matsuo}}, \bibinfo {author}
		{\bibfnamefont {T.}~\bibnamefont {Byrnes}}, \bibinfo {author} {\bibfnamefont
			{N.}~\bibnamefont {Ishida}}, \bibinfo {author} {\bibfnamefont
			{A.}~\bibnamefont {L{\"o}ffler}}, \bibinfo {author} {\bibfnamefont
			{S.}~\bibnamefont {H{\"o}fling}}, \bibinfo {author} {\bibfnamefont
			{Y.}~\bibnamefont {Shikano}}, \bibinfo {author} {\bibfnamefont
			{T.}~\bibnamefont {Ogawa}}, \bibinfo {author} {\bibfnamefont
			{A.}~\bibnamefont {Forchel}},\ and\ \bibinfo {author} {\bibfnamefont
			{Y.}~\bibnamefont {Yamamoto}},\ }\bibfield  {title} {\bibinfo {title}
		{High-energy side-peak emission of exciton-polariton condensates in high
			density regime},\ }\href {https://doi.org/10.1038/srep25655} {\bibfield
		{journal} {\bibinfo  {journal} {Sci. Rep.}\ }\textbf {\bibinfo {volume}
			{6}},\ \bibinfo {pages} {25655} (\bibinfo {year} {2016})}\BibitemShut
	{NoStop}%
	\bibitem [{\citenamefont {Mahan}(2000)}]{mahan.00}%
	\BibitemOpen
	\bibfield  {author} {\bibinfo {author} {\bibfnamefont {G.~D.}\ \bibnamefont
			{Mahan}},\ }\href {https://doi.org/10.1007/978-1-4757-5714-9} {\emph
		{\bibinfo {title} {Many-Particle Physics}}},\ \bibinfo {edition} {3rd}\ ed.,\
	Physics of Solids and Liquids\ (\bibinfo  {publisher} {Springer {US}},\
	\bibinfo {address} {New York},\ \bibinfo {year} {2000})\BibitemShut {NoStop}%
	\bibitem [{\citenamefont {Jepsen}\ \emph {et~al.}(2011)\citenamefont {Jepsen},
		\citenamefont {Cooke},\ and\ \citenamefont {Koch}}]{jepsen-etal.11}%
	\BibitemOpen
	\bibfield  {author} {\bibinfo {author} {\bibfnamefont {P.~U.}\ \bibnamefont
			{Jepsen}}, \bibinfo {author} {\bibfnamefont {D.~G.}\ \bibnamefont {Cooke}},\
		and\ \bibinfo {author} {\bibfnamefont {M.}~\bibnamefont {Koch}},\ }\bibfield
	{title} {\bibinfo {title} {Terahertz spectroscopy and imaging --- {Modern}
			techniques and applications},\ }\href
	{https://doi.org/https://doi.org/10.1002/lpor.201000011} {\bibfield
		{journal} {\bibinfo  {journal} {Laser \& Photonics Reviews}\ }\textbf
		{\bibinfo {volume} {5}},\ \bibinfo {pages} {124} (\bibinfo {year}
		{2011})}\BibitemShut {NoStop}%
	\bibitem [{\citenamefont {Kuwata-Gonokami}\ \emph {et~al.}(2004)\citenamefont
		{Kuwata-Gonokami}, \citenamefont {Kubouchi}, \citenamefont {Shimano},\ and\
		\citenamefont {Mysyrowicz}}]{gonokami-etal.04}%
	\BibitemOpen
	\bibfield  {author} {\bibinfo {author} {\bibfnamefont {M.}~\bibnamefont
			{Kuwata-Gonokami}}, \bibinfo {author} {\bibfnamefont {M.}~\bibnamefont
			{Kubouchi}}, \bibinfo {author} {\bibfnamefont {R.}~\bibnamefont {Shimano}},\
		and\ \bibinfo {author} {\bibfnamefont {A.}~\bibnamefont {Mysyrowicz}},\
	}\bibfield  {title} {\bibinfo {title} {Time-resolved {Excitonic} {Lyman}
			{Spectroscopy} of {Cu$_{2}$O}},\ }\href
	{https://doi.org/10.1143/JPSJ.73.1065} {\bibfield  {journal} {\bibinfo
			{journal} {J. Phys. Soc. Jpn.}\ }\textbf {\bibinfo {volume} {73}},\ \bibinfo
		{pages} {1065} (\bibinfo {year} {2004})}\BibitemShut {NoStop}%
	\bibitem [{\citenamefont {Kuwata-Gonokami}(2005)}]{gonokami.05}%
	\BibitemOpen
	\bibfield  {author} {\bibinfo {author} {\bibfnamefont {M.}~\bibnamefont
			{Kuwata-Gonokami}},\ }\bibfield  {title} {\bibinfo {title} {Observation of
			ortho and para-excitons by time-resolved excitonic {Lyman} spectroscopy},\
	}\href {https://doi.org/10.1016/j.ssc.2004.10.036} {\bibfield  {journal}
		{\bibinfo  {journal} {Solid State Commun.}\ }\bibinfo {series} {Spontaneous
			{Coherence} in {Excitonic} {Systems}},\ \textbf {\bibinfo {volume} {134}},\
		\bibinfo {pages} {127} (\bibinfo {year} {2005})}\BibitemShut {NoStop}%
	\bibitem [{\citenamefont {Kira}\ \emph {et~al.}(2001)\citenamefont {Kira},
		\citenamefont {Hoyer}, \citenamefont {Stroucken},\ and\ \citenamefont
		{Koch}}]{kira-etal.01}%
	\BibitemOpen
	\bibfield  {author} {\bibinfo {author} {\bibfnamefont {M.}~\bibnamefont
			{Kira}}, \bibinfo {author} {\bibfnamefont {W.}~\bibnamefont {Hoyer}},
		\bibinfo {author} {\bibfnamefont {T.}~\bibnamefont {Stroucken}},\ and\
		\bibinfo {author} {\bibfnamefont {S.~W.}\ \bibnamefont {Koch}},\ }\bibfield
	{title} {\bibinfo {title} {Exciton formation in semiconductors and the
			influence of a photonic environment},\ }\href
	{https://doi.org/10.1103/physrevlett.87.176401} {\bibfield  {journal}
		{\bibinfo  {journal} {Phys. Rev. Lett.}\ }\textbf {\bibinfo {volume} {87}},\
		\bibinfo {pages} {176401} (\bibinfo {year} {2001})}\BibitemShut {NoStop}%
	\bibitem [{\citenamefont {Danielson}\ \emph {et~al.}(2007)\citenamefont
		{Danielson}, \citenamefont {Lee}, \citenamefont {Prineas}, \citenamefont
		{Steiner}, \citenamefont {Kira},\ and\ \citenamefont
		{Koch}}]{danielson-etal.07}%
	\BibitemOpen
	\bibfield  {author} {\bibinfo {author} {\bibfnamefont {J.~R.}\ \bibnamefont
			{Danielson}}, \bibinfo {author} {\bibfnamefont {Y.-S.}\ \bibnamefont {Lee}},
		\bibinfo {author} {\bibfnamefont {J.~P.}\ \bibnamefont {Prineas}}, \bibinfo
		{author} {\bibfnamefont {J.~T.}\ \bibnamefont {Steiner}}, \bibinfo {author}
		{\bibfnamefont {M.}~\bibnamefont {Kira}},\ and\ \bibinfo {author}
		{\bibfnamefont {S.~W.}\ \bibnamefont {Koch}},\ }\bibfield  {title} {\bibinfo
		{title} {Interaction of {Strong} {Single}-{Cycle} {Terahertz} {Pulses} with
			{Semiconductor} {Quantum} {Wells}},\ }\href
	{https://doi.org/10.1103/PhysRevLett.99.237401} {\bibfield  {journal}
		{\bibinfo  {journal} {Phys. Rev. Lett.}\ }\textbf {\bibinfo {volume} {99}},\
		\bibinfo {pages} {237401} (\bibinfo {year} {2007})}\BibitemShut {NoStop}%
	\bibitem [{\citenamefont {Kaindl}\ \emph {et~al.}(2009)\citenamefont {Kaindl},
		\citenamefont {H{\"a}gele}, \citenamefont {Carnahan},\ and\ \citenamefont
		{Chemla}}]{kaindl-etal.09}%
	\BibitemOpen
	\bibfield  {author} {\bibinfo {author} {\bibfnamefont {R.~A.}\ \bibnamefont
			{Kaindl}}, \bibinfo {author} {\bibfnamefont {D.}~\bibnamefont {H{\"a}gele}},
		\bibinfo {author} {\bibfnamefont {M.~A.}\ \bibnamefont {Carnahan}},\ and\
		\bibinfo {author} {\bibfnamefont {D.~S.}\ \bibnamefont {Chemla}},\ }\bibfield
	{title} {\bibinfo {title} {Transient terahertz spectroscopy of excitons and
			unbound carriers in quasi-two-dimensional electron-hole gases},\ }\href
	{https://doi.org/10.1103/physrevb.79.045320} {\bibfield  {journal} {\bibinfo
			{journal} {Phys. Rev. B}\ }\textbf {\bibinfo {volume} {79}},\ \bibinfo
		{pages} {045320} (\bibinfo {year} {2009})}\BibitemShut {NoStop}%
	\bibitem [{\citenamefont {Kira}\ and\ \citenamefont
		{Koch}(2011)}]{kira-koch.11}%
	\BibitemOpen
	\bibfield  {author} {\bibinfo {author} {\bibfnamefont {M.}~\bibnamefont
			{Kira}}\ and\ \bibinfo {author} {\bibfnamefont {S.~W.}\ \bibnamefont
			{Koch}},\ }\href {http://www.sqobook.org/} {\emph {\bibinfo {title}
			{{Semiconductor Quantum Optics}}}}\ (\bibinfo  {publisher} {Cambridge
		University Press},\ \bibinfo {address} {Cambridge, UK},\ \bibinfo {year}
	{2011})\BibitemShut {NoStop}%
	\bibitem [{\citenamefont {Teich}\ \emph {et~al.}(2014)\citenamefont {Teich},
		\citenamefont {Wagner}, \citenamefont {Stehr}, \citenamefont {Schneider},
		\citenamefont {Helm}, \citenamefont {B{\"{o}}ttge}, \citenamefont {Klettke},
		\citenamefont {Chatterjee}, \citenamefont {Kira}, \citenamefont {Koch},
		\citenamefont {Khitrova},\ and\ \citenamefont {Gibbs}}]{teich-etal.14}%
	\BibitemOpen
	\bibfield  {author} {\bibinfo {author} {\bibfnamefont {M.}~\bibnamefont
			{Teich}}, \bibinfo {author} {\bibfnamefont {M.}~\bibnamefont {Wagner}},
		\bibinfo {author} {\bibfnamefont {D.}~\bibnamefont {Stehr}}, \bibinfo
		{author} {\bibfnamefont {H.}~\bibnamefont {Schneider}}, \bibinfo {author}
		{\bibfnamefont {M.}~\bibnamefont {Helm}}, \bibinfo {author} {\bibfnamefont
			{C.~N.}\ \bibnamefont {B{\"{o}}ttge}}, \bibinfo {author} {\bibfnamefont
			{A.~C.}\ \bibnamefont {Klettke}}, \bibinfo {author} {\bibfnamefont
			{S.}~\bibnamefont {Chatterjee}}, \bibinfo {author} {\bibfnamefont
			{M.}~\bibnamefont {Kira}}, \bibinfo {author} {\bibfnamefont {S.~W.}\
			\bibnamefont {Koch}}, \bibinfo {author} {\bibfnamefont {G.}~\bibnamefont
			{Khitrova}},\ and\ \bibinfo {author} {\bibfnamefont {H.~M.}\ \bibnamefont
			{Gibbs}},\ }\bibfield  {title} {\bibinfo {title} {Systematic investigation of
			terahertz-induced excitonic rabi splitting},\ }\href
	{https://doi.org/10.1103/physrevb.89.115311} {\bibfield  {journal} {\bibinfo
			{journal} {Phys. Rev. B}\ }\textbf {\bibinfo {volume} {89}},\ \bibinfo
		{pages} {115311} (\bibinfo {year} {2014})}\BibitemShut {NoStop}%
	\bibitem [{\citenamefont {Ulbricht}\ \emph {et~al.}(2011)\citenamefont
		{Ulbricht}, \citenamefont {Hendry}, \citenamefont {Shan}, \citenamefont
		{Heinz},\ and\ \citenamefont {Bonn}}]{ulbricht-etal.11}%
	\BibitemOpen
	\bibfield  {author} {\bibinfo {author} {\bibfnamefont {R.}~\bibnamefont
			{Ulbricht}}, \bibinfo {author} {\bibfnamefont {E.}~\bibnamefont {Hendry}},
		\bibinfo {author} {\bibfnamefont {J.}~\bibnamefont {Shan}}, \bibinfo {author}
		{\bibfnamefont {T.~F.}\ \bibnamefont {Heinz}},\ and\ \bibinfo {author}
		{\bibfnamefont {M.}~\bibnamefont {Bonn}},\ }\bibfield  {title} {\bibinfo
		{title} {Carrier dynamics in semiconductors studied with time-resolved
			terahertz spectroscopy},\ }\href {https://doi.org/10.1103/revmodphys.83.543}
	{\bibfield  {journal} {\bibinfo  {journal} {Rev. Mod. Phys.}\ }\textbf
		{\bibinfo {volume} {83}},\ \bibinfo {pages} {543} (\bibinfo {year}
		{2011})}\BibitemShut {NoStop}%
	\bibitem [{\citenamefont {Kampfrath}\ \emph {et~al.}(2013)\citenamefont
		{Kampfrath}, \citenamefont {Tanaka},\ and\ \citenamefont
		{Nelson}}]{kampfrath-etal.13}%
	\BibitemOpen
	\bibfield  {author} {\bibinfo {author} {\bibfnamefont {T.}~\bibnamefont
			{Kampfrath}}, \bibinfo {author} {\bibfnamefont {K.}~\bibnamefont {Tanaka}},\
		and\ \bibinfo {author} {\bibfnamefont {K.~A.}\ \bibnamefont {Nelson}},\
	}\bibfield  {title} {\bibinfo {title} {Resonant and nonresonant control over
			matter and light by intense terahertz transients},\ }\href
	{https://doi.org/10.1038/nphoton.2013.184} {\bibfield  {journal} {\bibinfo
			{journal} {Nat. Photonics}\ }\textbf {\bibinfo {volume} {7}},\ \bibinfo
		{pages} {680} (\bibinfo {year} {2013})}\BibitemShut {NoStop}%
	\bibitem [{\citenamefont {Maag}\ \emph {et~al.}(2016)\citenamefont {Maag},
		\citenamefont {Bayer}, \citenamefont {Baierl}, \citenamefont {Hohenleutner},
		\citenamefont {Korn}, \citenamefont {Sch{\"u}ller}, \citenamefont {Schuh},
		\citenamefont {Bougeard}, \citenamefont {Lange}, \citenamefont {Huber},
		\citenamefont {Mootz}, \citenamefont {Sipe}, \citenamefont {Koch},\ and\
		\citenamefont {Kira}}]{maag-etal.16}%
	\BibitemOpen
	\bibfield  {author} {\bibinfo {author} {\bibfnamefont {T.}~\bibnamefont
			{Maag}}, \bibinfo {author} {\bibfnamefont {A.}~\bibnamefont {Bayer}},
		\bibinfo {author} {\bibfnamefont {S.}~\bibnamefont {Baierl}}, \bibinfo
		{author} {\bibfnamefont {M.}~\bibnamefont {Hohenleutner}}, \bibinfo {author}
		{\bibfnamefont {T.}~\bibnamefont {Korn}}, \bibinfo {author} {\bibfnamefont
			{C.}~\bibnamefont {Sch{\"u}ller}}, \bibinfo {author} {\bibfnamefont
			{D.}~\bibnamefont {Schuh}}, \bibinfo {author} {\bibfnamefont
			{D.}~\bibnamefont {Bougeard}}, \bibinfo {author} {\bibfnamefont
			{C.}~\bibnamefont {Lange}}, \bibinfo {author} {\bibfnamefont
			{R.}~\bibnamefont {Huber}}, \bibinfo {author} {\bibfnamefont
			{M.}~\bibnamefont {Mootz}}, \bibinfo {author} {\bibfnamefont {J.~E.}\
			\bibnamefont {Sipe}}, \bibinfo {author} {\bibfnamefont {S.~W.}\ \bibnamefont
			{Koch}},\ and\ \bibinfo {author} {\bibfnamefont {M.}~\bibnamefont {Kira}},\
	}\bibfield  {title} {\bibinfo {title} {Coherent cyclotron motion beyond
			{Kohn}'s theorem},\ }\href {https://doi.org/10.1038/nphys3559} {\bibfield
		{journal} {\bibinfo  {journal} {Nat. Phys.}\ }\textbf {\bibinfo {volume}
			{12}},\ \bibinfo {pages} {119} (\bibinfo {year} {2016})}\BibitemShut
	{NoStop}%
	\bibitem [{\citenamefont {Virk}\ and\ \citenamefont
		{Sipe}(2011)}]{virk-sipe.11}%
	\BibitemOpen
	\bibfield  {author} {\bibinfo {author} {\bibfnamefont {K.~S.}\ \bibnamefont
			{Virk}}\ and\ \bibinfo {author} {\bibfnamefont {J.~E.}\ \bibnamefont
			{Sipe}},\ }\bibfield  {title} {\bibinfo {title} {Optical {Injection} and
			{Terahertz} {Detection} of the {Macroscopic} {Berry} {Curvature}},\ }\href
	{https://doi.org/10.1103/PhysRevLett.107.120403} {\bibfield  {journal}
		{\bibinfo  {journal} {Phys. Rev. Lett.}\ }\textbf {\bibinfo {volume} {107}},\
		\bibinfo {pages} {120403} (\bibinfo {year} {2011})}\BibitemShut {NoStop}%
	\bibitem [{\citenamefont {Dawlaty}\ \emph {et~al.}(2008)\citenamefont
		{Dawlaty}, \citenamefont {Shivaraman}, \citenamefont {Strait}, \citenamefont
		{George}, \citenamefont {Chandrashekhar}, \citenamefont {Rana}, \citenamefont
		{Spencer}, \citenamefont {Veksler},\ and\ \citenamefont
		{Chen}}]{dawlaty-etal.08}%
	\BibitemOpen
	\bibfield  {author} {\bibinfo {author} {\bibfnamefont {J.~M.}\ \bibnamefont
			{Dawlaty}}, \bibinfo {author} {\bibfnamefont {S.}~\bibnamefont {Shivaraman}},
		\bibinfo {author} {\bibfnamefont {J.}~\bibnamefont {Strait}}, \bibinfo
		{author} {\bibfnamefont {P.}~\bibnamefont {George}}, \bibinfo {author}
		{\bibfnamefont {M.}~\bibnamefont {Chandrashekhar}}, \bibinfo {author}
		{\bibfnamefont {F.}~\bibnamefont {Rana}}, \bibinfo {author} {\bibfnamefont
			{M.~G.}\ \bibnamefont {Spencer}}, \bibinfo {author} {\bibfnamefont
			{D.}~\bibnamefont {Veksler}},\ and\ \bibinfo {author} {\bibfnamefont
			{Y.}~\bibnamefont {Chen}},\ }\bibfield  {title} {\bibinfo {title}
		{Measurement of the optical absorption spectra of epitaxial graphene from
			terahertz to visible},\ }\href {https://doi.org/10.1063/1.2990753} {\bibfield
		{journal} {\bibinfo  {journal} {Appl. Phys. Lett.}\ }\textbf {\bibinfo
			{volume} {93}},\ \bibinfo {pages} {131905} (\bibinfo {year}
		{2008})}\BibitemShut {NoStop}%
	\bibitem [{\citenamefont {Rao}\ and\ \citenamefont {Sipe}(2014)}]{rao-sipe.14}%
	\BibitemOpen
	\bibfield  {author} {\bibinfo {author} {\bibfnamefont {K.~M.}\ \bibnamefont
			{Rao}}\ and\ \bibinfo {author} {\bibfnamefont {J.~E.}\ \bibnamefont {Sipe}},\
	}\bibfield  {title} {\bibinfo {title} {Terahertz radiation as a probe of the
			dynamics of coherently injected photocurrents in quantum well and graphene
			systems},\ }\href {https://doi.org/10.1103/PhysRevB.90.155313} {\bibfield
		{journal} {\bibinfo  {journal} {Phys. Rev. B}\ }\textbf {\bibinfo {volume}
			{90}},\ \bibinfo {pages} {155313} (\bibinfo {year} {2014})}\BibitemShut
	{NoStop}%
	\bibitem [{\citenamefont {Wietzke}\ \emph {et~al.}(2009)\citenamefont
		{Wietzke}, \citenamefont {Jansen}, \citenamefont {Jung}, \citenamefont
		{Reuter}, \citenamefont {Baudrit}, \citenamefont {Bastian}, \citenamefont
		{Chatterjee},\ and\ \citenamefont {Koch}}]{wietzke-etal.09}%
	\BibitemOpen
	\bibfield  {author} {\bibinfo {author} {\bibfnamefont {S.}~\bibnamefont
			{Wietzke}}, \bibinfo {author} {\bibfnamefont {C.}~\bibnamefont {Jansen}},
		\bibinfo {author} {\bibfnamefont {T.}~\bibnamefont {Jung}}, \bibinfo {author}
		{\bibfnamefont {M.}~\bibnamefont {Reuter}}, \bibinfo {author} {\bibfnamefont
			{B.}~\bibnamefont {Baudrit}}, \bibinfo {author} {\bibfnamefont
			{M.}~\bibnamefont {Bastian}}, \bibinfo {author} {\bibfnamefont
			{S.}~\bibnamefont {Chatterjee}},\ and\ \bibinfo {author} {\bibfnamefont
			{M.}~\bibnamefont {Koch}},\ }\bibfield  {title} {\bibinfo {title} {Terahertz
			time-domain spectroscopy as a tool to monitor the glass transition in
			polymers},\ }\href {https://doi.org/10.1364/OE.17.019006} {\bibfield
		{journal} {\bibinfo  {journal} {Opt. Express}\ }\textbf {\bibinfo {volume}
			{17}},\ \bibinfo {pages} {19006} (\bibinfo {year} {2009})}\BibitemShut
	{NoStop}%
	\bibitem [{\citenamefont {Kira}\ and\ \citenamefont
		{Koch}(2004)}]{kira-koch.04}%
	\BibitemOpen
	\bibfield  {author} {\bibinfo {author} {\bibfnamefont {M.}~\bibnamefont
			{Kira}}\ and\ \bibinfo {author} {\bibfnamefont {S.}~\bibnamefont {Koch}},\
	}\bibfield  {title} {\bibinfo {title} {Exciton-population inversion and
			terahertz gain in semiconductors excited to resonance},\ }\href
	{https://doi.org/10.1103/physrevlett.93.076402} {\bibfield  {journal}
		{\bibinfo  {journal} {Phys. Rev. Lett.}\ }\textbf {\bibinfo {volume} {93}},\
		\bibinfo {pages} {076402} (\bibinfo {year} {2004})}\BibitemShut {NoStop}%
	\bibitem [{\citenamefont {Huber}\ \emph {et~al.}(2006)\citenamefont {Huber},
		\citenamefont {Schmid}, \citenamefont {Shen}, \citenamefont {Chemla},\ and\
		\citenamefont {Kaindl}}]{huber-etal.06}%
	\BibitemOpen
	\bibfield  {author} {\bibinfo {author} {\bibfnamefont {R.}~\bibnamefont
			{Huber}}, \bibinfo {author} {\bibfnamefont {B.~A.}\ \bibnamefont {Schmid}},
		\bibinfo {author} {\bibfnamefont {Y.~R.}\ \bibnamefont {Shen}}, \bibinfo
		{author} {\bibfnamefont {D.~S.}\ \bibnamefont {Chemla}},\ and\ \bibinfo
		{author} {\bibfnamefont {R.~A.}\ \bibnamefont {Kaindl}},\ }\bibfield  {title}
	{\bibinfo {title} {Stimulated {Terahertz} {Emission} from {Intraexcitonic}
			{Transitions} in {$\mathrm{Cu}_{2} \mathrm{O}$}},\ }\href
	{https://doi.org/10.1103/PhysRevLett.96.017402} {\bibfield  {journal}
		{\bibinfo  {journal} {Phys. Rev. Lett.}\ }\textbf {\bibinfo {volume} {96}},\
		\bibinfo {pages} {017402} (\bibinfo {year} {2006})}\BibitemShut {NoStop}%
	\bibitem [{\citenamefont {Kavokin}\ \emph {et~al.}(2010)\citenamefont
		{Kavokin}, \citenamefont {Kaliteevski}, \citenamefont {Abram}, \citenamefont
		{Kavokin}, \citenamefont {Sharkova},\ and\ \citenamefont
		{Shelykh}}]{kavokin-etal.10}%
	\BibitemOpen
	\bibfield  {author} {\bibinfo {author} {\bibfnamefont {K.~V.}\ \bibnamefont
			{Kavokin}}, \bibinfo {author} {\bibfnamefont {M.~A.}\ \bibnamefont
			{Kaliteevski}}, \bibinfo {author} {\bibfnamefont {R.~A.}\ \bibnamefont
			{Abram}}, \bibinfo {author} {\bibfnamefont {A.~V.}\ \bibnamefont {Kavokin}},
		\bibinfo {author} {\bibfnamefont {S.}~\bibnamefont {Sharkova}},\ and\
		\bibinfo {author} {\bibfnamefont {I.~A.}\ \bibnamefont {Shelykh}},\
	}\bibfield  {title} {\bibinfo {title} {Stimulated emission of terahertz
			radiation by exciton-polariton lasers},\ }\href
	{https://doi.org/10.1063/1.3519978} {\bibfield  {journal} {\bibinfo
			{journal} {Appl. Phys. Lett.}\ }\textbf {\bibinfo {volume} {97}},\ \bibinfo
		{pages} {201111} (\bibinfo {year} {2010})}\BibitemShut {NoStop}%
	\bibitem [{\citenamefont {del Valle}\ and\ \citenamefont
		{Kavokin}(2011)}]{valle-kavokin.11}%
	\BibitemOpen
	\bibfield  {author} {\bibinfo {author} {\bibfnamefont {E.}~\bibnamefont {del
				Valle}}\ and\ \bibinfo {author} {\bibfnamefont {A.}~\bibnamefont {Kavokin}},\
	}\bibfield  {title} {\bibinfo {title} {Terahertz lasing in a polariton
			system: {Quantum} theory},\ }\href
	{https://doi.org/10.1103/PhysRevB.83.193303} {\bibfield  {journal} {\bibinfo
			{journal} {Phys. Rev. B}\ }\textbf {\bibinfo {volume} {83}},\ \bibinfo
		{pages} {193303} (\bibinfo {year} {2011})}\BibitemShut {NoStop}%
	\bibitem [{\citenamefont {Savenko}\ \emph {et~al.}(2011)\citenamefont
		{Savenko}, \citenamefont {Shelykh},\ and\ \citenamefont
		{Kaliteevski}}]{savenko-etal.11}%
	\BibitemOpen
	\bibfield  {author} {\bibinfo {author} {\bibfnamefont {I.~G.}\ \bibnamefont
			{Savenko}}, \bibinfo {author} {\bibfnamefont {I.~A.}\ \bibnamefont
			{Shelykh}},\ and\ \bibinfo {author} {\bibfnamefont {M.~A.}\ \bibnamefont
			{Kaliteevski}},\ }\bibfield  {title} {\bibinfo {title} {Nonlinear {Terahertz}
			{Emission} in {Semiconductor} {Microcavities}},\ }\href
	{https://doi.org/10.1103/PhysRevLett.107.027401} {\bibfield  {journal}
		{\bibinfo  {journal} {Phys. Rev. Lett.}\ }\textbf {\bibinfo {volume} {107}},\
		\bibinfo {pages} {027401} (\bibinfo {year} {2011})}\BibitemShut {NoStop}%
	\bibitem [{\citenamefont {Kavokin}\ \emph {et~al.}(2012)\citenamefont
		{Kavokin}, \citenamefont {Shelykh}, \citenamefont {Taylor},\ and\
		\citenamefont {Glazov}}]{kavokin-etal.12}%
	\BibitemOpen
	\bibfield  {author} {\bibinfo {author} {\bibfnamefont {A.~V.}\ \bibnamefont
			{Kavokin}}, \bibinfo {author} {\bibfnamefont {I.~A.}\ \bibnamefont
			{Shelykh}}, \bibinfo {author} {\bibfnamefont {T.}~\bibnamefont {Taylor}},\
		and\ \bibinfo {author} {\bibfnamefont {M.~M.}\ \bibnamefont {Glazov}},\
	}\bibfield  {title} {\bibinfo {title} {Vertical {Cavity} {Surface} {Emitting}
			{Terahertz} {Laser}},\ }\href
	{https://doi.org/10.1103/PhysRevLett.108.197401} {\bibfield  {journal}
		{\bibinfo  {journal} {Phys. Rev. Lett.}\ }\textbf {\bibinfo {volume} {108}},\
		\bibinfo {pages} {197401} (\bibinfo {year} {2012})}\BibitemShut {NoStop}%
	\bibitem [{\citenamefont {Tomaino}\ \emph {et~al.}(2012)\citenamefont
		{Tomaino}, \citenamefont {Jameson}, \citenamefont {Lee}, \citenamefont
		{Khitrova}, \citenamefont {Gibbs}, \citenamefont {Klettke}, \citenamefont
		{Kira},\ and\ \citenamefont {Koch}}]{tomaino-etal.12}%
	\BibitemOpen
	\bibfield  {author} {\bibinfo {author} {\bibfnamefont {J.~L.}\ \bibnamefont
			{Tomaino}}, \bibinfo {author} {\bibfnamefont {A.~D.}\ \bibnamefont
			{Jameson}}, \bibinfo {author} {\bibfnamefont {Y.-S.}\ \bibnamefont {Lee}},
		\bibinfo {author} {\bibfnamefont {G.}~\bibnamefont {Khitrova}}, \bibinfo
		{author} {\bibfnamefont {H.~M.}\ \bibnamefont {Gibbs}}, \bibinfo {author}
		{\bibfnamefont {A.~C.}\ \bibnamefont {Klettke}}, \bibinfo {author}
		{\bibfnamefont {M.}~\bibnamefont {Kira}},\ and\ \bibinfo {author}
		{\bibfnamefont {S.~W.}\ \bibnamefont {Koch}},\ }\bibfield  {title} {\bibinfo
		{title} {Terahertz excitation of a coherent $\ensuremath{\Lambda}$-type
			three-level system of exciton-polariton modes in a quantum-well
			microcavity},\ }\href {https://doi.org/10.1103/PhysRevLett.108.267402}
	{\bibfield  {journal} {\bibinfo  {journal} {Phys. Rev. Lett.}\ }\textbf
		{\bibinfo {volume} {108}},\ \bibinfo {pages} {267402} (\bibinfo {year}
		{2012})}\BibitemShut {NoStop}%
	\bibitem [{\citenamefont {De~Liberato}\ \emph {et~al.}(2013)\citenamefont
		{De~Liberato}, \citenamefont {Ciuti},\ and\ \citenamefont
		{Phillips}}]{deliberato-etal.13}%
	\BibitemOpen
	\bibfield  {author} {\bibinfo {author} {\bibfnamefont {S.}~\bibnamefont
			{De~Liberato}}, \bibinfo {author} {\bibfnamefont {C.}~\bibnamefont {Ciuti}},\
		and\ \bibinfo {author} {\bibfnamefont {C.~C.}\ \bibnamefont {Phillips}},\
	}\bibfield  {title} {\bibinfo {title} {Terahertz lasing from intersubband
			polariton-polariton scattering in asymmetric quantum wells},\ }\href
	{https://doi.org/10.1103/PhysRevB.87.241304} {\bibfield  {journal} {\bibinfo
			{journal} {Phys. Rev. B}\ }\textbf {\bibinfo {volume} {87}},\ \bibinfo
		{pages} {241304} (\bibinfo {year} {2013})}\BibitemShut {NoStop}%
	\bibitem [{\citenamefont {Schmutzler}\ \emph {et~al.}(2014)\citenamefont
		{Schmutzler}, \citenamefont {A{\ss}mann}, \citenamefont {Czerniuk},
		\citenamefont {Kamp}, \citenamefont {Schneider}, \citenamefont
		{H{\"o}fling},\ and\ \citenamefont {Bayer}}]{schmutzler-etal.14}%
	\BibitemOpen
	\bibfield  {author} {\bibinfo {author} {\bibfnamefont {J.}~\bibnamefont
			{Schmutzler}}, \bibinfo {author} {\bibfnamefont {M.}~\bibnamefont
			{A{\ss}mann}}, \bibinfo {author} {\bibfnamefont {T.}~\bibnamefont
			{Czerniuk}}, \bibinfo {author} {\bibfnamefont {M.}~\bibnamefont {Kamp}},
		\bibinfo {author} {\bibfnamefont {C.}~\bibnamefont {Schneider}}, \bibinfo
		{author} {\bibfnamefont {S.}~\bibnamefont {H{\"o}fling}},\ and\ \bibinfo
		{author} {\bibfnamefont {M.}~\bibnamefont {Bayer}},\ }\bibfield  {title}
	{\bibinfo {title} {Nonlinear spectroscopy of exciton-polaritons in a
			{GaAs}-based microcavity},\ }\href
	{https://doi.org/10.1103/PhysRevB.90.075103} {\bibfield  {journal} {\bibinfo
			{journal} {Phys. Rev. B}\ }\textbf {\bibinfo {volume} {90}},\ \bibinfo
		{pages} {075103} (\bibinfo {year} {2014})}\BibitemShut {NoStop}%
	\bibitem [{\citenamefont {Huppert}\ \emph {et~al.}(2014)\citenamefont
		{Huppert}, \citenamefont {Lafont}, \citenamefont {Baudin}, \citenamefont
		{Tignon},\ and\ \citenamefont {Ferreira}}]{huppert-etal.14}%
	\BibitemOpen
	\bibfield  {author} {\bibinfo {author} {\bibfnamefont {S.}~\bibnamefont
			{Huppert}}, \bibinfo {author} {\bibfnamefont {O.}~\bibnamefont {Lafont}},
		\bibinfo {author} {\bibfnamefont {E.}~\bibnamefont {Baudin}}, \bibinfo
		{author} {\bibfnamefont {J.}~\bibnamefont {Tignon}},\ and\ \bibinfo {author}
		{\bibfnamefont {R.}~\bibnamefont {Ferreira}},\ }\bibfield  {title} {\bibinfo
		{title} {Terahertz emission from multiple-microcavity exciton-polariton
			lasers},\ }\href {https://doi.org/10.1103/PhysRevB.90.241302} {\bibfield
		{journal} {\bibinfo  {journal} {Phys. Rev. B}\ }\textbf {\bibinfo {volume}
			{90}},\ \bibinfo {pages} {241302} (\bibinfo {year} {2014})}\BibitemShut
	{NoStop}%
	\bibitem [{\citenamefont {Lem{\'e}nager}\ \emph {et~al.}(2014)\citenamefont
		{Lem{\'e}nager}, \citenamefont {Pisanello}, \citenamefont {Bloch},
		\citenamefont {Kavokin}, \citenamefont {Amo}, \citenamefont {Lemaitre},
		\citenamefont {Galopin}, \citenamefont {Sagnes}, \citenamefont {Vittorio},
		\citenamefont {Giacobino},\ and\ \citenamefont
		{Bramati}}]{lemenager-etal.14}%
	\BibitemOpen
	\bibfield  {author} {\bibinfo {author} {\bibfnamefont {G.}~\bibnamefont
			{Lem{\'e}nager}}, \bibinfo {author} {\bibfnamefont {F.}~\bibnamefont
			{Pisanello}}, \bibinfo {author} {\bibfnamefont {J.}~\bibnamefont {Bloch}},
		\bibinfo {author} {\bibfnamefont {A.}~\bibnamefont {Kavokin}}, \bibinfo
		{author} {\bibfnamefont {A.}~\bibnamefont {Amo}}, \bibinfo {author}
		{\bibfnamefont {A.}~\bibnamefont {Lemaitre}}, \bibinfo {author}
		{\bibfnamefont {E.}~\bibnamefont {Galopin}}, \bibinfo {author} {\bibfnamefont
			{I.}~\bibnamefont {Sagnes}}, \bibinfo {author} {\bibfnamefont {M.~D.}\
			\bibnamefont {Vittorio}}, \bibinfo {author} {\bibfnamefont {E.}~\bibnamefont
			{Giacobino}},\ and\ \bibinfo {author} {\bibfnamefont {A.}~\bibnamefont
			{Bramati}},\ }\bibfield  {title} {\bibinfo {title} {Two-photon injection of
			polaritons in semiconductor microstructures},\ }\href
	{https://doi.org/10.1364/OL.39.000307} {\bibfield  {journal} {\bibinfo
			{journal} {Opt. Lett.}\ }\textbf {\bibinfo {volume} {39}},\ \bibinfo {pages}
		{307} (\bibinfo {year} {2014})}\BibitemShut {NoStop}%
	\bibitem [{\citenamefont {Barachati}\ \emph {et~al.}(2015)\citenamefont
		{Barachati}, \citenamefont {De~Liberato},\ and\ \citenamefont
		{K{\'{e}}na-Cohen}}]{barachati-etal.15}%
	\BibitemOpen
	\bibfield  {author} {\bibinfo {author} {\bibfnamefont {F.}~\bibnamefont
			{Barachati}}, \bibinfo {author} {\bibfnamefont {S.}~\bibnamefont
			{De~Liberato}},\ and\ \bibinfo {author} {\bibfnamefont {S.}~\bibnamefont
			{K{\'{e}}na-Cohen}},\ }\bibfield  {title} {\bibinfo {title} {Generation of
			{Rabi}-frequency radiation using exciton-polaritons},\ }\href
	{https://doi.org/10.1103/PhysRevA.92.033828} {\bibfield  {journal} {\bibinfo
			{journal} {Phys. Rev. A}\ }\textbf {\bibinfo {volume} {92}},\ \bibinfo
		{pages} {033828} (\bibinfo {year} {2015})}\BibitemShut {NoStop}%
	\bibitem [{\citenamefont {Kibis}\ \emph {et~al.}(2009)\citenamefont {Kibis},
		\citenamefont {Slepyan}, \citenamefont {Maksimenko},\ and\ \citenamefont
		{Hoffmann}}]{kibis-etal.09}%
	\BibitemOpen
	\bibfield  {author} {\bibinfo {author} {\bibfnamefont {O.~V.}\ \bibnamefont
			{Kibis}}, \bibinfo {author} {\bibfnamefont {G.~Y.}\ \bibnamefont {Slepyan}},
		\bibinfo {author} {\bibfnamefont {S.~A.}\ \bibnamefont {Maksimenko}},\ and\
		\bibinfo {author} {\bibfnamefont {A.}~\bibnamefont {Hoffmann}},\ }\bibfield
	{title} {\bibinfo {title} {Matter {Coupling} to {Strong} {Electromagnetic}
			{Fields} in {Two}-{Level} {Quantum} {Systems} with {Broken} {Inversion}
			{Symmetry}},\ }\href {https://doi.org/10.1103/PhysRevLett.102.023601}
	{\bibfield  {journal} {\bibinfo  {journal} {Phys. Rev. Lett.}\ }\textbf
		{\bibinfo {volume} {102}},\ \bibinfo {pages} {023601} (\bibinfo {year}
		{2009})}\BibitemShut {NoStop}%
	\bibitem [{\citenamefont {Shammah}\ \emph {et~al.}(2014)\citenamefont
		{Shammah}, \citenamefont {Phillips},\ and\ \citenamefont
		{De~Liberato}}]{shammah-etal.14}%
	\BibitemOpen
	\bibfield  {author} {\bibinfo {author} {\bibfnamefont {N.}~\bibnamefont
			{Shammah}}, \bibinfo {author} {\bibfnamefont {C.~C.}\ \bibnamefont
			{Phillips}},\ and\ \bibinfo {author} {\bibfnamefont {S.}~\bibnamefont
			{De~Liberato}},\ }\bibfield  {title} {\bibinfo {title} {Terahertz emission
			from ac {Stark}-split asymmetric intersubband transitions},\ }\href
	{https://doi.org/10.1103/PhysRevB.89.235309} {\bibfield  {journal} {\bibinfo
			{journal} {Phys. Rev. B}\ }\textbf {\bibinfo {volume} {89}},\ \bibinfo
		{pages} {235309} (\bibinfo {year} {2014})}\BibitemShut {NoStop}%
	\bibitem [{\citenamefont {Chestnov}\ \emph {et~al.}(2017)\citenamefont
		{Chestnov}, \citenamefont {Shahnazaryan}, \citenamefont {Alodjants},\ and\
		\citenamefont {Shelykh}}]{chestnov-etal.17}%
	\BibitemOpen
	\bibfield  {author} {\bibinfo {author} {\bibfnamefont {I.~Y.}\ \bibnamefont
			{Chestnov}}, \bibinfo {author} {\bibfnamefont {V.~A.}\ \bibnamefont
			{Shahnazaryan}}, \bibinfo {author} {\bibfnamefont {A.~P.}\ \bibnamefont
			{Alodjants}},\ and\ \bibinfo {author} {\bibfnamefont {I.~A.}\ \bibnamefont
			{Shelykh}},\ }\bibfield  {title} {\bibinfo {title} {Terahertz {Lasing} in
			{Ensemble} of {Asymmetric} {Quantum} {Dots}},\ }\href
	{https://doi.org/10.1021/acsphotonics.7b00575} {\bibfield  {journal}
		{\bibinfo  {journal} {ACS Photonics}\ }\textbf {\bibinfo {volume} {4}},\
		\bibinfo {pages} {2726} (\bibinfo {year} {2017})}\BibitemShut {NoStop}%
	\bibitem [{\citenamefont {De~Liberato}(2018)}]{deliberato.18}%
	\BibitemOpen
	\bibfield  {author} {\bibinfo {author} {\bibfnamefont {S.}~\bibnamefont
			{De~Liberato}},\ }\bibfield  {title} {\bibinfo {title} {Lasing from dressed
			dots},\ }\href {https://doi.org/10.1038/s41566-017-0074-3} {\bibfield
		{journal} {\bibinfo  {journal} {Nat. Photonics}\ }\textbf {\bibinfo {volume}
			{12}},\ \bibinfo {pages} {4} (\bibinfo {year} {2018})}\BibitemShut {NoStop}%
	\bibitem [{\citenamefont {Mandal}\ \emph {et~al.}(2019)\citenamefont {Mandal},
		\citenamefont {Dini}, \citenamefont {Kibis},\ and\ \citenamefont
		{Liew}}]{mandal-etal.19}%
	\BibitemOpen
	\bibfield  {author} {\bibinfo {author} {\bibfnamefont {S.}~\bibnamefont
			{Mandal}}, \bibinfo {author} {\bibfnamefont {K.}~\bibnamefont {Dini}},
		\bibinfo {author} {\bibfnamefont {O.~V.}\ \bibnamefont {Kibis}},\ and\
		\bibinfo {author} {\bibfnamefont {T.~C.~H.}\ \bibnamefont {Liew}},\
	}\bibfield  {title} {\bibinfo {title} {On the possibility of a terahertz
			light emitting diode based on a dressed quantum well},\ }\href
	{https://doi.org/10.1038/s41598-019-52704-6} {\bibfield  {journal} {\bibinfo
			{journal} {Sci. Rep.}\ }\textbf {\bibinfo {volume} {9}},\ \bibinfo {pages}
		{16320} (\bibinfo {year} {2019})}\BibitemShut {NoStop}%
	\bibitem [{\citenamefont {Gu}\ \emph {et~al.}(2013)\citenamefont {Gu},
		\citenamefont {Kwong},\ and\ \citenamefont {Binder}}]{gu-etal.13}%
	\BibitemOpen
	\bibfield  {author} {\bibinfo {author} {\bibfnamefont {B.}~\bibnamefont
			{Gu}}, \bibinfo {author} {\bibfnamefont {N.}~\bibnamefont {Kwong}},\ and\
		\bibinfo {author} {\bibfnamefont {R.}~\bibnamefont {Binder}},\ }\bibfield
	{title} {\bibinfo {title} {Relation between the interband dipole and momentum
			matrix elements in semiconductors},\ }\href
	{https://doi.org/10.1103/physrevb.87.125301} {\bibfield  {journal} {\bibinfo
			{journal} {Phys. Rev. B}\ }\textbf {\bibinfo {volume} {87}},\ \bibinfo
		{pages} {125301} (\bibinfo {year} {2013})}\BibitemShut {NoStop}%
	\bibitem [{\citenamefont {Mahon}\ \emph {et~al.}(2019)\citenamefont {Mahon},
		\citenamefont {Muniz},\ and\ \citenamefont {Sipe}}]{mahon-etal.19}%
	\BibitemOpen
	\bibfield  {author} {\bibinfo {author} {\bibfnamefont {P.~T.}\ \bibnamefont
			{Mahon}}, \bibinfo {author} {\bibfnamefont {R.~A.}\ \bibnamefont {Muniz}},\
		and\ \bibinfo {author} {\bibfnamefont {J.~E.}\ \bibnamefont {Sipe}},\
	}\bibfield  {title} {\bibinfo {title} {Microscopic polarization and
			magnetization fields in extended systems},\ }\href
	{https://doi.org/10.1103/PhysRevB.99.235140} {\bibfield  {journal} {\bibinfo
			{journal} {Phys. Rev. B}\ }\textbf {\bibinfo {volume} {99}},\ \bibinfo
		{pages} {235140} (\bibinfo {year} {2019})}\BibitemShut {NoStop}%
	\bibitem [{\citenamefont {Steiner}(2008)}]{steiner.08}%
	\BibitemOpen
	\bibfield  {author} {\bibinfo {author} {\bibfnamefont {J.~T.}\ \bibnamefont
			{Steiner}},\ }\emph {\bibinfo {title} {Microscopic Theory of Linear and
			Nonlinear Terahertz Spectroscopy of Semiconductors}},\ \href
	{https://doi.org/https://doi.org/10.17192/z2009.0275} {Ph.D. thesis},\
	\bibinfo  {school} {Philipps-Universitat Marburg} (\bibinfo {year}
	{2008})\BibitemShut {NoStop}%
	\bibitem [{\citenamefont {Jahnke}\ and\ \citenamefont
		{Henneberger}(1992)}]{jahnke-henneberger.92}%
	\BibitemOpen
	\bibfield  {author} {\bibinfo {author} {\bibfnamefont {F.}~\bibnamefont
			{Jahnke}}\ and\ \bibinfo {author} {\bibfnamefont {K.}~\bibnamefont
			{Henneberger}},\ }\bibfield  {title} {\bibinfo {title} {Light-induced effects
			in the interband absorption of semiconductors},\ }\href
	{https://doi.org/10.1103/PhysRevB.45.4077} {\bibfield  {journal} {\bibinfo
			{journal} {Phys. Rev. B}\ }\textbf {\bibinfo {volume} {45}},\ \bibinfo
		{pages} {4077} (\bibinfo {year} {1992})}\BibitemShut {NoStop}%
	\bibitem [{\citenamefont {Rammer}(2007)}]{rammer.07}%
	\BibitemOpen
	\bibfield  {author} {\bibinfo {author} {\bibfnamefont {J.}~\bibnamefont
			{Rammer}},\ }\href {https://doi.org/10.1017/CBO9780511618956} {\emph
		{\bibinfo {title} {Quantum Field Theory of Non-Equilibrium States}}}\
	(\bibinfo  {publisher} {Cambridge University Press},\ \bibinfo {address}
	{Cambridge, UK},\ \bibinfo {year} {2007})\BibitemShut {NoStop}%
	\bibitem [{\citenamefont {Sernelius}(1991)}]{sernelius.91}%
	\BibitemOpen
	\bibfield  {author} {\bibinfo {author} {\bibfnamefont {B.~E.}\ \bibnamefont
			{Sernelius}},\ }\bibfield  {title} {\bibinfo {title} {Intraband relaxation
			time in highly excited semiconductors},\ }\href
	{https://doi.org/10.1103/PhysRevB.43.7136} {\bibfield  {journal} {\bibinfo
			{journal} {Phys. Rev. B}\ }\textbf {\bibinfo {volume} {43}},\ \bibinfo
		{pages} {7136} (\bibinfo {year} {1991})}\BibitemShut {NoStop}%
	\bibitem [{\citenamefont {Leitenstorfer}\ \emph {et~al.}(2000)\citenamefont
		{Leitenstorfer}, \citenamefont {Hunsche}, \citenamefont {Shah}, \citenamefont
		{Nuss},\ and\ \citenamefont {Knox}}]{leitenstorfer-etal.00}%
	\BibitemOpen
	\bibfield  {author} {\bibinfo {author} {\bibfnamefont {A.}~\bibnamefont
			{Leitenstorfer}}, \bibinfo {author} {\bibfnamefont {S.}~\bibnamefont
			{Hunsche}}, \bibinfo {author} {\bibfnamefont {J.}~\bibnamefont {Shah}},
		\bibinfo {author} {\bibfnamefont {M.~C.}\ \bibnamefont {Nuss}},\ and\
		\bibinfo {author} {\bibfnamefont {W.~H.}\ \bibnamefont {Knox}},\ }\bibfield
	{title} {\bibinfo {title} {Femtosecond high-field transport in compound
			semiconductors},\ }\href {https://doi.org/10.1103/PhysRevB.61.16642}
	{\bibfield  {journal} {\bibinfo  {journal} {Phys. Rev. B}\ }\textbf {\bibinfo
			{volume} {61}},\ \bibinfo {pages} {16642} (\bibinfo {year}
		{2000})}\BibitemShut {NoStop}%
	\bibitem [{\citenamefont {Beard}\ \emph {et~al.}(2000)\citenamefont {Beard},
		\citenamefont {Turner},\ and\ \citenamefont {Schmuttenmaer}}]{beard-etal.00}%
	\BibitemOpen
	\bibfield  {author} {\bibinfo {author} {\bibfnamefont {M.~C.}\ \bibnamefont
			{Beard}}, \bibinfo {author} {\bibfnamefont {G.~M.}\ \bibnamefont {Turner}},\
		and\ \bibinfo {author} {\bibfnamefont {C.~A.}\ \bibnamefont
			{Schmuttenmaer}},\ }\bibfield  {title} {\bibinfo {title} {Transient
			photoconductivity in {GaAs} as measured by time-resolved terahertz
			spectroscopy},\ }\href {https://doi.org/10.1103/PhysRevB.62.15764} {\bibfield
		{journal} {\bibinfo  {journal} {Phys. Rev. B}\ }\textbf {\bibinfo {volume}
			{62}},\ \bibinfo {pages} {15764} (\bibinfo {year} {2000})}\BibitemShut
	{NoStop}%
	\bibitem [{\citenamefont {Shi}\ \emph {et~al.}(2008)\citenamefont {Shi},
		\citenamefont {Zhou}, \citenamefont {Zhang},\ and\ \citenamefont
		{Jin}}]{shi-etal.08}%
	\BibitemOpen
	\bibfield  {author} {\bibinfo {author} {\bibfnamefont {Y.}~\bibnamefont
			{Shi}}, \bibinfo {author} {\bibfnamefont {Q.-l.}\ \bibnamefont {Zhou}},
		\bibinfo {author} {\bibfnamefont {C.}~\bibnamefont {Zhang}},\ and\ \bibinfo
		{author} {\bibfnamefont {B.}~\bibnamefont {Jin}},\ }\bibfield  {title}
	{\bibinfo {title} {Ultrafast high-field carrier transport in {GaAs} measured
			by femtosecond pump-terahertz probe spectroscopy},\ }\href
	{https://doi.org/10.1063/1.2992067} {\bibfield  {journal} {\bibinfo
			{journal} {Appl. Phys. Lett.}\ }\textbf {\bibinfo {volume} {93}},\ \bibinfo
		{pages} {121115} (\bibinfo {year} {2008})}\BibitemShut {NoStop}%
	\bibitem [{\citenamefont {Press}\ \emph {et~al.}(1992)\citenamefont {Press},
		\citenamefont {Teukolsky}, \citenamefont {Vetterling},\ and\ \citenamefont
		{Flannery}}]{press-etal.92}%
	\BibitemOpen
	\bibfield  {author} {\bibinfo {author} {\bibfnamefont {W.~H.}\ \bibnamefont
			{Press}}, \bibinfo {author} {\bibfnamefont {S.~A.}\ \bibnamefont
			{Teukolsky}}, \bibinfo {author} {\bibfnamefont {W.~T.}\ \bibnamefont
			{Vetterling}},\ and\ \bibinfo {author} {\bibfnamefont {B.~P.}\ \bibnamefont
			{Flannery}},\ }\href {http://numerical.recipes/F210} {\emph {\bibinfo {title}
			{Numerical Recipes in Fortran 77: The Art of Scientific Computing}}},\
	\bibinfo {edition} {2nd}\ ed.,\ \bibinfo {series} {Fortran Numerical
		Recipes}, Vol.~\bibinfo {volume} {1}\ (\bibinfo  {publisher} {Cambridge
		University Press},\ \bibinfo {address} {Cambridge, UK},\ \bibinfo {year}
	{1992})\BibitemShut {NoStop}%
	\bibitem [{\citenamefont {Meystre}\ and\ \citenamefont
		{Sargent}(2007)}]{meystre-sargent.07}%
	\BibitemOpen
	\bibfield  {author} {\bibinfo {author} {\bibfnamefont {P.}~\bibnamefont
			{Meystre}}\ and\ \bibinfo {author} {\bibfnamefont {M.}~\bibnamefont
			{Sargent}},\ }\href {https://doi.org/10.1007/978-3-540-74211-1} {\emph
		{\bibinfo {title} {Elements of Quantum Optics}}}\ (\bibinfo  {publisher}
	{Springer},\ \bibinfo {address} {Berlin, Heidelberg},\ \bibinfo {year}
	{2007})\BibitemShut {NoStop}%
	\bibitem [{\citenamefont {Kalt}\ and\ \citenamefont
		{Klingshirn}(2019)}]{kalt-klingshirn.19}%
	\BibitemOpen
	\bibfield  {author} {\bibinfo {author} {\bibfnamefont {H.}~\bibnamefont
			{Kalt}}\ and\ \bibinfo {author} {\bibfnamefont {C.~F.}\ \bibnamefont
			{Klingshirn}},\ }\href {https://doi.org/10.1007/978-3-030-24152-0} {\emph
		{\bibinfo {title} {Semiconductor Optics 1: Linear Optical Properties of
				Semiconductors}}},\ \bibinfo {edition} {5th}\ ed.,\ Graduate Texts in
	Physics\ (\bibinfo  {publisher} {Springer},\ \bibinfo {address} {Cham,
		Switzerland},\ \bibinfo {year} {2019})\BibitemShut {NoStop}%
	\bibitem [{\citenamefont {Ashcroft}\ and\ \citenamefont
		{Mermin}(1976)}]{ashcroft-mermin.76}%
	\BibitemOpen
	\bibfield  {author} {\bibinfo {author} {\bibfnamefont {N.~W.}\ \bibnamefont
			{Ashcroft}}\ and\ \bibinfo {author} {\bibfnamefont {N.~D.}\ \bibnamefont
			{Mermin}},\ }\href
	{https://www.cengage.com/c/solid-state-physics-1e-ashcroft/9780030839931/}
	{\emph {\bibinfo {title} {Solid state physics}}}\ (\bibinfo  {publisher}
	{Holt, Rinehart and Winston},\ \bibinfo {address} {New York},\ \bibinfo
	{year} {1976})\ \bibinfo {note} {oCLC: 934604}\BibitemShut {NoStop}%
	\bibitem [{\citenamefont {Mihaila}(2011)}]{mihaila.11}%
	\BibitemOpen
	\bibfield  {author} {\bibinfo {author} {\bibfnamefont {B.}~\bibnamefont
			{Mihaila}},\ }\bibfield  {title} {\bibinfo {title} {{Lindhard} function of a
			$d$-dimensional {Fermi} gas},\ }\href@noop {} {\  (\bibinfo {year} {2011})},\
	\Eprint {https://arxiv.org/abs/1111.5337} {arXiv:1111.5337
		[cond-mat.quant-gas]} \BibitemShut {NoStop}%
\end{thebibliography}
\end{document}